\documentclass[journal,a4paper,10pt]{IEEEtran}
\usepackage[T1]{fontenc}
\usepackage[latin9]{inputenc}
\usepackage{epsfig,psfrag}
\usepackage{amsmath,amsfonts,amssymb,amsxtra,bm}
\usepackage{graphicx}
\usepackage{color}
\usepackage{cite}
\usepackage{tabularx}
\usepackage{pst-node}

\newcommand{\eq}{\,=\,}
\newcommand{\be}{\begin{equation}}
\newcommand{\ee}{\end{equation}}
\newcommand{\ist}{\hspace*{.3mm}}
\newcommand{\rmv}{\hspace*{-.3mm}}
\newcommand{\iist}{\hspace*{1mm}}

\newcommand{\bd}[1]{\mathbf{#1}}
\newcommand{\cl}[1]{\mathcal{#1}}
\newcommand{\nn}{\nonumber}

\allowdisplaybreaks

\begin{document}

\title{Distributed Estimation with Information-Seeking\\[-.3mm]Control in Agent
Networks
\thanks{Final manuscript, \today}
\thanks{F.\ Meyer and F.\ Hlawatsch are with the Institute of 
Telecommunications, Vienna University of Technology, 
Vienna, Austria (e-mail: \{florian.meyer, franz.hlawatsch\}@nt.tuwien.ac.at). 
H.\ Wymeersch and M.\ Fr\"ohle are with the Department of Signals and Systems, Chalmers University of Technology, Gothenburg, Sweden 
(e-mail: \{henkw, frohle\}@chalmers.se).
This work was supported by the Austrian Science Fund (FWF) under Grants S10603 and P27370  
and by the European Commission under ERC Grant No. 258418 (COOPNET), EU FP7 Marie Curie
Initial Training Network MULTI-POS under Grant 316528, and the Newcom\# Network of Excellence in Wireless Communications.
Parts of this paper were previously presented at ICASSP 2015, Brisbane, Australia, April 2015.
}
\vspace*{2.5mm}}

\author{Florian Meyer, \emph{Student Member, IEEE}, 
Henk Wymeersch, \emph{Member, IEEE}, 
Markus Fr\"ohle,\\ 
and Franz Hlawatsch, \emph{Fellow, IEEE}
\vspace*{-2mm}
 }

\maketitle

\begin{abstract}
We introduce a distributed, cooperative framework and method for Bayesian estimation and control in decentralized agent networks.
Our framework combines joint estimation of time-varying global and local states 
with information-seeking control optimizing the behavior of the agents. It is suited to nonlinear and non-Gaussian problems
and, in particular, to location-aware networks.
For cooperative estimation, a combination of belief propagation message passing and consensus is used. For cooperative control, the negative posterior joint entropy of all states is maximized via a gradient ascent. The estimation layer provides the control layer with probabilistic information in the form of sample 
representations of probability distributions.
Simulation results demonstrate intelligent behavior of the agents and excellent estimation performance for a simultaneous self-localization and target tracking problem. In a cooperative localization scenario with only one anchor,
mobile agents can localize themselves after a short time with an accuracy that is higher than the accuracy of the performed distance measurements.
\end{abstract}


\section{Introduction}

\subsection{Motivation and State of the Art} \label{sec:motiv}

Recent research on distributed estimation and control in mobile agent networks \cite{shima2009uav,bullo2009distributed,nayak2010wireless,zhao2004wsn} has 
frequently been motivated by location-aware scenarios and problems including environmental and agricultural monitoring \cite{corke10}, healthcare monitoring \cite{ko10}, target tracking \cite{hlinkaMag13}, pollution source localization \cite{zhao07}, chemical plume tracking \cite{nayak2010wireless}, and surveillance \cite{aghajan2009multi}. The agents in a mobile agent network are generally equipped with sensors, wireless communication interfaces, a processing unit, and actuators, all together forming a cyber-physical system \cite{kim12, haykin12} with a tight coupling between sensing, computing, communication, and control.
A common task in mobile agent networks is seeking information, either about external phenomena or about the network itself. This task relies on
\textit{estimation} (quantifying, fusing, and disseminating information) and \textit{control} (configuring the network to increase information). 
A common theme in previous works is the reliance on position information for estimation and/or control.

Estimation methods for mobile agent networks (our focus will be on distributed Bayesian estimation) address estimation of common global states \cite{hlinkaMag13,ryan07,hoffmann10,schwager11,julian12,atanasov14}, estimation of local states \cite{wymeersch,sathyan13,wu11,meyer13sync,etzlinger13asilomar,etzlinger14}, or combined estimation of local and global states \cite{teng2012distr, meyer12, meyer2014coslat}. In the first case, the agents obtain local measurements with respect to external objects or the surrounding environment, which are 
fused across the network. Global fusion methods that require only local communication include consensus \cite{olfati07} and gossip \cite{dimakis10}. 
Example applications are distributed target tracking \cite{hlinkaMag13}, cooperative exploration of the environment \cite{julian12}, and chemical plume tracking \cite{nayak2010wireless}. 
In the second case (estimation of local states), the agents cooperate such that each agent is better able to estimate its own local state. Here, the dimensionality of the total state grows with the network size, which leads to more complex factorizations of the joint posterior probability density function (pdf). 
When the factor graph \cite{kschischang} of this factorization matches the network topology, efficient message passing methods for distributed inference can be used, such as the belief propagation (BP) \cite{kschischang} and mean field \cite{bishop2006pattern} methods.
Example applications are cooperative localization \cite{wymeersch}, synchronization \cite{wu11,etzlinger14}, and simultaneous 
localization and synchronization \cite{meyer13sync,etzlinger13asilomar}. 
In the third case (estimation of both global states and local states), a message passing algorithm can be combined with a network-wide information
dissemination technique. An example application is cooperative simultaneous self-localization and target tracking \cite{teng2012distr, meyer2014coslat}.

In many cooperative estimation scenarios, it is 
advantageous to 
control certain properties of the agent network,
such as the agent positions or the measurement characteristics (``controlled sensing'') \cite{shima2009uav,bullo2009distributed,nayak2010wireless,zhao2004wsn, corke10, ko10, hlinkaMag13, zhao07, aghajan2009multi}. 
In particular, here we will address the problem of combining distributed estimation and distributed control in mobile agent networks. We will
limit our discussion to \emph{information-seeking} control, 
which seeks to maximize the 
information carried by the measurements of all agents about the global and/or local states to be estimated.
The use of information measures for the control of a single agent or a network of agents was introduced in \cite{burgard97} and \cite{grocholsky02}, respectively.
Suitable measures of information include negative posterior entropy \cite{cover06}, mutual information \cite{cover06}, and scalar-valued functions of the Fisher information matrix \cite{kay}.
In particular, the determinant, trace, and spectral norm of the Fisher information matrix were considered in \cite{morbidi13}, where the control objective is to maximize the information related to the positions of the agents and of a target.
The maximization of negative posterior entropy was considered in \cite{ryan07, hoffmann10, schwager11, julian12, atanasov14}. In \cite{schwager11}, a central controller steers agents with known positions along the gradient of negative posterior entropy to optimally sense a global state. A distributed solution for global state estimation was proposed in \cite{ryan07, hoffmann10} based on a pairwise neighboring-agents approximation of mutual information and in \cite{julian12, atanasov14} by using a consensus algorithm.
However, the methods proposed in \cite{ryan07, hoffmann10, schwager11, julian12, atanasov14} did not use BP, 
did not allow for multiple time-varying states, and did not include estimation of local (controlled) states.

\subsection{Contribution and Paper Organization} \label{sec:contrib}

Here, we present a unified Bayesian framework and method for (i) distributed, cooperative joint estimation of time-varying global and local states 
and (ii) distributed, cooperative information-seeking control. Our framework and method are suited to nonlinear and non-Gaussian problems, they require only communication with neighboring agents, and they are able to cope with a changing network topology. Thereby, they are particularly suited to localization and tracking tasks in location-aware scenarios involving mobile networks and nonlinear models. 

For distributed estimation, following \cite{meyer12,meyer2014coslat}, we combine BP message passing, consensus, 
and sample-based representations of the involved probability distributions.
For distributed control, we define a global (holistic) objective function as the negative joint posterior entropy of all states in the network at the next time step conditioned on all measurements at the next time step. This objective function is optimized jointly by all agents via a gradient ascent.
This reduces to the evaluation of local gradients at each agent, which is performed by using Monte Carlo integration based on 
the sample representations provided by the estimation stage and a distributed evaluation of the joint (networkwide) likelihood function.
Our method advances beyond \cite{ryan07, hoffmann10, schwager11, julian12, atanasov14} in the following \vspace{1mm}respects:
\begin{itemize}
\item It constitutes a more general information-maximizing control framework based on BP for estimation problems involving multiple time-varying states. 
\vspace{1mm}
\item It includes estimation of the local (controlled) states of the agents, thus enabling its use in a wider range of applications.
\vspace{1mm}
\end{itemize}

Contrary to \cite{meyer12,meyer2014coslat}, which introduces the distributed joint \emph{estimation} of time-varying local and global states in agent networks, 
here we focus on the information-maximizing \emph{control} of the agents. Our main contribution is a derivation and sample-based formulation of a new information-seeking controller that maximizes the negative joint posterior entropy of time-varying local and global states. Compared to the information-seeking controller 
proposed in \cite{ryan07, hoffmann10, schwager11, julian12, atanasov14}, where maximization of the negative posterior entropy reduces to maximization of the mutual information between observations and states, our controller includes an additional term that arises because the posterior entropy
involves also the local (controlled) states of the agents. Due to this more general formulation, our controller is suited to 
decentralized 
estimation tasks where the agents cooperatively infer also their own states.

This paper is organized as follows. In Section \ref{sec:sysModel}, the system model is described and the joint estimation and control problem is formulated.
Section \ref{sec:estimation} reviews joint local and global state estimation \cite{meyer2014coslat}. In Section \ref{sec:control_layer}, we introduce the proposed gradient-based controller. 
The distributed computation of the gradient is discussed in Sections \ref{sec:firstGradient}--\ref{sec:secondGradient}. Section \ref{sec:special} considers two special cases of the joint estimation and control framework. Finally, in Section \ref{sec:simRes}, we present simulation results demonstrating the performance of our method
for a simultaneous self-localization and target tracking problem.

\section{System Model and Problem Formulation}
\label{sec:sysModel}

\begin{figure}
\vspace{1mm}
\centering 
\psfrag{S1}[l][l][.7]{\raisebox{2mm}{\hspace{0mm}CA}} 
\psfrag{S2}[l][l][.7]{\raisebox{-3mm}{\hspace{0mm}target}}
\psfrag{S3}[l][l][.7]{\raisebox{1mm}{\hspace{0mm}communication link}} 
\psfrag{S4}[l][l][.7]{\raisebox{.5mm}{\hspace{0mm}measurement}}
\psfrag{A1}[l][l][.7]{\raisebox{2mm}{\hspace{2mm}}} 
\psfrag{A2}[l][l][.7]{\raisebox{2.1mm}{\hspace{-1mm}$\cl{C}^{(n)}_l$}}
\psfrag{A3}[l][l][.7]{\raisebox{4.1mm}{\hspace{-1.3mm}$\cl{T}^{(n)}_l$}} 
\psfrag{A4}[l][l][.7]{\raisebox{3mm}{\hspace{-1.5mm}$\cl{A}^{(n)}_l$}}
\psfrag{A5}[l][l][.7]{\raisebox{5mm}{\hspace{-9mm}CA $l$}}
\psfrag{A6}[l][l][.67]{\raisebox{-9.6mm}{\hspace{-5.1mm}target$\, m$}}
\psfrag{A8}[l][l][.7]{\raisebox{2.5mm}{\hspace{-1mm}$\cl{C}^{(n)}_m$}}
\psfrag{A7}[l][l][.7]{\raisebox{0mm}{\hspace{1mm}}}
\hspace*{-16mm}\includegraphics[scale=0.62]{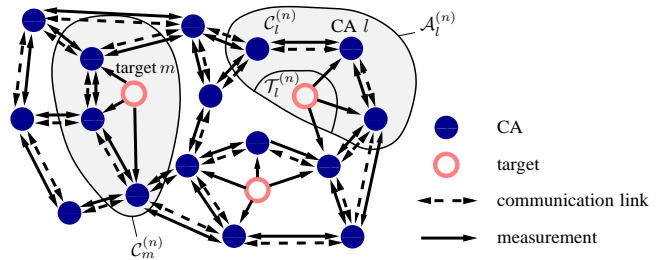} \vspace{-1mm}
\vspace{-.5mm}
\renewcommand{\baselinestretch}{1.05}\small\normalsize
\caption{Agent network with CAs and targets. The neighborhood sets $\cl{A}^{(n)}_l\!$, $\cl{C}^{(n)}_l\!$, $\cl{T}^{(n)}_l\!$, and $\cl{C}^{(n)}_m$ for one CA $l \!\in\rmv \cl{C}$ and one target $m \!\in\! \cl{T}$ are also shown.} 
\label{fig:an} 
\vspace{3mm}
\end{figure}

\begin{figure}
\centering 

\psfrag{S1}[l][l][.7]{\raisebox{3mm}{\hspace{-16mm}$\bd{x}_{k}^{(n)}\rmv\rmv$, $k \!\in\! \cl{A}_l^{(n)}$}} 

\psfrag{S2}[l][l][.7]{\raisebox{2mm}{\hspace{-2.2mm}$\bd{x}_{l}^{(n)}$}} 

\psfrag{S3}[l][l][.7]{\raisebox{4.5mm}{\hspace{-3.8mm}$\bd{x}_{l}^{(n+1)}$}} 

\psfrag{S4}[l][l][.7]{\raisebox{3.5mm}{\hspace{-7.2mm}$\bd{y}_{l,k}^{(n)}$, $k \!\in\! \cl{A}_l^{(n)}$}} 

\psfrag{S5}[l][l][.7]{\raisebox{3mm}{\hspace{-5mm}$\bd{x}_{l}^{(n)}$}} 

\psfrag{S6}[l][l][.55]{\raisebox{1.5mm}{\hspace{-6.5mm}}} 

\psfrag{S7}[l][l][.65]{\raisebox{4mm}{\hspace{-7mm}$\bd{u}_{l}^{(n+1)}$}} 

\psfrag{S9}[l][l][.7]{\raisebox{3.8mm}{\hspace{-2mm}$f\big(\bd{x}_{k}^{(n)} \big|\ist\bd{y}^{(1:n)};\bd{u}^{(1:n)}\big),\ist$}}

\psfrag{S9a}[l][l][.7]{\raisebox{2.7mm}{\hspace{-1.5mm}$\hat{\bd{x}}_{k}^{(n)}\rmv,\ist k\in\{l\}\rmv\cup\rmv\cl{T}$}} 

\psfrag{S20}[l][l][.7]{\raisebox{2mm}{\hspace{0mm}$k\in\{l\}\rmv\cup\rmv\cl{T}$}} 

\psfrag{S10}[l][l][.55]{\raisebox{1.2mm}{\hspace{1mm}}} 

\psfrag{S11}[l][l][.7]{\raisebox{2.5mm}{\hspace{0mm}$\bd{x}_{l}^{(n)}$}} 

\psfrag{S12}[l][l][.65]{\raisebox{4mm}{\hspace{-3.5mm}$\bd{q}_{l}^{(n+1)}$}} 

\psfrag{S13}[l][l][.65]{\raisebox{4mm}{\hspace{-3mm}$\bd{x}_{l}^{(n)}$}} 

\psfrag{S14}[l][l][.7]{\raisebox{-3mm}{\hspace{-1mm}$\bd{q}_{m}^{(n+1)}$}} 

\psfrag{S15}[l][l][.7]{\raisebox{1.8mm}{\hspace{-1mm}$\bd{x}_{m}^{(n)}$}} 

\psfrag{S16}[l][l][.7]{\raisebox{3.5mm}{\hspace{-5mm}$\bd{x}_{m}^{(n+1)}$}} 

\psfrag{T1}[l][l][0.83]{\raisebox{1mm}{\hspace{3mm}{\bf Sensor}} }

\psfrag{T2}[l][l][0.83]{\raisebox{-6mm}{\hspace{-.1mm}{\bf Estimation}}}

\psfrag{T6}[l][l][0.83]{\raisebox{-6mm}{\hspace{4.1mm}{\bf Layer}}} 

\psfrag{T4}[l][l][0.85]{\raisebox{-5.4mm}{\hspace{1.7mm}{\bf Control}}} 

\psfrag{T7}[l][l][0.85]{\raisebox{-4.7mm}{\hspace{3.3mm}{\bf Layer}}} 

\psfrag{T3a}[l][l][0.83]{\raisebox{-2mm}{\hspace{6mm}{\bf CA}}} 

\psfrag{T3}[l][l][0.83]{\raisebox{-4mm}{\hspace{.5mm}{\bf Dynamics}}} 

\psfrag{T5a}[l][l][0.83]{\raisebox{-5.3mm}{\hspace{3.3mm}{\bf Target}}}

\psfrag{T5}[l][l][0.83]{\raisebox{-6.5mm}{\hspace{.7mm}{\bf Dynamics}}} 

\psfrag{A1}[l][l][.85]{\raisebox{2mm}{\hspace{0mm}(a)}} 

\psfrag{A2}[l][l][.85]{\raisebox{2mm}{\hspace{0mm}(b)}} 

\psfrag{Z1}[l][l][.7]{\raisebox{.8mm}{\hspace{-1.2mm}$\bd{z}^{-1}$}} 

\psfrag{Z2}[l][l][.7]{\raisebox{.8mm}{\hspace{-1.2mm}$\bd{z}^{-1}$}} 

\psfrag{Z3}[l][l][.7]{\raisebox{0.3mm}{\hspace{-1.2mm}$\bd{z}^{-1}$}} 

\psfrag{Z4}[l][l][.7]{\raisebox{0.3mm}{\hspace{-1.2mm}$\bd{z}^{-1}$}}

\hspace{-1mm}\includegraphics[scale=0.38]{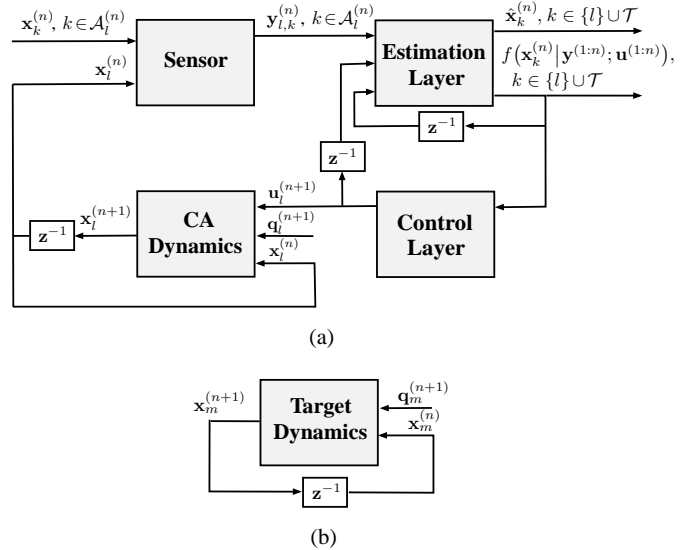} 

\renewcommand{\baselinestretch}{1.05}\small\normalsize
\vspace{-1.3mm}
\caption{\label{fig:blockDiagram}Block diagram of the overall ``signal processing system'' for (a) a CA $l\in\cl{C}$ and (b) a target $m\in\cl{T}$.}
\vspace{-2mm}
\end{figure}

We consider a network of mobile agents $k\rmv\in\!\cl{A}$ as shown in Fig.\ \ref{fig:an}.
The set of all agents, $\cl{A}$, consists of the set of cooperative agents (CAs), $\cl{C}\rmv\subseteq\rmv\cl{A}$, and the set of ``targets,'' $\cl{T}\rmv=\rmv\cl{A}\rmv\setminus\rmv\cl{C}$. Here, a target may be anything that does not cooperate and cannot be controlled, such as a noncooperative agent or a relevant feature of the environment. 
We will typically use the indices $k\rmv\in\rmv\cl{A}$, $l\rmv\in\rmv\cl{C}$, and $m\rmv\in\rmv\cl{T}$ to denote a generic agent, a CA, and a target, respectively. A block diagram of the overall ``signal processing system'' is shown in Fig. \ref{fig:blockDiagram} for a CA (including the estimation and control layers of the proposed method) and for a target. This system is described below.

\subsection{Agent States 
and Sensor Measurements} \label{sec:agents}

The \emph{state} of agent $k\rmv\in\rmv\cl{A}$ at discrete time $n\rmv\in\rmv\{0,1,\ldots\}$ is denoted by the vector $\bd{x}_{k}^{(n)}\!$.
For example, in a localization scenario, $\bd{x}_{k}^{(n)}$ may consist of the current position and motion-related quantities such as velocity, acceleration, and angular velocity \cite{rong}. The states evolve according 
to 
\begin{equation}
\bd{x}_{l}^{(n)} =\ist g_{l}\big(\bd{x}_{l}^{(n-1)}\!,\bd{u}_{l}^{(n)}\!,\bd{q}_{l}^{(n)}\big) \ist, \quad l \rmv\in\rmv \cl{C}
\label{eq:statrans_l}
\vspace{-2mm}
\end{equation}
or
\begin{equation}
\bd{x}_{m}^{(n)} =\ist g_{m}\big(\bd{x}_{m}^{(n-1)}\!,\bd{q}_{m}^{(n)}\big) \ist, \quad m \rmv\in\rmv\cl{T} \rmv,
\label{eq:statrans_m}
\end{equation}
where $g_{k}(\cdot)$ is a possibly nonlinear function, $\bd{q}_{k}^{(n)}$ is process (driving) noise, and $\bd{u}_{l}^{(n)} \!\in\rmv \cl{U}_l$ 
is a deterministic control vector 
that controls the $l$th CA. 
Since $\bd{u}_l^{(n)}$ is deterministic \cite[Chap.\ 5]{barShalom01}, it is either completely unknown (before it is determined) or perfectly known (after being determined by the control layer).
Note that also targets may have control variables. However, as these are hidden from the CAs, 
we will subsume any control for target $m$ in the noise $\bd{q}_{m}^{(n)}\rmv$. 
For the derivation of the controller, we assume that for $l \rmv\in\rmv \cl{C}$, $g_{l}\big(\bd{x}_{l}^{(n-1)}\!,\bd{u}_{l}^{(n)}\!,\bd{q}_{l}^{(n)}\big)$ is bijective with respect to $\bd{x}_{l}^{(n-1)}$ and differentiable with respect to $\bd{u}_{l}^{(n)}\!$. 
The statistical relation between $\bd{x}_{k}^{(n-1)}$ and $\bd{x}_{k}^{(n)}$ as defined by \eqref{eq:statrans_l} or \eqref{eq:statrans_m} 
can also be described by the \emph{state-transition pdf} $f\big(\bd{x}_{l}^{(n)}\big|\ist \bd{x}_{l}^{(n-1)};\bd{u}_{l}^{(n)}\big)$ for $l \rmv\in\rmv \cl{C}$ and $f\big(\bd{x}_{m}^{(n)}\big|\ist \bd{x}_{m}^{(n-1)}\big)$ for $m \rmv\in\rmv \cl{T}$.



The measurement and communication topology of the network is described by the neighborhood sets $\cl{C}_{l}^{(n)}\!$, $\cl{T}_{l}^{(n)}\!$, and $\cl{A}_{l}^{(n)}$ as follows. 
CA $l$ acquires a measurement $\bd{y}_{l,l'}^{(n)}$ relative to CA $l'$ if $l' \!\in\rmv \cl{C}_{l}^{(n)}\!$.
This relation is symmetric, i.e., $l'\!\in\rmv \cl{C}_{l}^{(n)}$ implies $l\rmv\in\rmv\cl{C}_{l'}^{(n)}\!$.
It is assumed that CAs that acquire measurements relative to each other are able to communicate, i.e., to transmit data via a communication link.
Furthermore, CA $l\rmv\in\rmv\cl C$ acquires a measurement $\bd{y}_{l,m}^{(n)}$ relative to target $m$ if $m\rmv\in\rmv\cl{T}_{l}^{(n)} \!\subseteq\rmv\cl{T}$. The targets are noncooperative in that they do not communicate
and do not acquire any measurements. We also define $\cl{A}_{l}^{(n)} \rmv\triangleq\ist \cl{C}_{l}^{(n)}\cup\cl{T}_{l}^{(n)}\!$. Finally, the set $\cl{C}_{m}^{(n)}$ contains all CAs measuring target $m$, i.e., all $l \!\in\! \cl{C}$ with $m \!\in\! \cl{T}^{(n)}_l\!$. The sets $\cl{C}_{l}^{(n)}\!$, $\cl{T}_{l}^{(n)}\!$, $\cl{A}_{l}^{(n)}\!$, and $\cl{C}_{m}^{(n)}$ are generally time-dependent. An example of a measurement and communication topology is given in Fig.\ \ref{fig:an}.

We consider ``pairwise'' measurements\footnote{The proposed framework can be easily extended to self-measurements (measurements that involve only the own state) and cluster measurements (measurements that involve the states of several other agents).} $\bd{y}_{l,k}^{(n)}$ that depend on the state $\bd{x}_{l}^{(n)}$ of a measuring
CA $l\rmv\in\rmv\cl{C}$ and the state $\bd{x}_{k}^{(n)}$ of a measured agent (CA or target) $k\rmv\in\rmv\cl{A}_{l}^{(n)}$ according to 
\begin{equation}
\bd{y}_{l,k}^{(n)}=\ist d_l\big(\bd{x}_{l}^{(n)}\!,\bd{x}_{k}^{(n)}\!,\bd{v}_{l,k}^{(n)}\big)\,, \quad\! l\rmv\in\rmv\cl{C}, \; k\rmv\in\rmv\cl{A}_{l}^{(n)},
\label{eq:meas_mod}
\end{equation}
where $d_l(\cdot)$ is a possibly nonlinear function and $\bd{v}_{l,k}^{(n)}$ is measurement noise. An example is the scalar measurement
\begin{equation}
y_{l,k}^{(n)} \ist=\ist \big\|\bd{x}_{l}^{(n)}\!-\rmv\bd{x}_{k}^{(n)}\big\| + v_{l,k}^{(n)}\,,\label{eq:mess}
\end{equation}
where $\bd{x}_{k}^{(n)}$ equals the position of agent $k$ and, hence, $\big\|\bd{x}_{l}^{(n)}\rmv-\bd{x}_{k}^{(n)}\big\|$ is the spatial distance between agents $l$ and $k$. 
The statistical relation between the measurement $\bd{y}_{l,k}^{(n)}$ and the involved states $\bd{x}_{l}^{(n)}$ and $\bd{x}_{k}^{(n)}$ is also described by the \textit{local likelihood function} $f\big(\bd{y}_{l,k}^{(n)}\big|\ist \bd{x}_{l}^{(n)}\!,\bd{x}_{k}^{(n)}\big)$. For the derivation of the controller, $f\big(\bd{y}_{l,k}^{(n)}\big|\ist \bd{x}_{l}^{(n)}\!,\bd{x}_{k}^{(n)}\big)$ is assumed differentiable with respect to $\bd{x}_{l}^{(n)}\!$. 

We also make the following assumptions. The number of targets is known, and the targets can be identified by the CAs, i.e., target-to-measurement assignments are known. 
Furthermore, CA $l$ knows the state evolution models $g_{k}(\cdot)$ and 
process noise pdfs $f\big(\bd{q}_{k}^{(n)}\big)$ for $k \in \{l\} \cup\ist \cl{C}_l^{(n)} \cup \cl{T}$;
the initial prior pdfs of the agent states, $f\big(\bd{x}_{k}^{(0)}\big)$, for $k \in \{l\} \cup\ist \cl{C}_l^{(0)} \cup \cl{T}$;
the measurement models $d_{l'}(\cdot)$ for $l' \!\in \{l\} \cup \cl{C}_l^{(n)}$; and the measurement noise pdfs 
$f\big(\bd{v}_{l,k}^{(n)}\big)$, $k\rmv\in\rmv\cl{A}_{l}^{(n)}$ and $f\big(\bd{v}_{l'\!,l}^{(n)}\big)$, $l' \!\in \cl{C}^{(n)}_l\!$.

\vspace{-1mm}

\subsection{Problem Formulation} \label{sec:prob}

The following 
tasks are to be performed at each time $n$:

\begin{enumerate}

\vspace{1mm}

\item Each CA $l \rmv\in\rmv \cl{C}$ estimates the states $\bd{x}_{k}^{(n)}\!$, $k \in \{l\} \cup \cl{T}$ (i.e., its own local state and the states of all targets)
from prior information and all past and present measurements in the network.

\vspace{1mm}

\item The state of each CA is controlled such that the negative joint posterior entropy of all states in the network at the next time, conditioned on all measurements in the network at the next time, is maximized. 

\vspace{1mm}

\end{enumerate}

We solve these two problems in a distributed and recursive manner. Our method consists of an \emph{estimation layer} and a \emph{control layer}, as shown in Fig.\ \ref{fig:blockDiagram}(a). In the estimation layer, CA $l$ computes an approximation of the marginal posterior pdfs of the states $\bd{x}_{k}^{(n)}\!$, $k \in \{l\} \cup \cl{T}$ given all the past and present measurements and control vectors in the entire network. In the control layer, CA $l$ uses these marginal posteriors and the statistical model to determine a quasi-optimal control variable $\bd{u}_{l}^{(n+1)}\!$. In both layers, the CAs communicate with neighbor CAs.
    
\vspace{-4mm}

\section{Estimation Layer}
\label{sec:estimation}

\setcounter{subsubsection}{0}
The estimation layer performs distributed estimation of the local and global states by using the BP- and consensus-based method introduced in \cite{meyer12,meyer2014coslat}.
We will review this method in our present context. Let us denote by $\bd{x}^{(n)}\rmv\triangleq {\big[\bd{x}_{k}^{(n)}\big]}_{k\in\cl{A}}$, $\bd{u}^{(n)}\rmv\triangleq\rmv \big[\bd{u}_{l}^{(n)}\big]_{l\in\cl{C}}$, and $\bd{y}^{(n)}\rmv\triangleq\rmv\big[\bd{y}_{l,k}^{(n)}\big]_{l\in\cl{C},\, k\in\cl{A}_{l}^{(n)}}$
the stacked vectors of, respectively, all states, control variables, and measurements at time $n$. Furthermore, let $\bd{x}^{(1:n)}\rmv\triangleq\big[\bd{x}^{(1)\text{T}}\rmv,\dots,\bd{x}^{(n)\text{T}}\big]^{\text{T}}$, $\bd{u}^{(1:n)}\rmv\triangleq\big[\bd{u}^{(1)\text{T}}\rmv,\dots,\bd{u}^{(n)\text{T}}\big]^{\text{T}}$, and $\bd{y}^{(1:n)}\rmv\triangleq\bd[\bd{y}^{(1)\text{T}}\rmv,\dots,\bd{y}^{(n)\text{T}}\big]^{\text{T}}$. Each CA $l \rmv\in\rmv\cl{C}$ estimates its local state $\bd{x}_{l}^{(n)}$ and all the target states $\bd{x}_{m}^{(n)}\rmv$, $m \rmv\in\rmv \cl{T}$ from the measurements of all CAs up to time $n$, $\bd{y}^{(1:n)}\rmv$. This estimation is based on the posteriors $f\big(\bd{x}_{k}^{(n)} \big|\ist \bd{y}^{(1:n)};\bd{u}^{(1:n)}\big)$, $k\in\{l\}\cup\cl{T}$, which are marginals of the joint posterior $f\big(\bd{x}^{(1:n)}\big|\ist \bd{y}^{(1:n)};\bd{u}^{(1:n)}\big)$.


Using Bayes' rule and common assumptions \cite{wymeersch}, the joint posterior can be factorized as
\begin{align}
& f\big(\bd{x}^{(1:n)} \big|\ist \bd{y}^{(1:n)};\bd{u}^{(1:n)}\big) \nonumber \\[0mm]
 & \;\propto\rmv \Bigg( \prod_{k\in\cl{A}}\rmv f\big(\bd{x}_{k}^{{(0)}}\big) \! \Bigg)
 \!\prod_{n'=1}^{n} \!\rmv\Bigg(\prod_{m\in\cl{T}}\! f\big(\bd{x}_{m}^{(n')}\big|\ist \bd{x}_{m}^{(n'-1)}\big) \! \Bigg)\nonumber \\[.5mm]
 & \hspace*{5.5mm}\times \prod_{l\in\cl{C}} f\big(\bd{x}_{l}^{(n')}\big|\ist \bd{x}_{l}^{(n'-1)};\bd{u}_{l}^{(n')}\big)
   \!\!\prod_{k'\in\cl{A}_{l}^{(n')}}\!\! f\big(\bd{y}_{l,k'}^{(n')} \big|\ist \bd{x}_{l}^{(n')}\!,\bd{x}_{k'}^{(n')}\big) \,. \nonumber\\[-3.5mm]
\label{eq:factorization} \\[-7.5mm]
\nonumber
\end{align}
The marginal posterior of state $\bd{x}_{k}^{(n)}$ is then given by \begin{equation}
f\big(\bd{x}_{k}^{(n)} \big|\ist \bd{y}^{(1:n)};\bd{u}^{(1:n)}\big) \ist= \int \rmv f\big(\bd{x}^{(1:n)} \big|\ist \bd{y}^{(1:n)};\bd{u}^{(1:n)}\big) \,\mathrm{d} \bd{x}_{\sim k,n}^{(1:n)} \,,
\label{eq:marginalization} 
\vspace{-.4mm}
\end{equation}
where $\bd{x}_{\sim k,n}^{(1:n)}$ denotes $\bd{x}^{(1:n)}$ with $\bd{x}_{k}^{(n)}$ removed. Using $f\big(\bd{x}_{k}^{(n)} \big|\ist \bd{y}^{(1:n)};\bd{u}^{(1:n)}\big)$, the minimum mean-square error (MMSE) estimator \cite{kay} of $\mathbf{x}^{(n)}_{k}$ is obtained as
\begin{equation}
\label{eq:mmse}
\hat{\mathbf{x}}^{(n)}_{k,\text{MMSE}} \ist\triangleq\rmv 
\int \! \mathbf{x}^{(n)}_{k} f\big(\bd{x}_{k}^{(n)} \big|\ist \bd{y}^{(1:n)};\bd{u}^{(1:n)}\big)  \, \mathrm{d}\mathbf{x}^{(n)}_{k}, \;\,\,
  k \rmv \in \rmv \mathcal{A} \,.
\vspace{-2mm}
\end{equation}

\subsection{Sequential Calculation} \label{sec:sequ}
For a review of the sequential calculation of \eqref{eq:marginalization} proposed in \cite{meyer12,meyer2014coslat}, we now switch to the following simplified notation for the sake of readability. In the conditions of the various conditional pdfs, we omit the measurements up to time $n \rmv-\! 1$,\linebreak i.e., $\bd{y}^{(1:n-1)}\rmv$, and the control vectors up to time $n$, i.e., $\bd{u}^{(1:n)}\rmv$, because $\bd{y}^{(1:n-1)}$ has already been observed and $\bd{u}^{(1:n)}$ has already been determined; hence both are considered fixed. Furthermore, we suppress the time index $n$, and we write the current and previous states of CA $l$ as $\bd{x}_{l}^{(n)} \rmv= \bd{x}_{l}$ and $\bd{x}_{l}^{(n-1)} \rmv= \bd{x}_{l}^{-}\rmv$, respectively. Similarly, we write $\bd{u}_{l}^{(n+1)} \rmv= \bd{u}_{l}^{+}\rmv$. For sequential calculation of \eqref{eq:marginalization}, CA $l \rmv\in\rmv \cl{C}$ employs a basic Bayesian recursive filtering method \cite{wymeersch} consisting of a prediction step and a correction step. In the prediction step, CA $l \rmv\in\rmv \cl{C}$ computes a predictive posterior of its current state, 
\begin{equation}
\label{eq:pred}
f(\bd{x}_{l}) \,=\int \rmv f(\bd{x}_{l}|\ist \bd{x}_{l}^{-})
\ist\ist f(\bd{x}_{l}^{-}) \,\mathrm{d}\bd{x}_{l}^{-} .
\vspace{-.7mm}
\end{equation}
Here, $f(\bd{x}_{l})$ and $f(\bd{x}^{-}_{l})$ are short for $f\big(\bd{x}^{(n)}_{l}\big|\ist\bd{y}^{(1:n-1)} ; \bd{u}^{(1:n)} \big)$ and $f\big(\bd{x}^{(n-1)}_{l}\big|\ist\bd{y}^{(1:n-1)} ; \bd{u}^{(1:n-1)}\big)$, respectively. Furthermore, CA $l$ computes predictive posteriors of the target states
\begin{equation}
\label{eq:pred_f-recurs}
f(\bd{x}_{m}) \,=\int \rmv f(\bd{x}_{m}|\ist \bd{x}_{m}^{-}) \ist\ist f(\bd{x}_{m}^{-}) \,\mathrm{d}\bd{x}_{m}^{-} \,, \quad m \rmv\in\rmv \cl{T} .
\end{equation}
In the correction step, CA $l$ determines the marginal posteriors $f(\bd{x}_{l}|\ist \bd{y})$ and $f(\bd{x}_{m}|\ist \bd{y})$, $m \rmv\in\rmv \cl{T}$, which are given by
\begin{align}
\hspace{-3mm}f(\bd{x}_{k}|\ist \bd{y}) &\,\propto\ist \int \prod_{k'\in \cl{A}} \! f(\bd{x}_{k'}) \ist \prod_{l'\in \cl{C}} \prod_{k_1\in \cl{A}_{l'}} \!\! f(\bd{y}_{l'\!,k_1}|\ist \bd{x}_{l'},\bd{x}_{k_1}) 
  \,\mathrm{d}\bd{x}_{\sim k} \,, \nonumber\\[-2mm]
&  \hspace{50mm} k \in \{l\} \cup \cl{T} .
\label{eq:CoSLATlocal} 
\end{align}
Here, $\bd{x}_{\sim k}$ denotes $\bd{x}$ with $\bd{x}_{k}$ removed and $f(\bd{x}_{k}|\ist \bd{y})$ is short for 
$f\big(\bd{x}_{k}^{(n)} \big|\ist \bd{y}^{(1:n)};\bd{u}^{(1:n)}\big)$.
As shown in Fig.\ \ref{fig:blockDiagram}(a), these marginal posteriors are handed over to the control layer, which determines the control input for the next time step, 
$\bd{u}_{l}^{+}\rmv$. 

\vspace{-1mm}

\subsection{BP Message Passing and Consensus} \label{sec:BPandLC}
Calculation of $f(\bd{x}_{k}|\ist \bd{y})$ according to
\eqref{eq:CoSLATlocal} is generally infeasible,
due to the reliance on nonlocal information and the inherent complexity of the marginalization process. 
Based on the factorization of the joint posterior in \eqref{eq:factorization}, a
computationally feasible approximation of \eqref{eq:CoSLATlocal} is provided by a distributed, cooperative algorithm that combines BP message passing and the average consensus scheme
\cite{meyer12,meyer2014coslat}. 
This algorithm computes an approximation (``belief'') $b(\bd{x}_{k})\approx f(\bd{x}_{k}|\ist \bd{y})$ in an iterative manner, using only communication with neighbor CAs 
$l'\!\in\rmv \cl{C}_{l}$. Its complexity scales only linearly with the number of states in the network, $|\cl{A}|$ (for a fixed number of iterations). 

According to \cite{meyer12,meyer2014coslat}, the belief of local state $\bd{x}_{l}$ at message passing iteration $p \rmv\rmv\in\rmv\rmv \{1,\dots,P\}$ is given by 
\begin{align}
\hspace*{-2mm}b^{(p)}(\bd{x}_{l}) &\,\propto\ist f(\bd{x}_{l}) \rmv\prod_{l'\in\cl{C}_{l}}\int\! f(\bd{y}_{l,l'}|\ist\bd{x}_{l},\bd{x}_{l'}) \,b^{(p-1)}(\bd{x}_{l'}) \,\mathrm{d}\bd{x}_{l'}
  \nonumber\\[.5mm]
&\hspace*{5mm}\times\!\prod_{m\in\cl{T}_{l}}\int\! f(\bd{y}_{l,m}|\ist\bd{x}_{l},\bd{x}_{m}) \,\psi_{m\rightarrow l}^{(p-1)}(\bd{x}_{m}) \,\mathrm{d}\bd{x}_{m} \,,
\label{eq:CoSLATlocalBP}\\[-6mm]
\nonumber
\end{align}
which is initialized as $b^{(0)}(\bd{x}_{l})=f(\bd{x}_{l})$. 
Similarly, the belief of target state $\bd{x}_{m}$ at message passing iteration $p$ is given by 
\begin{align}
b^{(p)}(\bd{x}_{m}) \,\propto\ist f(\bd{x}_{m}) \rmv\prod_{l\in\cl{C}_{m}} \rmv\int\! f(\bd{y}_{l,m}|\ist\bd{x}_{l},\bd{x}_{m}) \,\psi_{l\rightarrow m}^{(p-1)}(\bd{x}_{l}) 
\,\mathrm{d}\bd{x}_{l} \,, \nonumber\\[-3mm]
\label{eq:CoSLATglobal}\\[-7mm]
\nonumber
\end{align}
with initialization $b^{(0)}(\bd{x}_{m})=f(\bd{x}_{m})$. Here, $\psi_{m\rightarrow l}^{(p-1)}(\bd{x}_{m})$ and $\psi_{l\rightarrow m}^{(p-1)}(\bd{x}_{l})$ are the 
``extrinsic informations'' from target $m$ to CA $l$ and from CA $l$ to target $m$, respectively, at the previous message passing iteration $p \rmv-\rmv 1$.
These extrinsic informations are calculated recursively as 
\begin{align}
\psi_{m\rightarrow l}^{(p)}(\bd{x}_{m}) &\,=\, \frac{b^{(p)}(\bd{x}_{m})}{\int f(\bd{y}_{l,m}|\ist\bd{x}_{l},\bd{x}_{m}) \,\psi_{l\rightarrow m}^{(p-1)}(\bd{x}_{l}) \,\mathrm{d}\bd{x}_{l}}\\[.5mm]
\psi_{l\rightarrow m}^{(p)}(\bd{x}_{l}) &\,=\, \frac{b^{(p)}(\bd{x}_{l})}{\int f(\bd{y}_{l,m}|\ist\bd{x}_{l},\bd{x}_{m}) \,\psi_{m\rightarrow l}^{(p-1)}(\bd{x}_{m}) \,\mathrm{d}\bd{x}_{m}} 
  \,, \label{eq:extrinsicInfo}\\[-4.5mm]
\nn
\end{align}
with initialization $\psi_{m\rightarrow l}^{(0)}(\bd{x}_{m}) \rmv=\! f(\bd{x}_{m})$ and $\psi_{l\rightarrow m}^{(0)}(\bd{x}_{l}) \rmv=\! f(\bd{x}_{l})$, 
respectively. 
In \eqref{eq:CoSLATlocalBP}, the beliefs $b^{(p-1)}(\bd{x}_{l'})$ of neighbor CAs $l' \in \cl{C}_l$ are used instead of the extrinsic informations. 
This is part of the \emph{sum-product algorithm over a wireless network} (SPAWN) scheme for cooperative localization \cite{wymeersch}, 
which results from a specific choice of the message schedule in loopy BP and has been observed to provide highly accurate estimates \cite{wymeersch, meyer12, meyer2014coslat, lien}.

A sample-based distributed implementation of \eqref{eq:CoSLATlocalBP}--\eqref{eq:extrinsicInfo} has been proposed in \cite{meyer12,meyer2014coslat}. A problem for a distributed implementation is that the products $\prod_{l\in\cl{C}_{m}} \rmv\int\rmv f(\bd{y}_{l,m}|\ist\bd{x}_{l},\bd{x}_{m}) \,\psi_{l\rightarrow m}^{(p-1)}(\bd{x}_{l}) \,\mathrm{d}\bd{x}_{l}$, $m\!\in\!\cl{T}$ involved in \eqref{eq:CoSLATglobal} are not available at the CAs. However, an approximation of these products can be provided to each CA in a distributed manner (i.e., using only local communications) either by a consensus algorithm performed in parallel for each sample weight \cite{farahmand, savic14, meyer2014coslat} or by the likelihood consensus scheme \cite{meyer12, hlinkaMag13, hlinka14adaptation}. 
For the calculations in the control layer---to be described in Sections 
\ref{sec:control_layer}--\ref{sec:secondGradient}---all CAs require a common set of samples
representing the target beliefs $b^{(p)}(\bd{x}_{m})$. This can be ensured by additionally using, e.g., a max-consensus and providing all CAs with the same seed for random number generation \cite{savic14,lindberg13}.

The output of the estimation layer is the set of beliefs $b^{(P)}(\bd{x}_k)$, $k \in \{l\} \cup \cl{T}$ at the final message passing iteration $p \rmv= P$. These beliefs are handed over to the control layer, which calculates the control variables $\bd{u}^{+}_{l}$ (see Fig.\ \ref{fig:blockDiagram}(a)). Each belief $b^{(P)}(\bd{x}_k)$ is represented by $J$ samples $\big\{\bd{x}^{(j)}_k\big\}^{J}_{j=1}$, which is briefly denoted as $\big\{\bd{x}^{(j)}_k\big\}^{J}_{j=1} \!\sim\rmv b^{(P)}(\bd{x}_k)$. From these \vspace{.5mm} samples, an approximation of the MMSE estimate \eqref{eq:mmse} is calculated as
\[
\hat{\mathbf{x}}_{k} =\ist \frac{1}{J} \sum_{j=1}^J \rmv\bd{x}^{(j)}_{k}.
\]

\section{Control Layer}
\label{sec:control_layer}

Next, we present the information-seeking controller. We temporarily revert to the full notation.

\vspace{-1.5mm}

\subsection{Objective Function and Controller}
\label{sec:control_layer:obj}

According to our definition at the beginning of Section \ref{sec:estimation}, the vector comprising all the measurements in the network at the next time $n \rmv+\rmv 1$ is $\bd{y}^{(n+1)} = \big[ \bd{y}_{l,k}^{(n+1)} \big]_{l\in\cl{C},\ist k\in\cl{A}_{l}^{(n+1)}}$. However, in this definition of $\bd{y}^{(n+1)}\rmv$, we now formally replace $\cl{A}_{l}^{(n+1)}$ by $\cl{A}_{l}^{(n)}$ since at the current time $n$, the sets $\cl{A}_{l}^{(n+1)}$ are not yet known. Thus, with an abuse of notation, $\bd{y}^{(n+1)}$ is redefined 
\vspace{-.8mm}
as
\be
\bd{y}^{(n+1)} \ist\triangleq\ist \big[ \bd{y}_{l,k}^{(n+1)} \big]_{l\in\cl{C},\ist k\in\cl{A}_{l}^{(n)}} \,.
\label{eq:futuremeas}
\ee
In the proposed control approach, each CA $l \rmv\in\rmv \cl{C}$ determines its next control variable $\bd{u}_{l}^{(n+1)}$ such that the information about the next joint state $\bd{x}^{(n+1)}$ given $\bd{y}^{(1:n+1)}$ is maximized. We quantify this information by the negative conditional differential entropy \cite[Chap.\ 8]{cover06} of $\bd{x}^{(n+1)}$ given $\bd{y}^{(n+1)}\rmv$, with $\bd{y}^{(1:n)}$ being an additional condition that has been observed previously and is thus \vspace{1mm}
fixed: 
\begin{align}
 & \hspace*{-2mm}-h\big(\bm{{\sf{x}}}^{(n+1)} \big|\ist \bm{{\sf{y}}}^{(n+1)};\bd{y}^{(1:n)}\!, \bd{u}^{(1:n+1)}\big)\nonumber \\[1mm]
 & \hspace*{8mm}
 =\, \int\!\! \int\rmv  f\big( \bd{x}^{(n+1)}\!,\bd{y}^{(n+1)} \big|\ist \bd{y}^{(1:n)};\bd{u}^{(1:n+1)} \big)\nonumber \\[-1mm]
 & \hspace*{16mm}\times\log f\big(\bd{x}^{(n+1)} \big|\ist \bd{y}^{(n+1)}\!, \bd{y}^{(1:n)}; \bd{u}^{(1:n+1)}\big) \nonumber \\[.5mm]
 & \hspace*{16mm}\times 
   \mathrm{d}\bd{x}^{(n+1)}\ist \mathrm{d}\bd{y}^{(n+1)} \ist ,
 \label{eq:entropy1}
\end{align}
where $\log$ denotes the natural logarithm. Note that $h\big(\bm{{\sf{x}}}^{(n+1)} \big|\ist \bm{{\sf{y}}}^{(n+1)}; \bd{y}^{(1:n)}\!,\bd{u}^{(1:n+1)}\big)$ 
depends on the \emph{random vectors} $\bd{x}^{(n+1)}$ and $\bd{y}^{(n+1)}\rmv$, i.e., on their joint distribution but not on their values. Our notation indicates this fact by using a sans serif font for $\bm{{\sf{x}}}^{(n+1)}$ and $\bm{{\sf{y}}}^{(n+1)}$ in $h\big(\bm{{\sf{x}}}^{(n+1)} \big|\ist \bm{{\sf{y}}}^{(n+1)};\bd{y}^{(1:n)}\!,\bd{u}^{(1:n+1)}\big)$.

According to expression \eqref{eq:entropy1}, $-h\big(\bm{{\sf{x}}}^{(n+1)} \big|\ist \bm{{\sf{y}}}^{(n+1)};\bd{y}^{(1:n)}\!,$\linebreak $\bd{u}^{(1:n+1)}\big)$ is 
a function of the control vector $\bd{u}^{(n+1)}\rmv$. This function will be denoted as $D_h\big(\bd{u}^{(n+1)}\big)$, i.e.,
\be
D_h\big(\bd{u}^{(n+1)}\big) \,\triangleq\, -\,h\big(\bm{{\sf{x}}}^{(n+1)} \big|\ist \bm{{\sf{y}}}^{(n+1)};\bd{y}^{(1:n)}\!,\bd{u}^{(1:n+1)}\big) \,,
\label{eq:objfunc_Dh}
\ee
and it will be used as the objective function for control at each CA. This objective function is holistic in that it 
involves \emph{all} the next states (of both the CAs and the targets), $\bd{x}^{(n+1)}\rmv$, and \emph{all} the next measurements, $\bd{y}^{(n+1)}\rmv$.
Instead of a full-blown maximization of $D_h\big(\bd{u}^{(n+1)}\big)$, we perform one step of a gradient ascent \cite{fletcher87} at each time $n$. Thus, 
$\bd{u}^{(n+1)}$ is determined as 
\begin{equation}
\hat{\bd{u}}^{(n+1)} \ist=\, \bd{u}^{(n+1)}_{\text{r}} +\ist 
  c^{(n+1)} \, \nabla D_h\big(\bd{u}^{(n+1)}\big) \big{|}_{\bd{u}^{(n+1)}=\ist\bd{u}^{(n+1)}_{\text{r}}} \, ,
\label{eq:grad-asc}
\end{equation}
where $\bd{u}^{(n+1)}_{\text{r}}$ is a reference vector and $c^{(n+1)} \rmv>\rmv 0$ is a step size. The choice of $\bd{u}^{(n+1)}_{\text{r}}$ depends on the manner in which the local control vectors $\bd{u}_l^{(n)}$ (which are subvectors of $\bd{u}^{(n)}$) appear in
the state evolution functions $g_{l}\big(\bd{x}_{l}^{(n-1)}\!,\bd{u}_{l}^{(n)}\!,\bd{q}_{l}^{(n)}\big)$ in \eqref{eq:statrans_l}; two common choices are $\bd{u}^{(n+1)}_{\text{r}} \!=\rmv \bd{u}^{(n)}$ and $\bd{u}^{(n+1)}_{\text{r}} \!=\rmv \bd{0}$ (cf. Section \ref{sec:sim:setup}).

Since $\bd{u}^{(n+1)} = \big[\bd{u}_{l}^{(n+1)}\big]_{l\in\cl{C}}\ist$,
we have 
\[
\nabla D_h(\bd{u}^{(n+1)})
= \bigg[ \frac{\partial D_h(\bd{u}^{(n+1)})}{\partial\bd{u}_l^{(n+1)}} \bigg]_{l\in\cl{C}}\ist,
\]
and thus the gradient ascent \eqref{eq:grad-asc} with respect to $\bd{u}^{(n+1)}$ is equivalent to local gradient ascents at the individual CAs $l$ with respect to the local control vectors $\bd{u}_l^{(n+1)}\!$. At CA $l$, the local gradient ascent is performed as
\begin{equation}
\hat{\bd{u}}_{l}^{(n+1)} \ist=\, \bd{u}^{(n+1)}_{\text{r},l} + \ist c_{l}^{(n+1)} \,\frac{\partial D_h\big(\bd{u}^{(n+1)}\big)}{\partial\bd{u}_{l}^{(n+1)}}
  \Bigg{|}_{\bd{u}^{(n+1)}=\ist\bd{u}^{(n+1)}_{\text{r}}} \,,
\label{eq:controller_output}
\end{equation}
where $\bd{u}^{(n+1)}_{\text{r},l}$ is the part of $\bd{u}^{(n+1)}_{\text{r}}$ that corresponds to CA $l$ (we have $\bd{u}^{(n+1)}_{\text{r}} \!=\rmv \big[\bd{u}^{(n+1)}_{\text{r},l}\big]_{l \in \cl{C}}$). 
The local step sizes $c_{l}^{(n+1)}$ are constrained by the condition $\bd{u}_{l}^{(n+1)} \!\rmv\in\rmv \cl{U}_l$ for given sets $\cl{U}_l$. In practice, this condition can be easily satisfied by an appropriate scaling of the $c_{l}^{(n+1)}\!$. Note that, as in \cite{julian12}, we use different local step sizes $c_{l}^{(n+1)}$ at the individual CAs $l$. This heuristic modification is made to account for the possibly different sets $\cl{U}_l$ and to avoid the necessity of reaching a consensus on a common step size across all the CAs; it was observed to yield good results. Because the objective function $D_h(\cdot)$ changes over time $n$, the local ascent described by \eqref{eq:controller_output} generally is not guaranteed to converge; this is similar to existing information-seeking control algorithms \cite{hoffmann10, julian12}. Indeed, the goal of the proposed control algorithm is to make available informative measurements to the estimation layer; because of the dynamic scenario, this is generally not compatible with convergence.

\vspace{-1mm}

\subsection{Expansion of the Objective Function}

A central contribution of this paper is a distributed sample-based technique for calculating the gradients $\frac{\partial D_h(\bd{u}^{(n+1)})}{\partial\bd{u}_{l}^{(n+1)}}\Big{|}_{\bd{u}^{(n+1)}=\ist\bd{u}^{(n+1)}_{\text{r}}}$ in \eqref{eq:controller_output}. 
As a starting point for developing this technique, we next derive an expansion of the objective function $D_h\big(\bd{u}^{(n+1)}\big)$. We will use the following simplified notation. We do not indicate the conditioning on $\bd{y}^{(1:n)}$ and the dependence on $\bd{u}^{(1:n)}$ because at time $n+1$, $\bd{y}^{(1:n)}$ has already been observed and $\bd{u}^{(1:n)}$ has already been determined; hence both are fixed. (Note that in Section \ref{sec:estimation}, we did not indicate the conditioning on $\bd{y}^{(1:n-1)}\rmv$, rather than $\bd{y}^{(1:n)}\rmv$.) Also, we suppress the time index $n$ and designate variables at time $n \rmv+\! 1$ by the superscript ``$+$''. For example, we write $h(\bm{{\sf{x}}}^{+} |\ist \bm{{\sf{y}}}^{+};\bd{u}^{+})$ instead of $h\big(\bm{{\sf{x}}}^{(n+1)} \big|\ist \bm{{\sf{y}}}^{(n+1)};\bd{y}^{(1:n)}\!,\bd{u}^{(1:n+1)}\big)$.

For calculating the gradient, following \cite{hoffmann10} and \cite{julian12}, we disregard the unknown driving noise $\bd{q}_{l}$ in \eqref{eq:statrans_l} 
by formally replacing it with its expectation, $\bar{\bd{q}}_{l} \triangleq \int \bd{q}_{l} \, f(\bd{q}_{l})  \,\mathrm{d}\bd{q}_{l}$. We can then rewrite \eqref{eq:statrans_l} 
(with $n$ replaced by $n+1$) as
\vspace{-1mm}
\begin{equation}
\bd{x}_{l}^{+} =\ist\ist g_{l}(\bd{x}_{l},\bd{u}_{l}^{+}\!,\bar{\bd{q}}_{l}^{+})
  \ist\ist=\ist\ist \tilde{g}_{l}(\bd{x}_{l},\bd{u}_{l}^{+}) \,, \quad l \!\in\rmv \cl{C} \,.
\label{eq:statrans_control}
\end{equation}
As shown in Appendix \ref{sec:proof}, the conditional differential entropy in \eqref{eq:objfunc_Dh} can be expressed as
\be
h(\bm{{\sf{x}}}^{+} |\ist \bm{{\sf{y}}}^{+};\bd{u}^{+}) \,=\, h(\bm{{\sf{x}}}_{\cl{C}}, \bm{{\sf{x}}}_{\cl{T}}^{+} \ist|\ist \bm{{\sf{y}}}^{+};\bd{u}^{+}) 
  \ist\ist +\ist \sum_{l \in \cl{C}} G_l(\bd{u}_{l}^{+}) \,,
\label{eq:entropyStep1}
\vspace{-1mm}
\ee
where $\bd{x}_{\cl{C}} \triangleq \big[\bd{x}_l\big]_{l\in\cl{C}}\ist$, $\bd{x}_{\cl{T}}^{+} \triangleq \big[\bd{x}_m^{+}\big]_{m\in\cl{T}}\ist$, and
\begin{align}
&G_l(\bd{u}_{l}^{+}) \,\triangleq \int \! f(\bd{x}_{l}) \ist\log | J_{\tilde{g}_{l}}(\bd{x}_{l};\bd{u}^{+}_{l}) | \,\ist\mathrm{d}\bd{x}_{l} \,, \nonumber\\[-1mm] 
&\hspace{20mm}\text{with} \;\;
J_{\tilde{g}_{l}}(\bd{x}_{l};\bd{u}_{l}^{+}) \,\triangleq\, \det\rmv \frac{\partial\tilde{g}_{l}(\bd{x}_{l},\bd{u}_{l}^{+})}{\partial\bd{x}_{l}} \,.
\label{eq:G_def}
\end{align}

The first term on the right-hand side of \eqref{eq:entropyStep1} can be decomposed as \cite[Chap.\ 8]{cover06}
\begin{equation}
\label{eq:splitEntropy}
h(\bm{{\sf{x}}}_{\cl{C}}, \bm{{\sf{x}}}_{\cl{T}}^{+} \ist|\ist \bm{{\sf{y}}}^{+};\bd{u}^{+}) 
\,=\, h(\bm{{\sf{x}}}_{\cl{C}}, \bm{{\sf{x}}}_{\cl{T}}^{+}) - I(\bm{{\sf{x}}}_{\cl{C}}, \bm{{\sf{x}}}_{\cl{T}}^{+}\ist; \bm{{\sf{y}}}^{+};\bd{u}^{+}) \,.
\end{equation}
Here, $I(\bm{{\sf{x}}}_{\cl{C}}, \bm{{\sf{x}}}_{\cl{T}}^{+}\ist; \bm{{\sf{y}}}^{+};\bd{u}^{+})$ denotes the two-variable mutual information between 
$(\bd{x}_{\cl{C}}, \bd{x}_{\cl{T}}^{+})$ and $\bd{y}^{+}$ (with $\bd{u}^{+}$ being a deterministic parameter, i.e., not a third random variable), which is given 
by \cite[Chap.\ 8]{cover06}
\begin{align}
&\hspace{-1.7mm}I(\bm{{\sf{x}}}_{\cl{C}}, \bm{{\sf{x}}}_{\cl{T}}^{+}\ist; \bm{{\sf{y}}}^{+};\bd{u}^{+}) \nonumber\\[0mm]
&\hspace{-2mm}=\rmv \int \!\! \int \!\! \int \rmv\rmv f(\bd{x}_{\cl{C}},\bd{x}_{\cl{T}}^{+}, \bd{y}^{+};\bd{u}^{+}) \, \log \frac{f(\bd{x}_{\cl{C}},\bd{x}_{\cl{T}}^{+}, \bd{y}^{+};\bd{u}^{+})}{f(\bd{x}_{\cl{C}},\bd{x}_{\cl{T}}^{+}) \ist f(\bd{y}^{+};\bd{u}^{+})}\nonumber\\[.5mm]
&\hspace{53mm}\times \mathrm{d}\bd{x}_{\cl{C}} \, \mathrm{d}\bd{x}_{\cl{T}}^{+} \, \mathrm{d}\bd{y}^{+} \rmv.
\label{eq:mutualinfo}
\end{align}
Note that $h(\bm{{\sf{x}}}_{\cl{C}}, \bm{{\sf{x}}}_{\cl{T}}^{+})$ in \eqref{eq:splitEntropy} does not depend on $\bd{u}^{+}\!$, since neither the CA states $\bd{x}_{\cl{C}}$ nor the future target states $\bd{x}_{\cl{T}}^{+}$ are controlled by the future control variable $\bd{u}^{+}\!$. We explicitly express the dependence of $I(\bm{{\sf{x}}}_{\cl{C}}, \bm{{\sf{x}}}_{\cl{T}}^{+}\ist; \bm{{\sf{y}}}^{+};\bd{u}^{+})$ on $\bd{u}^{+}\rmv$ by defining the 
\vspace{-1.7mm}
function
\begin{equation}
\label{eq:D_I_def}
D_I(\bd{u}^{+}) \,\triangleq\, I(\bm{{\sf{x}}}_{\cl{C}}, \bm{{\sf{x}}}_{\cl{T}}^{+}\ist; \bm{{\sf{y}}}^{+};\bd{u}^{+}) \,.
\vspace{.4mm}
\end{equation}
Combining \eqref{eq:objfunc_Dh}, \eqref{eq:entropyStep1}, \eqref{eq:splitEntropy}, and \eqref{eq:D_I_def} then yields the following expansion of the objective function:
\begin{equation}
D_h(\bd{u}^{+}) \eq -h(\bm{{\sf{x}}}_{\cl{C}}, \bm{{\sf{x}}}_{\cl{T}}^{+}) \ist\ist+\ist\ist D_I(\bd{u}^{+})  \ist- \sum_{l\in \cl{C}} \rmv G_{l}(\bd{u}_{l}^{+}) \,.
\label{eq:objfunc_expans}
\end{equation}
This entails the following expansion of the gradient in \eqref{eq:controller_output}:
\begin{equation}
\label{eq:controlVariable}
\frac{\partial D_h(\bd{u}^{+})}{\partial\bd{u}_{l}^{+}} \,=\,\frac{\partial D_I(\bd{u}^{+})}{\partial\bd{u}_{l}^{+}}
  \,-\, \frac{\partial G_l(\bd{u}_{l}^{+})}{\partial\bd{u}_{l}^{+}} \,.
\end{equation}

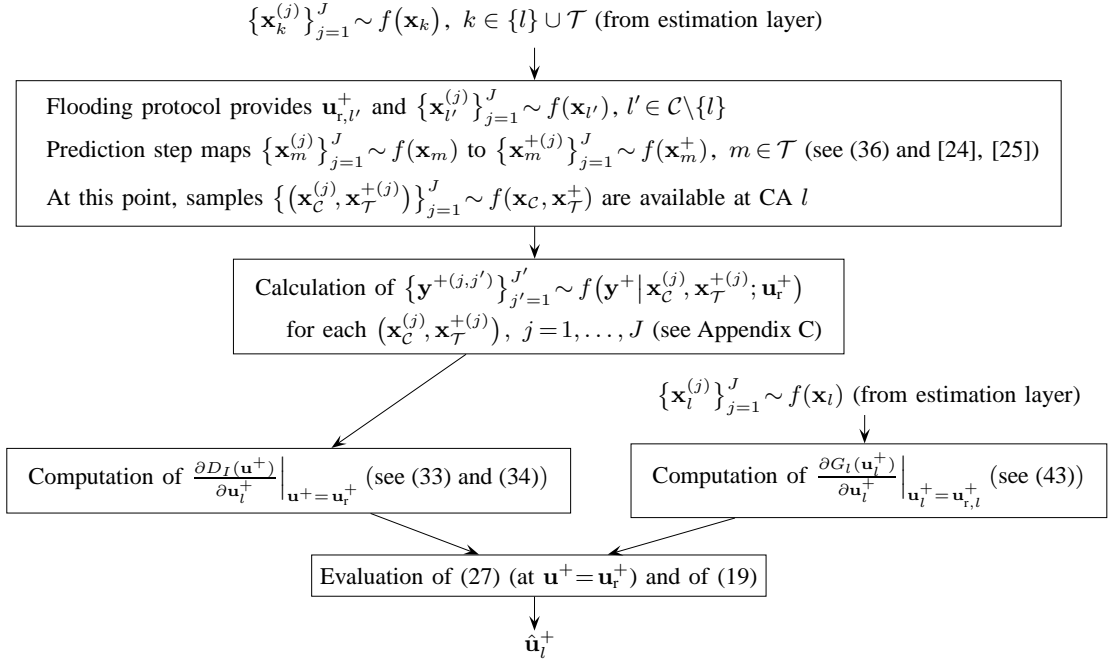
\begin{figure*}[t]
\vspace*{-2mm}
\centering \small \psset{xunit=7mm,yunit=7mm,runit=8mm}\psset{linewidth=0.3pt}
\begin{pspicture}(-3,1.8)(5.5,13.5)
\rput(0,13.1){\rnode{gb1}{ $\big\{ \bd{x}^{(j)}_{k} \big\}_{j=1}^{J} \rmv\rmv\sim\rmv f\big(\bd{x}_{k}\big)\ist$,\, $k \in \{l\} \cup \cl{T}$ (from estimation layer) }}

\rput(0,10.6){\rnode{gb1b}{ \psframebox{ $ \begin{array}{l} \text{Flooding protocol provides $\bd{u}_{\text{r},l'}^+$
and $\big\{\bd{x}^{(j)}_{l'} \big\}_{j=1}^{J} \rmv\rmv\sim\rmv f(\bd{x}_{l'})\ist$, $l' \!\in \cl{C} \backslash \{l\}$}\\[1.5mm]
\text{Prediction step maps $\big\{ \bd{x}^{(j)}_{m} \big\}_{j=1}^{J} \rmv\rmv\sim\rmv f(\bd{x}_{m})$ to 
$\big\{ \bd{x}^{+ (j)}_{m} \big\}_{j=1}^{J} \rmv\rmv\sim\rmv f(\bd{x}^{+}_{m})\ist$,\, $m \rmv\in\rmv \cl{T}$ (see\,\ist\eqref{eq:predT+}{\,\ist}and{\rmv} \cite{meyer12,meyer2014coslat})}\\[2mm] 
\text{At this point, samples $\big\{\big(\bd{x}_{\cl{C}}^{(j)}\!,\bd{x}_{\cl{T}}^{+(j)} \big)\big\}_{j=1}^{J} 
\rmv\rmv\sim\rmv f(\bd{x}_{\cl{C}},\bd{x}_{\cl{T}}^{+})$ are available at CA $l$ }
\end{array}$}}}

\rput(0,7.7){\rnode{gb2o}{ \psframebox{$ \begin{array}{l}\text{Calculation of }
  \big\{ \bd{y}^{+ (j,j')} \big\}_{j'=1}^{J'} \rmv\rmv\sim\rmv f\big(\bd{y}^{+} \big|\ist \bd{x}^{(j)}_{\cl{C}}\!,\bd{x}_{\cl{T}}^{+(j)}; \bd{u}^{+}_{\text{r}} \big)\\[1.3mm]
  \hspace*{3mm}\text{ for each $\big(\bd{x}_{\cl{C}}^{(j)}\!,\bd{x}_{\cl{T}}^{+(j)} \big) \ist, \iist j \rmv=\! 1,\dots,J$ (see$\rmv$ Appendix$\rmv$ \ref{sec:drawingCen})} \end{array}$}}}

\rput(-4.7,4.4){\rnode{gb6}{ \psframebox{$ \begin{array}{c} \text{Computation of } \frac{\partial D_I(\bd{u}^{+})}{\partial\bd{u}_{l}^{+}}\Big{|}_{\bd{u}^{+} =\ist \bd{u}^{+}_{\text{r}}} 
\ist\ist (\text{see}\,\ist\eqref{eq:MIderivative3}{\,\ist}\text{and}{\,\ist}\eqref{eq:fyparticle}) \end{array}$}}}

\rput(6.2,4.4){\rnode{gb9}{ \psframebox{$ \begin{array}{c} \text{Computation of } \frac{\partial G_l(\bd{u}_{l}^{+})}{\partial\bd{u}_{l}^{+}}\Big{|}_{\bd{u}_{l}^{+} =\ist \bd{u}^{+}_{\text{r},l}} \ist\ist (\text{see}\,\ist\eqref{eq:MCEntropyDiffInt})\end{array}$}}}

\rput(6.2,6){\rnode{gb10}{ $\big\{ \bd{x}^{(j)}_{l} \big\}_{j=1}^{J} \rmv\rmv\sim\rmv f(\bd{x}_{l})$ (from estimation layer)}} 

\rput(0,2.6){\rnode{gb8}{ \psframebox{Evaluation of \eqref{eq:controlVariable} (at $\bd{u}^{+} \!\rmv=\! \bd{u}^{+}_{\text{r}}$) and of \eqref{eq:controller_output}}}}
	
\rput(0,1.3){\rnode{gb10b}{ $\;\,\hat{\bd{u}}_{l}^{+}$ }}
\ncline[linecolor=black,nodesepB=0mm,nodesepA=1mm,arrowsize=4pt]{->}{gb1}{gb1b}
\ncline[linecolor=black,nodesepB=0mm,nodesepA=0mm,arrowsize=4pt]{->}{gb1b}{gb2o}
\ncline[linecolor=black,nodesepB=0mm,nodesepA=0mm,arrowsize=4pt]{->}{gb2o}{gb6}
\ncline[linecolor=black,nodesepB=0mm,nodesepA=0mm,arrowsize=4pt]{->}{gb6}{gb8}
\ncline[linecolor=black,nodesepB=0mm,nodesepA=0mm,arrowsize=4pt]{->}{gb9}{gb8}
\ncline[linecolor=black,nodesepB=0mm,nodesepA=0mm,arrowsize=4pt]{->}{gb10}{gb9}
\ncline[linecolor=black,nodesepB=0mm,nodesepA=0mm,arrowsize=4pt]{->}{gb8}{gb10b}

\end{pspicture}
\vspace{4mm}
\renewcommand{\baselinestretch}{1.15}\small\normalsize
\caption{Flow chart of the flooding-based implementation of the control layer at CA $l$ (see Section \ref{sec:quasiCentral}).} 
\label{fig:flowChartCen}
\vspace{4mm}
\end{figure*}

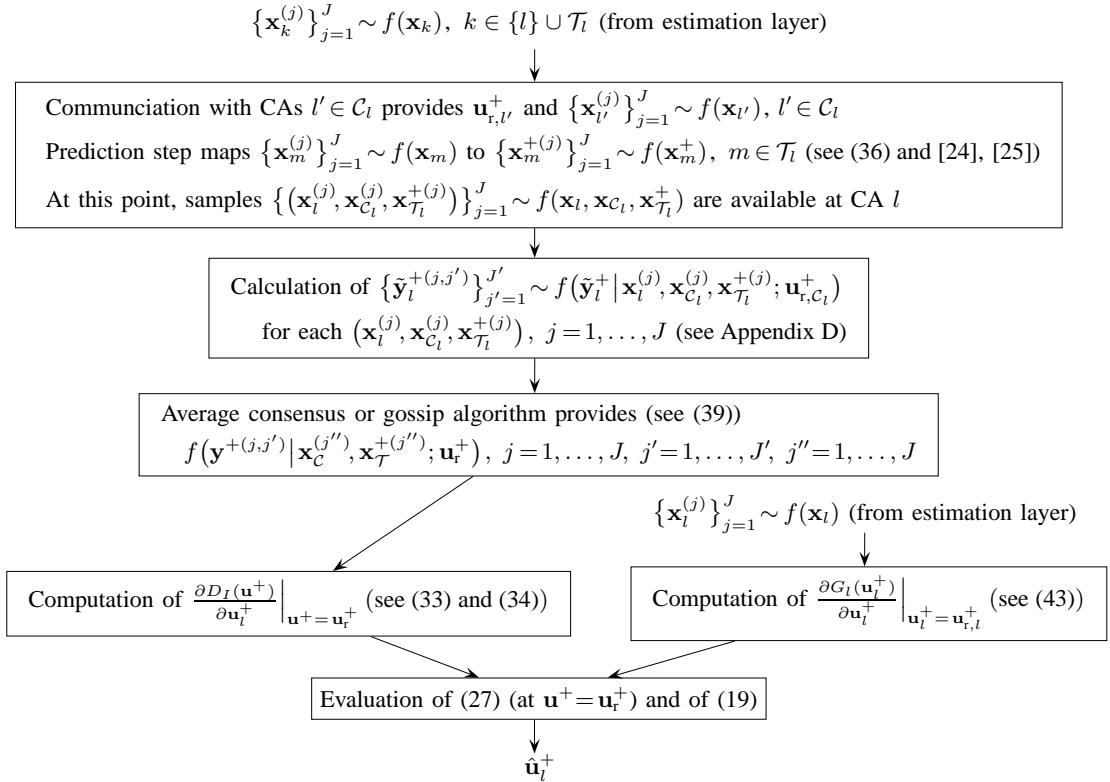
\begin{figure*}[t!]
\centering \small \psset{xunit=7mm,yunit=7mm,runit=8mm}\psset{linewidth=0.3pt}
\begin{pspicture}(-3,1.8)(5.5,13.5)
\rput(0,13.1){\rnode{gb1}{ $\big\{ \bd{x}^{(j)}_{k} \big\}_{j=1}^{J} \rmv\rmv\sim\rmv f(\bd{x}_{k})\ist$,\, $k \in \{l\} \cup \cl{T}_l\ist$ (from estimation layer)}}

\rput(0,10.6){\rnode{gb1b}{ \psframebox{ $ \begin{array}{l} 
\text{Communciation with CAs $l' \!\in \cl{C}_l$ provides $\bd{u}_{\text{r},l'}^+$
and $\big\{\bd{x}^{(j)}_{l'} \big\}_{j=1}^{J} \rmv\rmv\sim\rmv f(\bd{x}_{l'})\ist$, $l' \!\in \cl{C}_l$}\\[1.5mm]
\text{Prediction step maps $\big\{ \bd{x}^{(j)}_{m} \big\}_{j=1}^{J} \rmv\rmv\sim\rmv f(\bd{x}_{m})$ to 
$\big\{ \bd{x}^{+ (j)}_{m} \big\}_{j=1}^{J} \rmv\rmv\sim\rmv f(\bd{x}^{+}_{m})\ist$,\, $m \rmv\in\rmv \cl{T}_l$ (see\,\ist\eqref{eq:predT+}{\,\ist}and{\rmv} \cite{meyer12,meyer2014coslat})}\\[2mm] 
\text{At this point, samples $\big\{\big(\bd{x}_{l}^{(j)}\!,\bd{x}_{\cl{C}_l}^{(j)}\!,\bd{x}_{\cl{T}_l}^{+(j)} \big)\big\}_{j=1}^{J}
\rmv\rmv\sim\rmv f(\bd{x}_{l}, \bd{x}_{\cl{C}_l}, \bd{x}_{\cl{T}_l}^{+})$ are available at CA $l$ } \end{array}$}}}

\rput(0,7.7){\rnode{gb2o}{ \psframebox{$ \begin{array}{l} \text{Calculation of }
  \big\{ \tilde{\bd{y}}_{l}^{+ (j,j')} \big\}_{j'=1}^{J'} \rmv\rmv\sim\rmv f\big( \tilde{\bd{y}}_l^{+} \big|\ist \bd{x}^{(j)}_{l}\!, \bd{x}^{(j)}_{\cl{C}_{l}}\!,\bd{x}_{\cl{T}_{l}}^{+ (j)}; \bd{u}^{+}_{\text{r},\cl{C}_l}\big) 
  \\[1.7mm]
  \hspace*{3mm}\text{ for each $\big(\bd{x}_{l}^{(j)}\!,\bd{x}_{\cl{C}_l}^{(j)}\!,\bd{x}_{\cl{T}_l}^{+(j)} \big) \ist, \iist j \rmv=\! 1,\dots,J$ (see$\rmv$ Appendix$\rmv$ \ref{sec:drawingDis})}  \end{array}$}}}

\rput(0,5.3){\rnode{gb2}{  \psframebox{$  \begin{array}{l} \text{Average consensus or gossip algorithm provides (see \eqref{eq:f_2_F})} \\[.6mm] 
  \hspace*{3mm} f\big(\bd{y}^{+ (j,j')} \big|\ist  \bd{x}^{(j'')}_{\cl{C}}\!,\bd{x}_{\cl{T}}^{+(j'')}; \bd{u}^{+}_{\text{r}} \big)\ist, \iist j \rmv=\! 1,\dots,J, \iist j' \!\rmv=\! 1,\dots,J'\! , \iist j'' \!\rmv=\! 1,\dots,J \end{array}$}}}

\rput(-4.7,2.1){\rnode{gb6}{ \psframebox{$ \begin{array}{c} \text{Computation of } \frac{\partial D_I(\bd{u}^{+})}{\partial\bd{u}_{l}^{+}}\Big{|}_{\bd{u}^{+} =\ist \bd{u}^{+}_{\text{r}}} 
\ist\ist (\text{see}\,\ist\eqref{eq:MIderivative3}{\,\ist}\text{and}{\,\ist}\eqref{eq:fyparticle})\end{array}$}}}

\rput(6.2,2.1){\rnode{gb9}{ \psframebox{$ \begin{array}{c} \text{Computation of } \frac{\partial G_l(\bd{u}_{l}^{+})}{\partial\bd{u}_{l}^{+}}\Big{|}_{\bd{u}_{l}^{+} =\ist \bd{u}^{+}_{\text{r},l}} \ist\ist (\text{see}\,\ist\eqref{eq:MCEntropyDiffInt})\end{array}$}}}

\rput(6.2,3.8){\rnode{gb10}{ $\big\{ \bd{x}^{(j)}_{l} \big\}_{j=1}^{J} \rmv\rmv\sim\rmv f(\bd{x}_{l})$ (from estimation layer) }}

\rput(0,.3){\rnode{gb8}{ \psframebox{Evaluation of \eqref{eq:controlVariable} (at $\bd{u}^{+} \!\rmv=\! \bd{u}^{+}_{\text{r}}$) and of \eqref{eq:controller_output}}}}
	
\rput(0,-1){\rnode{gb10b}{ $\;\,\hat{\bd{u}}_{l}^{+}$ }}
\ncline[linecolor=black,nodesepB=0mm,nodesepA=1mm,arrowsize=4pt]{->}{gb1}{gb1b}
\ncline[linecolor=black,nodesepB=0mm,nodesepA=0mm,arrowsize=4pt]{->}{gb1b}{gb2o}
\ncline[linecolor=black,nodesepB=0mm,nodesepA=0mm,arrowsize=4pt]{->}{gb2o}{gb2}
\ncline[linecolor=black,nodesepB=0mm,nodesepA=0mm,arrowsize=4pt]{->}{gb2}{gb6}
\ncline[linecolor=black,nodesepB=0mm,nodesepA=0mm,arrowsize=4pt]{->}{gb6}{gb8}
\ncline[linecolor=black,nodesepB=0mm,nodesepA=0mm,arrowsize=4pt]{->}{gb9}{gb8}
\ncline[linecolor=black,nodesepB=0mm,nodesepA=0mm,arrowsize=4pt]{->}{gb10}{gb9}
\ncline[linecolor=black,nodesepB=0mm,nodesepA=0mm,arrowsize=4pt]{->}{gb8}{gb10b}

\end{pspicture}
\vspace{20mm}
\renewcommand{\baselinestretch}{1.15}\small\normalsize
\caption{Flow chart of the consensus-based implementation of the control layer at CA $l$ (see Section \ref{sub:Decentralized-processing}).} 
\label{fig:flowChartDis}
\vspace{2mm}
\end{figure*}

In Sections \ref{sec:firstGradient}--\ref{sec:secondGradient}, we will develop sample-based techniques for calculating $\frac{\partial D_I(\bd{u}^{+})}{\partial\bd{u}_{l}^{+}}\Big{|}_{\bd{u}^{+}=\ist \bd{u}^{+}_{\text{r}}}$ and $\frac{\partial G_l(\bd{u}_{l}^{+})}{\partial\bd{u}_{l}^{+}}\Big{|}_{\bd{u}_l^{+}=\ist \bd{u}^{+}_{\text{r},l}}$. The calculation of $\frac{\partial D_I(\bd{u}^{+})}{\partial\bd{u}_{l}^{+}}\Big{|}_{\bd{u}^{+}=\ist \bd{u}^{+}_{\text{r}}}$ is cooperative and distributed; it requires communication with neighbor CAs $l'\!\in\rmv\cl{C}_{l}$. The calculation of $\frac{\partial G_l(\bd{u}_{l}^{+})}{\partial\bd{u}_{l}^{+}}\Big{|}_{\bd{u}_l^{+}=\ist \bd{u}^{+}_{\text{r},l}}$ is 
performed locally at each CA $l$. Both calculations use the samples of relevant marginal posteriors that were computed by the estimation layer.
The operations performed in the control layer as described in this section and in Sections \ref{sec:firstGradient}--\ref{sec:secondGradient}
are summarized in Figs.\ \ref{fig:flowChartCen} and \ref{fig:flowChartDis}
for two alternative distributed implementations (discussed in Section \ref{sec:dist-proc}).

\section{Calculation of the Gradient of $D_I(\bd{u}^{+})$}
\label{sec:firstGradient} 



\vspace{.8mm}

In this section, we develop a Monte Carlo approximation of $\rmv\frac{\partial D_I(\bd{u}^{+})}{\partial\bd{u}_{l}^{+}}\Big{|}_{\bd{u}^{+}=\ist\bd{u}^{+}_{\text{r}}}$ that uses 
importance sampling.
The distributed computation of this approximation will be addressed in Section \ref{sec:dist-proc}. 

The mutual information in \eqref{eq:mutualinfo} can be rewritten as
\begin{align*}
D_I(\bd{u}^{+})
&\,= \int\!\! \int \!\! \int\rmv\rmv f(\bd{y}^{+}|\ist \bd{x}_{\cl{C}},\bd{x}_{\cl{T}}^{+};\bd{u}^{+}) \, f(\bd{x}_{\cl{C}},\bd{x}_{\cl{T}}^{+}) \\[-.8mm]
&\hspace{15mm}  \times \log \frac{f(\bd{y}^{+}|\ist \bd{x}_{\cl{C}},\bd{x}_{\cl{T}}^{+};\bd{u}^{+})}{f(\bd{y}^{+};\bd{u}^{+})} \,\ist  
  \mathrm{d}\bd{x}_{\cl{C}} \, \mathrm{d}\bd{x}_{\cl{T}}^{+} \, \mathrm{d}\bd{y}^{+} \rmv.
\end{align*}
Invoking \cite[Th.\ 1]{julian12}, we obtain
\begin{align}
\hspace{-1mm}\frac{\partial D_I(\bd{u}^{+})}{\partial\bd{u}_{l}^{+}}
&\,= \int\!\! \int \!\! \int \frac{\partial  f(\bd{y}^{+}|\ist \bd{x}_{\cl{C}},\bd{x}_{\cl{T}}^{+};\bd{u}^{+})}{\partial\bd{u}_{l}^{+}} \,
  f(\bd{x}_{\cl{C}},\bd{x}_{\cl{T}}^{+}) \nonumber\\[-.5mm]
&\hspace{6mm}  \times \log \frac{f(\bd{y}^{+}|\ist \bd{x}_{\cl{C}},\bd{x}_{\cl{T}}^{+};\bd{u}^{+})}{f(\bd{y}^{+};\bd{u}^{+})} 
  \,\ist\mathrm{d}\bd{x}_{\cl{C}} \, \mathrm{d}\bd{x}_{\cl{T}}^{+} \, \mathrm{d}\bd{y}^{+} \rmv.
  \label{eq:MIderivative}
\end{align}
The likelihood function $f(\bd{y}^{+}|\ist \bd{x}_{\cl{C}},\bd{x}_{\cl{T}}^{+};\bd{u}^{+})$ involved in \eqref{eq:MIderivative} can be written as
\begin{align}
\hspace{-2mm}f(\bd{y}^{+}|\ist \bd{x}_{\cl{C}},\bd{x}_{\cl{T}}^{+};\bd{u}^{+}) 
&\,=\ist \prod_{l\in\cl{C}} \prod_{l'\in\cl{C}_{l}}  \rmv f(\bd{y}^+_{l,l'}|\ist \bd{x}_{l},\bd{x}_{l'};\bd{u}_{l}^{+}\!,\bd{u}_{l'}^{+})\nonumber\\[0mm]
&\hspace{12mm}  \times\!\prod_{m\in\cl{T}_{l}}  \! f(\bd{y}^+_{l,m}|\ist \bd{x}_{l},\bd{x}^+_{m};\bd{u}_{l}^{+}) \,, 
  \label{eq:CoSLATJointLikelihood}\\[-9mm]
\nonumber
\end{align}
with
\begin{align}
&f(\bd{y}_{l,l'}^{+} |\ist \bd{x}_{l},\bd{x}_{l'}; \bd{u}^{+}_{l}\!, \bd{u}^{+}_{l'} ) \nn\\[.5mm]
  &\ist=\ist f(\bd{y}_{l,l'}^{+}|\ist\bd{x}^{+}_{l}\!,\bd{x}^+_{l'}) \big|_{\bd{x}^{+}_{l} =\, \tilde{g}_{l}(\bd{x}_{l},\bd{u}^{+}_{l}),\ist\ist\bd{x}^{+}_{l'} =\, \tilde{g}_{l'}(\bd{x}_{l'},\bd{u}^{+}_{l'})} \,,
 \;\,  l \!\in\rmv \cl{C} , \,l' \!\rmv\in\rmv \cl{C}_l \nn\\[-2mm]
 \label{eq:Likelihood_evol}\\[-2.5mm]
&f(\bd{y}_{l,m}^{+} |\ist \bd{x}_{l},\bd{x}^{+}_{m};\bd{u}^{+}_{l}) \nn\\[.5mm]
  &\ist=\ist f(\bd{y}_{l,m}^{+}|\ist\bd{x}^{+}_{l}\!,\bd{x}^+_{m}) \big|_{\bd{x}^{+}_{l} =\,\tilde{g}_{l}(\bd{x}_{l},\bd{u}^{+}_{l})} \,,
  \;\, l \!\in\rmv \cl{C} , \, m \!\in\rmv \cl{T}_l \,. \nn
\end{align}
(The latter expressions are obtained using \eqref{eq:meas_mod} and \eqref{eq:statrans_control}.)

Let $\tilde{\bd{y}}_l^{+}$ denote the subvector of $\bd{y}^{+} = \big[ \bd{y}_{l,k}^{+} \big]_{l\in\cl{C},\ist k\in\cl{A}_{l}}$ (cf.\ \eqref{eq:futuremeas}) whose likelihood function
includes all those factors of $f(\bd{y}^{+}|\ist \bd{x}_{\cl{C}},\bd{x}_{\cl{T}}^{+};\bd{u}^{+})$ in \eqref{eq:CoSLATJointLikelihood} that depend on the local control vector $\bd{u}_{l}^{+}\rmv$. 
This subvector is given by
\be
\tilde{\bd{y}}_l^{+} \ist\triangleq\, \Big[ \big[ \bd{y}_{l,k}^{+} \big]_{k\in\cl{A}_{l}}^{\text{T}} \; 
  \big[ \bd{y}_{l'\rmv,l}^{+} \big]_{l'\in\cl{C}_{l}}^{\text{T}} \Big]^{\text{T}} ,
\label{eq:comp-meas-D-tilde}
\ee
and its likelihood function is obtained as
\begin{align}
&\hspace*{-2mm}f(\tilde{\bd{y}}_l^{+}|\ist \bd{x}_{l}, \bd{x}_{\cl{C}_l},\bd{x}_{\cl{T}_l}^{+}; \bd{u}^{+}_{\cl{C}_l}) \nonumber\\[1mm] 
&\hspace*{-2mm}\;\,= \prod_{l'\in\cl{C}_{l}} \! f(\bd{y}^+_{l,l'}|\ist \bd{x}_{l},\bd{x}_{l'};\bd{u}_{l}^{+}\!,\bd{u}_{l'}^{+}) \, 
f(\bd{y}^+_{l'\!,l}|\ist \bd{x}_{l'},\bd{x}_{l};\bd{u}_{l'}^{+},\bd{u}_{l}^{+}) \nn\\[-2mm]
 & \hspace{38mm}\times\! \prod_{m\in\cl{T}_{l}} \! f(\bd{y}^+_{l,m}|\ist \bd{x}_{l},\bd{x}^+_{m};\bd{u}_{l}^{+}) \,, \label{eq:firstLikeL} \\[-6mm] 
  \nn
\end{align}
with $\bd{u}^{+}_{\cl{C}_l} \!\triangleq\! \big[\bd{u}_{l'}^{+}\big]_{l'\in \{l\} \cup \cl{C}_{l}}$.
By comparing \eqref{eq:firstLikeL} with \eqref{eq:CoSLATJointLikelihood}, it is seen that $\tilde{\bd{y}}_l^{+}$ is also the subvector of $\bd{y}^{+}$ whose likelihood function 
includes all those factors of $f(\bd{y}^{+}|\ist \bd{x}_{\cl{C}},\bd{x}_{\cl{T}}^{+};\bd{u}^{+})$ in \eqref{eq:CoSLATJointLikelihood} that involve the state $\bd{x}_l^{+}\rmv$.




Using \eqref{eq:CoSLATJointLikelihood} and \eqref{eq:firstLikeL}, it is shown in Appendix \ref{sec:sampleBasedCalcGradient-A} that a Monte Carlo (i.e., sample-based) approximation of \eqref{eq:MIderivative} evaluated at $\bd{u}^{+} \!\rmv=\! \bd{u}^{+}_{\text{r}}$ is given
by\footnote{With an abuse of notation, the superscript $(j)$ now indicates the $j$th sample, whereas previously, in our full-blown notation, the superscript $(n)$ indicated the $n$th time step.}
\vspace{-1.5mm}
\begin{align}
&\hspace*{0mm}\frac{\partial D_I(\bd{u}^{+})}{\partial\bd{u}_{l}^{+}}\bigg{|}_{\bd{u}^{+}=\ist \bd{u}^{+}_{\text{r}}} \nn\\[.5mm]
&\hspace*{3mm} \approx \frac{1}{J \rmv J'}\sum_{j=1}^{J}\sum_{j'=1}^{J'} \ist 
  \frac{1}{f\big(\tilde{\bd{y}}_l^{+(j,j')} \big|\ist \bd{x}^{(j)}_{l}\!, \bd{x}^{(j)}_{\cl{C}_l}\!,\bd{x}^{+(j)}_{\cl{T}_l}; \bd{u}^{+}_{\text{r},\ist\cl{C}_l}\big)}
  \nonumber\\[.5mm]
& \hspace*{10mm}\times\frac{\partial f\big(\tilde{\bd{y}}_l^{+(j,j')} \big|\ist \bd{x}^{(j)}_{l}\!, \bd{x}^{(j)}_{\cl{C}_l}\!,\bd{x}^{+(j)}_{\cl{T}_l}; \bd{u}^{+}_{\ist\cl{C}_l}\big)}{\partial\bd{u}_{l}^{+}}\bigg{|}_{\bd{u}_{\cl{C}_l}^{+} =\ist \bd{u}^{+}_{\text{r},\cl{C}_l}} \nn\\[0mm]
 & \hspace*{10mm}\times \log\frac{f\big(\bd{y}^{+(j,j')} \big|\ist \bd{x}_{\cl{C}}^{(j)}\!,\bd{x}_{\cl{T}}^{+(j)}; \bd{u}^{+}_{\text{r}}\big)}{f\big(\bd{y}^{+(j,j')}; \bd{u}^{+}_{\text{r}}\big)} 
   \,,\label{eq:MIderivative3}\\[-8mm] \nn
\end{align}
with
\begin{equation}
\hspace*{-1.8mm}f\big(\bd{y}^{+(j,j')}; \bd{u}^{+}_{\text{r}}\big)  
  \approx \frac{1}{J} \rmv \sum_{j''=1}^J  \rmv f\big(\bd{y}^{+(j,j')} \big|\ist \bd{x}_{\cl{C}}^{(j'')}\!,\bd{x}_{\cl{T}}^{+(j'')}; \bd{u}^{+}_{\text{r}}\big) \ist.
\label{eq:fyparticle}
\end{equation}
Here, $\bd{y}^{+(j,j')}\rmv$, $\bd{x}_{\cl{C}}^{(j)}\!$, and $\bd{x}_{\cl{T}}^{+(j)}$ are samples of $\bd{y}^{+}\!$, $\bd{x}_{\cl{C}}$, and $\bd{x}_{\cl{T}}^{+}$,
respectively that are drawn from the importance density \cite{doucet}
$q(\bd{y}^{+}\!, \bd{x}_{\cl{C}},\bd{x}_{\cl{T}}^{+}) \triangleq f(\bd{x}_{\cl{C}},\bd{x}_{\cl{T}}^{+}) \ist f(\bd{y}^{+}| \ist \bd{x}_{\cl{C}},\bd{x}_{\cl{T}}^{+}; \bd{u}_{\text{r}}^{+})$ 
via the following two-stage procedure. First, samples $\big\{ \big(\bd{x}_{\cl{C}}^{(j)}\!,$\linebreak $\bd{x}_{\cl{T}}^{+(j)} \big) \big\}_{j=1}^J$ are drawn from 
\begin{equation}
f(\bd{x}_{\cl{C}},\bd{x}_{\cl{T}}^{+}) \eq \prod_{l\in\cl{C}}f(\bd{x}_{l}) \rmv\prod_{m\in\cl{T}} \! f(\bd{x}_{m}^{+}) \,.
\label{eq:statIndep}
\end{equation}
(This factorization expresses the conditional statistical independence of the $\bd{x}_{l}$, $l \in\cl{C}$ and the $\bd{x}_{m}^{+}$, $m\in\cl{T}$ given $\bd{y}^{(1:n)}\rmv$.
This is a common approximation, which was introduced in \cite{wymeersch} and is also used in the estimation layer \cite{meyer12,meyer2014coslat}.) Then, for each sample 
$\big(\bd{x}_{\cl{C}}^{(j)}\!,\bd{x}_{\cl{T}}^{+(j)} \big)$, samples $\big\{ \bd{y}^{+(j,j')} \big\}_{j'=1}^{J'}$ are drawn from the conditional pdf 
$f\big(\bd{y}^{+} \big|\ist  \bd{x}^{(j)}_{\cl{C}}\!,\bd{x}_{\cl{T}}^{+(j)}; \bd{u}^{+}_{\text{r}} \big)$. 
The distributed calculation of these samples will be discussed in Section \ref{sec:dist-proc} and in Appendix \ref{sec:drawingCen}. Finally, we note that using \eqref{eq:firstLikeL}, one easily obtains a (rather unwieldy) expression of the derivative 
$\frac{\partial f(\tilde{\bd{y}}_l^{+}|\ist \bd{x}_{l}, \ist\bd{x}_{\cl{C}_l},\ist\bd{x}_{\cl{T}_l}^{+}; \ist\bd{u}^{+}_{\cl{C}_l}) }{\partial\bd{u}_{l}^{+}}$ occurring in \eqref{eq:MIderivative3}. This expression involves the factors in \eqref{eq:firstLikeL} and the derivatives $\frac{\partial \tilde{g}_l(\bd{x}_{l},\bd{u}_{l}^{+})}{\partial\bd{u}_{l}^{+}}$, 
$\frac{\partial f(\bd{y}^+_{l,l'}|\ist \bd{x}^+_{l}\!,\ist\bd{x}_{l'}^{+})}{\partial\bd{x}_{l}^{+}}$ for $l' \!\in \cl{C}_l$, 
and $\frac{\partial f(\bd{y}^+_{l,m}|\ist \bd{x}^+_{l}\!,\ist\bd{x}^+_{m})}{\partial\bd{x}_{l}^{+}}$ for $m \rmv\in\rmv \cl{T}_l$.

\vspace{-1mm}

\section{Distributed Processing}
\label{sec:dist-proc} 

\vspace{.5mm}

In this section, we present two alternative schemes for a distributed, sample-based computation of $\frac{\partial D_I(\bd{u}^{+})}{\partial\bd{u}_{l}^{+}}\Big{|}_{\bd{u}^{+}=\ist \bd{u}^{+}_{\text{r}}}\!$ according to \eqref{eq:MIderivative3} and \eqref{eq:fyparticle}. 
\vspace{.3mm}
Both schemes are distributed in that they require only local communication, i.e., each CA $l \rmv\in\rmv \cl{C}$ transmits data only to its neighbors $l' \!\rmv\in\rmv \cl{C}_l$.


\subsection{Flooding-Based Processing} 
\label{sec:quasiCentral}

We first discuss a distributed scheme where each CA $l \in \cl{C}$ performs an exact (``quasi-centralized'') calculation of \eqref{eq:MIderivative3} and \eqref{eq:fyparticle}. As a result of the estimation layer, samples $\big\{\bd{x}^{(j)}_{k} \big\}_{j=1}^J \!\sim f(\bd{x}_{k})$, $k \in \{l\} \cup \cl{T}$ are available at CA $l$ (see \eqref{eq:CoSLATlocal}, noting that $f(\bd{x}_k)$ was denoted $f(\bd{x}_k\ist|\ist\bd{y})$ there). A flooding protocol \cite{lim00} is now used to
make available to each CA $l$ the reference vectors $\bd{u}^+_{\text{r},l'}$ and the samples $\big\{\bd{x}^{(j)}_{l'} \big\}_{j=1}^J \!\sim\rmv f(\bd{x}_{l'})$ 
of all the other CAs $l'\!\in \cl{C}\ist\backslash \{l\}$. 
The flooding protocol requires each CA $l$ to communicate with its neighbor CAs $l' \!\rmv\in\rmv \cl{C}_l$.
In addition, CA $l$ locally calculates predictive marginal posteriors for all target states 
via the following prediction step (which is \eqref{eq:pred_f-recurs} with $n$ replaced by $n+1$):
\begin{equation}
\label{eq:predT+}
f(\bd{x}^+_{m}) \,=\int \rmv\rmv f(\bd{x}^+_{m}|\ist \bd{x}_{m}) \ist\ist f(\bd{x}_{m}) \,\mathrm{d}\bd{x}_{m} \,, \quad m \rmv\in\rmv \cl{T} .
\end{equation}
An implementation of \eqref{eq:predT+} using the samples $\big\{\bd{x}^{(j)}_m\big\}^{J}_{j=1}$, $m \!\in\! \cl{T}$
produced by the estimation layer (which are available at CA $l$) and yielding samples $\big\{\bd{x}^{+(j)}_m\big\}^{J}_{j=1} \!\sim\rmv f(\bd{x}^+_{m})$, $m \!\in\! \cl{T}$ 
is described in \cite{meyer12,meyer2014coslat}. 
At this point, 
samples $\big\{\bd{x}^{(j)}_{l'} \big\}_{j=1}^J\!\sim f(\bd{x}_{l'})$ for $l'\! \in \cl{C}$ and $\big\{\bd{x}^{+(j)}_{m} \big\}_{j=1}^J\!\sim f(\bd{x}_{m}^+)$ for $m \in \cl{T}$ are available at CA $l$. 
Because all states $\bd{x}_{l}$, $l \in \cl{C}$ and $\bd{x}^{+}_{m}$, $m \in \cl{T}$ are conditionally independent given $\bd{y}^{(1:n)}$ (see \eqref{eq:statIndep}), 
CA $l$ can now obtain samples $\big\{\big(\bd{x}_{\cl{C}}^{(j)}\!,\bd{x}_{\cl{T}}^{+(j)} \big)\big\}_{j=1}^{J} \!\sim\rmv f\big(\bd{x}_{\cl{C}},\bd{x}_{\cl{T}}^{+} \big)$
by the simple stacking operations 
$\bd{x}_{\cl{C}}^{(j)} = \big[ \bd{x}^{(j)}_{l'} \big]_{l'\in \cl{C}}\ist$ and $\bd{x}_{\cl{T}}^{+(j)} = \big[ \bd{x}^{+(j)}_m \big]_{m\in \cl{T}}\ist$. 
Finally, for each $\big(\bd{x}_{\cl{C}}^{(j)}\!,\bd{x}_{\cl{T}}^{+(j)} \big)$, CA $l$ computes samples $\big\{ \bd{y}^{+(j,j')} \big\}_{j'=1}^{J'} \!\sim\rmv 
f\big(\bd{y}^{+} \big|\ist  \bd{x}^{(j)}_{\cl{C}}\!,\bd{x}_{\cl{T}}^{+(j)}; \bd{u}^{+}_{\text{r}} \big)$ as described in Appendix \ref{sec:drawingCen}.

Using the samples $\big(\bd{x}_{\cl{C}}^{(j)}\!,\bd{x}_{\cl{T}}^{+(j)} \big)$ and $\bd{y}^{+(j,j')}\rmv$, $j=1,\ldots,J$, $j' \rmv=1,\ldots,J'\rmv$,
as well as the reference vectors $\bd{u}^+_{\text{r},l'}$, $l'\!\in \cl{C}$, CA $l$ can compute
the gradient $\frac{\partial D_I(\bd{u}^{+})}{\partial\bd{u}_{l}^{+}}\Big{|}_{\bd{u}^{+}=\ist \bd{u}^{+}_{\text{r}}}$ locally
according to \eqref{eq:MIderivative3} and \eqref{eq:fyparticle}. Note, however, that this flooding-based scheme presupposes that each 
CA $l$ knows the state evolution models \eqref{eq:statrans_l} and the measurement models \eqref{eq:meas_mod} of all the other CAs $l'\!\in \cl{C}\ist\backslash \{l\}$.

The communication cost of the flooding-based scheme, in terms of the number of real values transmitted by each CA, is $(J M \rmv+\rmv M_u)\ist W \approx J M \ist W\rmv$. Here, $M$ and $M_u$ are the dimensions of the vectors $\bd{x}_{l}$ and $\bd{u}_l$, respectively, and $W$ depends on the network size and topology and is bounded as $1 \rmv\leq\rmv W \rmv\leq |\cl{C}|$. Thus, the number of transmissions scales linearly with $J$ and does not depend on $J'\!$. In large networks, flooding protocols tend to require a large memory and book-keeping overhead and introduce a significant delay \cite{xiao05}. If the network formed by the CAs is \emph{fully} connected, i.e., $\cl{C} = \{l\}\cup \cl{C}_l$, then all the samples $\big\{ \big(\bd{x}_{\cl{C}}^{(j)}\!,\bd{x}_{\cl{T}}^{+(j)} \big) \big\}_{j=1}^J$ can be obtained without flooding: CA $l$ simply broadcasts its reference vector $\bd{u}^+_{\text{r},l}$ and its samples $\big\{\bd{x}^{(j)}_{l} \big\}_{j=1}^J \!\sim\rmv f(\bd{x}_{l})$   
to all the other CAs in the network and receives their reference vectors and samples. Here, the number of real values transmitted by each CA is only $J M + M_u$. 

Finally, the computational complexity per CA of the flooding-based scheme---i.e., evaluation of \eqref{eq:MIderivative3} and \eqref{eq:fyparticle}, with $J$ and $J'$ fixed---scales linearly with the number of agents in the network. Because the computational complexity and the communication cost increase with the network size, the flooding-based distributed processing scheme is primarily suited to small networks.


\subsection{Consensus-Based Processing}
\label{sub:Decentralized-processing}

Next, we present an alternative distributed computation of \eqref{eq:MIderivative3} and \eqref{eq:fyparticle} that avoids the use of a flooding protocol and does not require each CA to know the state evolution and measurement models of all the other CAs. As a first step, CA $l$ broadcasts its own samples $\big\{\bd{x}^{(j)}_{l} \big\}_{j=1}^J \!\sim f(\bd{x}_{l})$ calculated in the estimation layer to all neighbor CAs $l' \!\in\rmv \cl{C}_l\ist$, and it receives from them their own samples $\big\{\bd{x}^{(j)}_{l'} \big\}_{j=1}^J \!\sim f(\bd{x}_{l'})$, $l' \!\in\rmv \cl{C}_l\ist$. In addition, CA $l$ locally calculates samples $\big\{\bd{x}^{+(j)}_{m} \big\}_{j=1}^J \!\sim\rmv f(\bd{x}_{m}^+)$ for all $m \in \cl{T}_l$ via the prediction step \eqref{eq:predT+} (with $\cl{T}$ replaced by $\cl{T}_l$), using the sample-based implementation described in \cite{meyer12,meyer2014coslat}.
Thus, after the stacking operations $\bd{x}^{(j)}_{\cl{C}_l} = \big[ \bd{x}^{(j)}_{l'} \big]_{l'\in \cl{C}_l}$ and $\bd{x}^{+ (j)}_{\cl{T}_l} = \big[ \bd{x}^{+(j)}_m \big]_{m\in \cl{T}_l}\ist$, samples $\big\{ \big(\bd{x}^{(j)}_{l}\!,\bd{x}^{(j)}_{\cl{C}_l},\bd{x}^{+ (j)}_{\cl{T}_l}\big) \big\}_{j=1}^{J} \!\rmv\sim\rmv f(\bd{x}_{l},\bd{x}_{\cl{C}_l},\bd{x}^{+}_{\cl{T}_l})$ are available at CA $l$. 
Then, for each sample $\big(\bd{x}^{(j)}_{l}\!,\bd{x}^{(j)}_{\cl{C}_l},\bd{x}^{+ (j)}_{\cl{T}_l}\big)$, $J'$ samples $\big\{ \tilde{\bd{y}}_l^{+(j,j')} \big\}_{j'=1}^{J'} \!\sim\rmv f\big(\tilde{\bd{y}}_l^{+} \big|\ist \bd{x}^{(j)}_{l}\!, \bd{x}^{(j)}_{\cl{C}_l}\!,\bd{x}_{\cl{T}_l}^{+ (j)}; \bd{u}^{+}_{\text{r},\cl{C}_l}\big)$ are computed as described in Appendix \ref{sec:drawingDis}. This only involves communication between neighboring CAs.

The key question at this point is as to whether the quantities 
$f\big(\tilde{\bd{y}}_l^{+ (j,j')} \big|\ist \bd{x}^{(j)}_{l}\!, \bd{x}^{(j)}_{\cl{C}_{l}}\!,\bd{x}^{+ (j)}_{\cl{T}_{l}}; \bd{u}^{+}_{\text{r},\cl{C}_l}\big)$, 
$\frac{\partial f(\tilde{\bd{y}}_l^{+ (j,j')} |\ist \bd{x}^{(j)}_{l}\!, \ist\bd{x}^{(j)}_{\cl{C}_{l}}\!,\ist\bd{x}^{+ (j)}_{\cl{T}_{l}}; \ist\bd{u}^{+}_{\cl{C}_l})}{\partial\bd{u}_{l}^{+}}\Big{|}_{\bd{u}_{\cl{C}_l}^{+} =\ist \bd{u}^{+}_{\text{r},\cl{C}_l}}$, and 
$f\big(\bd{y}^{+ (j,j')} \big|\ist  \bd{x}^{(j'')}_{\cl{C}}\!,$ $\bd{x}_{\cl{T}}^{+(j'')};\bd{u}^{+}_{\text{r}} \big)$ 
(and, in particular, $f\big(\bd{y}^{+ (j,j')} \big|\ist  \bd{x}^{(j)}_{\cl{C}}\!,$ $\bd{x}_{\cl{T}}^{+(j)};\bd{u}^{+}_{\text{r}} \big)$) 
involved in \eqref{eq:MIderivative3} and \eqref{eq:fyparticle} are locally available at CA $l$. 
The factors of $f\big(\tilde{\bd{y}}_l^{+} \big|\ist \bd{x}_{l}, \bd{x}_{\cl{C}_{l}},\bd{x}^{+}_{\cl{T}_{l}}; \bd{u}^{+}_{\text{r},\cl{C}_l}\big)$ (see \eqref{eq:firstLikeL}) correspond to measurements to be acquired by CA $l$ or by its neighbor CAs $l' \in \cl{C}_l$; they are known to CA $l$ since its own state evolution 
and measurement models and those of its neighbors are known to CA $l$ (cf.\ \eqref{eq:Likelihood_evol}). Thus, we conclude that the 
$f\big(\tilde{\bd{y}}_l^{+ (j,j')} \big|\ist \bd{x}^{(j)}_{l}\!, \bd{x}^{(j)}_{\cl{C}_{l}}\!,\bd{x}^{+ (j)}_{\cl{T}_{l}}; \bd{u}^{+}_{\text{r},\cl{C}_l}\big)$ are available at CA $l$.
On the other hand, many of the factors of $f(\bd{y}^{+} \big|\ist  \bd{x}_{\cl{C}}, \bd{x}_{\cl{T}}^{+}; \bd{u}^{+}_{\text{r}} )$ (see \eqref{eq:CoSLATJointLikelihood}) correspond to measurements to be acquired by CAs that are not in the neighborhood of CA $l$; they are not known to CA $l$ since, typically, the respective state evolution and measurement models are unknown to CA $l$. Therefore, the $f\big(\bd{y}^{+ (j,j')} \big|\ist  \bd{x}^{(j'')}_{\cl{C}}\!,\bd{x}_{\cl{T}}^{+(j'')}; \bd{u}^{+}_{\text{r}} \big)$ are not available at CA $l$.

We will now present a distributed computation of $f\big(\bd{y}^{+ (j,j')} \big|\ist  \bd{x}^{(j'')}_{\cl{C}}\!,\bd{x}_{\cl{T}}^{+(j'')}; \bd{u}^{+}_{\text{r}} \big)$. 
Let $\bd{y}_l^{+}$ denote the subvector of $\bd{y}^{+} = \big[ \bd{y}_{l,k}^{+} \big]_{l\in\cl{C},\ist k\in\cl{A}_{l}}$ in \eqref{eq:futuremeas}
that comprises the measurements acquired by CA $l$ at the next time, i.e.,
\be
\bd{y}_l^{+} \ist\triangleq\, \big[ \bd{y}_{l,k}^{+} \big]_{k\in\cl{A}_{l}}.
\label{eq:comp-meas-D-1}
\ee
The likelihood function of $\bd{y}_l^{+}$ combines all the factors in \eqref{eq:CoSLATJointLikelihood} that involve the entries of $\bd{y}_l^{+}$, i.e.,
\begin{align}
&\hspace{-2mm}f\big(\bd{y}_l^{+} \big|\ist \bd{x}_{l},\bd{x}_{\cl{C}_l},\bd{x}_{\cl{T}_l}^{+};\bd{u}^{+}_{\cl{C}_l} \big) \nn \\[.5mm]
&\hspace{-1.5mm}=\rmv \prod_{l'\in\cl{C}_{l}} \!\rmv f(\bd{y}^+_{l,l'}|\ist \bd{x}_{l},\bd{x}_{l'};\bd{u}_{l}^{+}\!,\bd{u}_{l'}^{+}) 
  \!\prod_{m\in\cl{T}_{l}} \!\rmv f(\bd{y}^+_{l,m}|\ist \bd{x}_{l},\bd{x}^+_{m};\bd{u}_{l}^{+}) \,.
 \label{eq:localLikelihoods} \\[-5.5mm] \nn
\end{align}
Using \eqref{eq:CoSLATJointLikelihood} and \eqref{eq:localLikelihoods}, one can 
show 
that 
\be
f\big(\bd{y}^{+ (j,j')} \big|\ist  \bd{x}^{(j'')}_{\cl{C}}\!,\bd{x}_{\cl{T}}^{+(j'')}; \bd{u}^{+}_{\text{r}} \big) \ist=\, \exp \rmv\big(|\cl{C}| \ist F_{j,j'\!,j''} \big) \,,\rule[-3mm]{0mm}{4mm}
\label{eq:f_2_F}
\vspace{-1mm}
\ee
where
\vspace{.5mm}
\be
F_{j,j'\!,j''} \triangleq\ist \frac{1}{|\cl{C}|}\sum_{l\in\mathcal{C}}F_{j,j'\!,j''}^{(l)} 
\label{eq:Fmatrix}
\vspace{-2mm}
\ee
with
\vspace*{.5mm}
\be
F_{j,j'\!,j''}^{(l)} \triangleq\, \log f\big(\bd{y}_l^{+(j,j')} \big|\ist \bd{x}^{(j'')}_{l}\!, \bd{x}^{(j'')}_{\cl{C}_l}\!, \bd{x}^{+ (j'')}_{\cl{T}_l}; \bd{u}^{+}_{\text{r},  \cl{C}_l}\big) \, ,
\rule[-3mm]{0mm}{4mm}
\label{eq:consInit}
\vspace{0mm}
\ee
for $j \ist=\ist 1,\dots,J$, $j' \!=\ist 1,\dots,J'\!$, and $j'' \!=\ist 1,\dots,J\!$. To compute $F_{j,j'\!,j''}^{(l)}$ in \eqref{eq:consInit}, CA $l$ 
needs samples 
$\big\{ \bd{y}_l^{+(j,j')} \big\}_{j'=1}^{J'} \!\sim f\big(\bd{y}_l^{+} \big|\ist \bd{x}^{(j)}_{l}\!, \bd{x}^{(j)}_{\cl{C}_l}\!,\bd{x}_{\cl{T}_l}^{+ (j)}; \bd{u}^{+}_{\text{r},\cl{C}_l}\big)$ 
and the reference vectors $\bd{u}_{\text{r},l'}^+$ for $l' \!\in \{l\} \cup \cl{C}_l$.
The samples $\big\{ \bd{y}_l^{+(j,j')} \big\}_{j'=1}^{J'}$ are already available at CA $l$ since $\bd{y}^{+}_l$ is a subvector of $\tilde{\bd{y}}^{+}_l$ 
(see \eqref{eq:comp-meas-D-tilde} and \eqref{eq:comp-meas-D-1}) and samples $\big\{ \tilde{\bd{y}}_l^{+(j,j')} \big\}_{j'=1}^{J'} \!\sim\rmv f\big(\tilde{\bd{y}}_l^{+} \big|\ist \bd{x}^{(j)}_{l}\!, \bd{x}^{(j)}_{\cl{C}_l}\!,\bd{x}_{\cl{T}_l}^{+ (j)}; \bd{u}^{+}_{\text{r},\cl{C}_l}\big)$ have already been computed as described above.
The $\bd{u}_{\text{r},l'}^+$ can be obtained at CA $l$ through communication with the neighbor CAs $l' \!\in \cl{C}_l$.

Once the $F_{j,j'\!,j''}^{(l)}$ have been calculated at CA $l$, their averages $F_{j,j'\!,j''}$ in \eqref{eq:Fmatrix} can be computed in a distributed way by using $J^2\rmv J'$ parallel instances of an average consensus or gossip scheme \cite{olfati07,dimakis10}. These schemes are iterative; they are initialized at each CA $l$ with $F_{j,j'\!,j''}^{(l)}$. They are robust to communication link failures \cite{olfati07,dimakis10} and use only communication between neighbor CAs (i.e., each CA $l \rmv\in\rmv \cl{C}$ transmits data to each neighbor $l' \!\rmv\in\rmv \cl{C}_l$).  After convergence of the consensus or gossip scheme, $F_{j,j'\!,j''}$ and, hence, $f\big(\bd{y}^{+ (j,j')} \big|\ist  \bd{x}^{(j'')}_{\cl{C}}\!,\bd{x}_{\cl{T}}^{+(j'')}; \bd{u}^{+}_{\text{r}} \big)$ for all $j, j' \!, j''\!$ is available at each CA $l$. 

At this point, CA $l$ has available 
$f\big(\tilde{\bd{y}}_l^{+ (j,j')} \big|\ist \bd{x}^{(j)}_{l}\!, \bd{x}^{(j)}_{\cl{C}_{l}}\!,$\linebreak 
$\bd{x}^{+ (j)}_{\cl{T}_{l}}; \bd{u}^{+}_{\text{r},\cl{C}_l}\big)$ and 
$\frac{\partial f(\tilde{\bd{y}}_l^{+ (j,j')} |\ist \bd{x}^{(j)}_{l}\!, \ist\bd{x}^{(j)}_{\cl{C}_{l}}\!,\ist\bd{x}^{+ (j)}_{\cl{T}_{l}}; \ist\bd{u}^{+}_{\cl{C}_l})}{\partial\bd{u}_{l}^{+}}\Big{|}_{\bd{u}_{\cl{C}_l}^{+} =\ist \bd{u}^{+}_{\text{r},\cl{C}_l}}$, and an approximation of
$f\big(\bd{y}^{+ (j,j')} \big|\ist  \bd{x}^{(j'')}_{\cl{C}}\!,\bd{x}_{\cl{T}}^{+(j'')}; \bd{u}^{+}_{\text{r}} \big)$ has been provided by the consensus or gossip scheme, for $j \ist=\ist 1,\dots,J$, $j' \!=\ist 1,\dots,J'\!$, and $j'' \!=\ist 1,\dots,J\!$. Therefore, CA $l$ is now able to evaluate \eqref{eq:MIderivative3} and \eqref{eq:fyparticle}.

In the course of the overall distributed computation, CA $l$ transmits $J^2J'|\cl{C}_l|R + JJ'|\cl{C}_l|M_y + JM + M_u \approx J^2J'|\cl{C}_l|R$ real values, where $R$ is the number of iterations used for one instance of the consensus or gossip scheme and $M_y$ is the dimension of the vectors $\bd{y}_{l,k}$. Asymptotically, for $R \rightarrow \infty$, this distributed computation of $\frac{\partial D_I(\bd{u}^{+})}{\partial\bd{u}_{l}^{+}}\Big{|}_{\bd{u}^{+}=\ist \bd{u}^{+}_{\text{r}}}\!$ \vspace{.3mm} converges to the exact centralized result given by \eqref{eq:MIderivative3} and \eqref{eq:fyparticle}. The speed of convergence depends on the topology and size of the network \cite{olfati07,dimakis10}. As $R$ increases, the information available at each agent converges, which means that local data is disseminated over large distances in the network. However, because the control vector of a given CA might not be strongly affected by information from far away CAs, a small $R$ might be sufficient for good performance.
Because the communication requirements are proportional to $J^2J'\rmv$, they are typically higher than those of the flooding-based scheme discussed in Section \ref{sec:quasiCentral} unless the network is large and $R$ is small.

Finally, the computational complexity of the distributed processing---i.e., evaluation of \eqref{eq:MIderivative3} and \eqref{eq:fyparticle}, with $J$, $J'$, and $R$ fixed---is constant in the number of agents in the network.


\section{Calculation of the Gradient of $G_l(\bd{u}_{l}^{+})$}
\label{sec:secondGradient} 

\vspace{.5mm}

Next, we consider the second gradient in the expansion \eqref{eq:controlVariable}, i.e., 
$\frac{\partial G_l(\bd{u}_{l}^{+})}{\partial\bd{u}_{l}^{+}} \Big{|}_{\bd{u}_{l}^{+}=\ist \bd{u}^{+}_{\text{r},l}}$.
Using \eqref{eq:G_def}, we obtain
\begin{align}
&\frac{\partial G_l(\bd{u}_{l}^{+})}{\partial\bd{u}_{l}^{+}}\bigg{|}_{\bd{u}_{l}^{+} =\ist \bd{u}^{+}_{\text{r},l}} \nonumber\\[.8mm]
&\;=  \int \! f(\bd{x}_{l}) \, \frac{\partial \log | J_{\tilde{g}_l}(\bd{x}_{l};\bd{u}_{l}^{+}) |}{\partial\bd{u}_{l}^{+}} 
  \bigg{|}_{\bd{u}_{l}^{+} =\ist \bd{u}^{+}_{\text{r},l}} \mathrm{d}\bd{x}_{l} \nn \\[.5mm]
&\;= \int \! f(\bd{x}_{l}) \, \frac{1}{|J_{\tilde{g}_l}(\bd{x}_{l};\bd{u}^{+}_{\text{r},l})|} \, \frac{\partial |J_{\tilde{g}_l}(\bd{x}_{l};\bd{u}_{l}^{+})|}{\partial\bd{u}_{l}^{+}} 
    \bigg{|}_{\bd{u}_{l}^{+} =\ist \bd{u}^{+}_{\text{r},l}} \mathrm{d}\bd{x}_{l} \,.
\label{eq:EntropyDiffInt}
\end{align}
Here, we assumed that $| J_{\tilde{g}_l}(\bd{x}_{l};\bd{u}_{l}^{+}) |$ is continuous and satisfies
$|f(\bd{x}_{l}) \, \partial \log | J_{\tilde{g}_l}(\bd{x}_{l};\bd{u}_{l}^{+}) |/\partial\bd{u}_{l}^{+}| \leq \alpha(\bd{x}_{l},\bd{u}_{l}^{+})$ for all $(\bd{x}_{l},\bd{u}_{l}^{+})$,
for some function $\alpha(\bd{x}_{l},\bd{u}_{l}^{+}) \geq 0$ that is integrable with respect to $\bd{x}_{l}$ for each $\bd{u}_{l}^{+}$ \cite[Cor.\ 5.9]{bartle95}. 
Furthermore, we assumed that for each value of $\bd{x}_{l}$, $| J_{\tilde{g}_l}(\bd{x}_{l};\bd{u}_{l}^{+}) |$ is differentiable with respect to $\bd{u}_{l}^{+}$ at $\bd{u}^{+}_{\text{r},l}$. A sufficient condition is that $J_{\tilde{g}_l}(\bd{x}_{l};\bd{u}_{l}^{+})$ is differentiable with respect to $\bd{u}_{l}^{+}$ at $\bd{u}^{+}_{\text{r},l}$ and nonzero for all $\bd{u}_{l}^{+}$ in some (arbitrarily small) neighborhood of $\bd{u}^{+}_{\text{r},l}$.

Based on the samples $\big\{ \bd{x}_{l}^{(j)} \big\}_{j=1}^J \!\sim\rmv f(\bd{x}_{l})$ that were calculated in the estimation layer, a Monte Carlo approximation of 
\eqref{eq:EntropyDiffInt} 
is obtained as 
\begin{align}
&\frac{\partial G_l(\bd{u}_{l}^{+})}{\partial\bd{u}_{l}^{+}}\bigg{|}_{\bd{u}_{l}^{+} =\ist \bd{u}^{+}_{\text{r},l}}\nonumber\\[0mm]
&\;\;\approx\, \frac{1}{J} \sum_{j=1}^{J} \frac{1}{\big|J_{\tilde{g}_l}\rmv\big(\bd{x}_{l}^{(j)};\bd{u}^{+}_{\text{r},l}\big)\big|} \, 
  \frac{\partial \big|J_{\tilde{g}_l}\rmv\big(\bd{x}_{l}^{(j)};\bd{u}_{l}^+\big)\big|}{\partial\bd{u}_{l}^+}\bigg{|}_{\bd{u}_{l}^{+} =\ist \bd{u}^{+}_{\text{r},l}} .
\label{eq:MCEntropyDiffInt}
\end{align}

\vspace{.5mm}

For many practically relevant state evolution models \eqref{eq:statrans_control}, the computation of \vspace{-.7mm} $\frac{\partial G_l(\bd{u}_{l}^{+})}{\partial\bd{u}_{l}^{+}}\Big{|}_{\bd{u}_{l}^{+} =\ist \bd{u}^{+}_{\text{r},l}}$ can be avoided altogether or $\frac{\partial G_l(\bd{u}_{l}^{+})}{\partial\bd{u}_{l}^{+}}\Big{|}_{\bd{u}_{l}^{+} =\ist \bd{u}^{+}_{\text{r},l}}$ can be calculated in closed form, without a sample-based approximation. Some examples are considered in the following.

\vspace{1.5mm}

\begin{enumerate}

\item \emph{$J_{\tilde{g}_l}(\bd{x}_{l};\bd{u}_{l}^{+})$ does not depend on $\bd{u}_{l}^{+}$}:
In this case, $\frac{\partial G_l(\bd{u}_{l}^{+})}{\partial\bd{u}_{l}^{+}} = \bd{0}$. An important example is the ``linear additive'' state evolution model 
$\tilde{g}_l(\bd{x}_{l},\bd{u}_{l}^{+}) = \bd{A}\bd{x}_{l}+\zeta(\bd{u}_{l}^{+})$ with some matrix $\bd{A}$ and function $\zeta(\cdot)$ of suitable dimensions. Here, we obtain $J_{\tilde{g}_l}(\bd{x}_{l};\bd{u}_{l}^{+}) = \det\bd{A}$ and thus $\frac{\partial G_l(\bd{u}_{l}^{+})}{\partial\bd{u}_{l}^{+}}=\bd{0}$. A second important example is the \emph{odometry motion model} \cite[Sec.\ 5.3]{thrun05}. Here, the local state $\bd{x}_{l}$ is the pose of a robot,
which consists of the 2D position $(x_{l,1},x_{l,2})$ and the orientation $\theta_{l}$, and the control vector $\bd{u}_{l}$ consists of the translational velocity $\nu_{l}$ and the rotational velocity $\omega_{l}$. The state evolution model is given 
\vspace{-1.5mm}
by
\[
\tilde{g}_l(\bd{x}_{l},\bd{u}_{l}^{+}) \,=\ist \begin{bmatrix}
x_{l,1} \ist+\ist \nu_{l}^+ \rmv\cos(\theta_{l} \rmv+ \omega_{l}^+) \\[.5mm]
x_{l,2} \ist+\ist \nu_{l}^+ \rmv\sin(\theta_{l} \rmv+ \omega_{l}^+) \\[.5mm]
\theta_{l} +\ist \omega_{l}^+
\end{bmatrix} \rmv.
\vspace{-.5mm}
\]
Here, $J_{\tilde{g}_l}(\bd{x}_{l};\bd{u}_{l}^{+}) = 1$ and thus $\frac{\partial G_l(\bd{u}_{l}^{+})}{\partial\bd{u}_{l}^{+}}=\bd{0}$.

\vspace{1.5mm}

\item \emph{$J_{\tilde{g}_l}(\bd{x}_{l};\bd{u}_{l}^{+})$ does not depend on $\bd{x}_{l}$}:
If $J_{\tilde{g}_l}(\bd{x}_{l};\bd{u}_{l}^{+})$\linebreak $ = J_{\tilde{g}_l}(\bd{u}_{l}^{+})$, then \eqref{eq:G_def} simplifies to
$G_l(\bd{u}_{l}^{+}) = \log | J_{\tilde{g}_l}(\bd{u}_{l}^{+}) |$. Thus, we have
\[
\frac{\partial G_l(\bd{u}_{l}^{+}) }{\partial\bd{u}_{l}^{+}} \,=\, \frac{1}{|J_{\tilde{g}_l}(\bd{u}^{+}_{l})|} \, \frac{\partial |J_{\tilde{g}_l}(\bd{u}_{l}^{+})|}{\partial\bd{u}_{l}^{+}} \,,
\]
which can be calculated in closed form.


\end{enumerate}


\section{Two Special Cases}
\label{sec:special} 


\vspace{.5mm}

\subsection{Cooperative Estimation of Local States} \label{sec:inf_local} 

Here, we assume that there are no targets, and thus the task considered is only the distributed, cooperative estimation of the local states. 

\vspace{1.5mm}

\subsubsection{Estimation Layer}
The marginal posteriors corresponding to the targets are no longer calculated. In the calculation of the marginal posterior of CA $l$, 
the correction step \eqref{eq:CoSLATlocal} simplifies to 
\be
f(\bd{x}_{l}|\ist \bd{y}) \,\propto\ist \int \rmv \prod_{l'\in \cl{C}} f(\bd{x}_{l'})  \!\prod_{l_1\in \cl{C}_{l'}} \!\! f(\bd{y}_{l_1,l'}|\ist \bd{x}_{l_1},\bd{x}_{l'}) \,\mathrm{d}\bd{x}_{\sim l} \,,
\label{eq:CSLlocal}
\ee
while the prediction step \eqref{eq:pred} remains unchanged. A feasible and, typically, accurate approximation of $f(\bd{x}_{l}|\ist \bd{y})$ in \eqref{eq:CSLlocal} can be obtained by evaluating
\be
\hspace*{-1.5mm}
b^{(p)}(\bd{x}_{l}) \,\propto\ist f(\bd{x}_{l}) \rmv\prod_{l'\in\cl{C}_{l}}\int\! f(\bd{y}_{l,l'}|\ist \bd{x}_{l},\bd{x}_{l'}) \,b^{(p-1)}(\bd{x}_{l'}) \,\mathrm{d}\bd{x}_{l'} \ist
\label{eq:CoSLATlocal-1}
\vspace{-1mm}
\ee
for iteration index $p \rmv=\rmv 1,\dots,P$, where $b^{(0)}(\bd{x}_{l'}) \rmv=\rmv f(\bd{x}_{l'})$, $l' \!\in \cl{C}_l$. This amounts to the BP-based 
SPAWN scheme presented in \cite{wymeersch}.
All quantities involved in \eqref{eq:CoSLATlocal-1} are locally available at CA $l$ or can be made available by communicating only with the neighbor CAs $l' \!\in \cl{C}_{l}$. A sample-based implementation of \eqref{eq:CoSLATlocal-1} is discussed in \cite{ihler} and \cite{lien}.

\subsubsection{Control Layer}
Since there are no targets, the component $D_I(\bd{u}^{+}) = I(\bm{{\sf{x}}}_{\cl{C}}, \bm{{\sf{x}}}_{\cl{T}}^{+}\ist; \bm{{\sf{y}}}^{+};\bd{u}^{+})$ of the objective function in \eqref{eq:objfunc_expans} simplifies to $D_I(\bd{u}^{+}) = I(\bm{{\sf{x}}}_{\cl{C}}\ist; \bm{{\sf{y}}}^{+}; \bd{u}^{+})$.
The expression of the gradient of $D_I(\bd{u}^{+})$ in \eqref{eq:MIderivative3} and \eqref{eq:fyparticle} simplifies as well because 
$f(\tilde{\bd{y}}_l^{+}|\ist \bd{x}_{l}, \bd{x}_{\cl{C}_l},\bd{x}_{\cl{T}_l}^{+}; \bd{u}^{+}_{\cl{C}_l}) = f(\tilde{\bd{y}}_l^{+}|\ist \bd{x}_{l}, \bd{x}_{\cl{C}_l}; \bd{u}^{+}_{\cl{C}_l})$ 
and $f(\bd{y}^{+}|\ist \bd{x}_{\cl{C}},\bd{x}_{\cl{T}}^{+};\bd{u}^{+}) = f(\bd{y}^{+}|\ist \bd{x}_{\cl{C}};\bd{u}^{+})$ (according to \eqref{eq:firstLikeL} and \eqref{eq:CoSLATJointLikelihood}, since $\cl{T} \!=\rmv \emptyset$); furthermore, sampling from $f(\bd{x}_{\cl{C}},\bd{x}_{\cl{T}}^{+})$ (see Section \ref{sec:dist-proc}) 
reduces to sampling from $f(\bd{x}_{\cl{C}})$.

\vspace{-1mm}

\subsection{Cooperative Estimation of Global States}
\label{sec:inf_global}

Next, we discuss the case where the local states of the CAs are known, and thus our task is only the distributed, cooperative estimation of the target states. 

\vspace{1.5mm}

\subsubsection{Estimation Layer}
The marginal posteriors corresponding to the CAs are no longer calculated, and the correction step \eqref{eq:CoSLATlocal}
in the calculation of the marginal posterior of the $m$th target  simplifies to
\be
\label{eq:seqbayGlobal}
f(\bd{x}_{m}|\ist \bd{y}) \,\propto\ist f(\bd{x}_{m}) \rmv\prod_{l\in\cl{C}_{m}} \! f(\bd{y}_{l,m}|\ist \bd{x}_{l},\bd{x}_{m}) \,,
\ee
where $f(\bd{x}_{m})$ is calculated according to \eqref{eq:pred_f-recurs}. A computationally feasible sample-based 
approximation of sequential state estimation as given by \eqref{eq:seqbayGlobal} and \eqref{eq:pred_f-recurs} is provided by the particle filter \cite{ristic,hlinkaMag13,hlinka14adaptation}. 

The product of local likelihood functions $\prod_{l\in\cl{C}_{m}} \!f(\bd{y}_{l,m}|\ist\bd{x}_{l},\bd{x}_{m})$ is not available at the CAs. 
However, as in Section \ref{sec:BPandLC}, an approximation of these products can be provided to each CA in a distributed manner
by a consensus (or gossip) algorithm performed in parallel for each sample weight \cite{farahmand, savic14, meyer2014coslat} or by
the likelihood consensus scheme \cite{meyer12, hlinkaMag13, hlinka14adaptation}.

For the calculations in the control layer (described presently), a common set of samples is required at each CA. 
This can be ensured by additionally using, e.g., a max-consensus and 
providing all CAs with the same seed for random number generation \cite{savic14,lindberg13}.

\vspace{1.5mm}

\subsubsection{Control Layer}
Since there are no unknown CA states, the objective function in \eqref{eq:objfunc_expans} simplifies in that $D_I(\bd{u}^{+}) = I(\bm{{\sf{x}}}_{\cl{T}}^{+}\ist ; \bm{{\sf{y}}}^{+}; \bd{u}^{+})$ and $G_l(\bd{u}_l^{+}) = 0$ for all $l \!\in\! \cl{C}$. The expression of the gradient of $D_I(\bd{u}^{+})$ in \eqref{eq:MIderivative3} and \eqref{eq:fyparticle} simplifies because 
$f(\tilde{\bd{y}}_l^{+}|\ist \bd{x}_{l}, \bd{x}_{\cl{C}_l},\bd{x}_{\cl{T}_l}^{+}; \bd{u}^{+}_{\cl{C}_l}) = f(\tilde{\bd{y}}_l^{+}|\ist \bd{x}_{l}, \bd{x}_{\cl{T}_l}^+; \bd{u}^{+}_{\cl{C}_l})$ and 
$f(\bd{y}^{+}|\ist \bd{x}_{\cl{C}},\bd{x}_{\cl{T}}^{+};\bd{u}^{+}) = f(\bd{y}^{+}|\ist \bd{x}_{\cl{T}}^{+};\bd{u}^{+})$; furthermore, sampling from $f(\bd{x}_{\cl{C}},\bd{x}_{\cl{T}}^{+})$ reduces to sampling from $f(\bd{x}_{\cl{T}}^{+})$.

This special case was previously considered in \cite{julian12}. More specifically, \cite{julian12} studied estimation of one static global state and proposed a distributed, gradient-based, information-seeking controller and a sample-based implementation. 

\section{Simulation Results}
\label{sec:simRes}

We demonstrate the performance of the proposed method for three different localization scenarios. In Section \ref{sec:simNoncooperative}, we study the behavior of the controller by considering noncooperative self-localization of four mobile CAs based on distance measurements relative to an anchor. In Section \ref{sec:simCooperative}, we consider cooperative self-localization of three mobile CAs. Finally, in Section \ref{sec:simCoslat}, two mobile CAs perform cooperative simultaneous self-localization and 
tracking of a target. Further simulation results demonstrating the performance of the estimation layer in larger networks are reported in \cite{meyer2014coslat}.
Simulation source files and animated plots are available at http://www.nt.tuwien.ac.at/about-us/staff/florian-meyer/.

\vspace{-1mm}

\subsection{Simulation Setup}
\label{sec:sim:setup}

The following aspects of the simulation setup are common to all three scenarios. The states of the CAs consist of their 2D position, i.e., $\mathbf{x}_{l}^{(n)} \rmv\triangleq \big[ x_{l,1}^{(n)} , x_{l,2}^{(n)} \big]^\text{T}\rmv$ in a global reference frame. In addition to the mobile CAs, there is one anchor CA (indexed by $l \rmv=\rmv 1$), i.e., a static CA that broadcasts its own (true) position to the mobile CAs but does not perform any measurements. The CA network is fully connected.
The states of the mobile CAs evolve independently according to \cite{rong}
\be
\mathbf{x}_{l}^{(n)} \ist=\, \mathbf{x}_{l}^{(n-1)} + \mathbf{u}_{l}^{(n)} + \mathbf{q}_{l}^{(n)} \ist, \quad n \!=\! 1,2,\dots \,.
\label{eq:simu_state-evol_CA}
\vspace{-.3mm}
\ee
Here, $\mathbf{q}_{l}^{(n)} \!\rmv\in\rmv \mathbb{R}^2\rmv$ is zero-mean Gaussian with independent and identically distributed entries, i.e., $\mathbf{q}_{l}^{(n)} \!\sim\rmv \mathcal{N}(\mathbf{0},\sigma_q^2\mathbf{I})$ with $\sigma_{q}^2 \!=\! 10^{-3}\rmv$, and $\mathbf{q}_{l}^{(n)}$ and $\mathbf{q}_{l'}^{(n')}$ are independent unless $(l,n) \!=\! (l'\!,n')$. The admissible set $\cl{U}_l$ of the control vector $\bd{u}^{(n)}_{l}$ is defined by the norm constraint $\big\|\bd{u}^{(n)}_{l}\big\| \rmv\leq\rmv u^{\text{max}}_l$. For the interpretation of $\mathbf{u}_{l}^{(n)}$ within \eqref{eq:simu_state-evol_CA}, it is assumed that the CAs know the orientation of the global reference frame. In the initialization of the algorithms, at time $n \rmv=\rmv 0$, we use a state prior that is uniform on $[-200,200] \!\times\! [-200,200]$.

The mobile CAs acquire distance measurements according to \eqref{eq:mess}, i.e., $y_{l,k}^{(n)} = \big\|\mathbf{x}_{l}^{(n)} \!-\rmv \mathbf{x}_{k}^{(n)} \big\| + v_{l,k}^{(n)}$, where the measurement noise $v_{l,k}^{(n)}$ is independent across $l$, $k$, and $n$ and Gaussian with variance 
\be
\label{eq:simVariance}
\sigma_{l,k}^{(n)2} = \begin{cases} 
     \sigma^2_0 \ist, & \!\rmv\big\|\mathbf{x}_{l}^{(n)} \!-\rmv \mathbf{x}_{k}^{(n)} \big\| \leq d_0 \\[.7mm]
     \sigma^2_0 \Big[ \Big(\frac{\|\mathbf{x}_{l}^{(n)} - \mathbf{x}_{k}^{(n)} \|}{d_0} - 1\Big)^{\!\kappa}  \rmv+ 1\Big]\ist, & 
       \!\rmv\big\|\mathbf{x}_{l}^{(n)} \!-\rmv \mathbf{x}_{k}^{(n)} \big\| > d_0 \, .
   \end{cases}
\vspace{.5mm}
\ee
That is, $\sigma_{l,k}^{(n)2}$ is a function of $\big\|\mathbf{x}_{l}^{(n)} \!-\rmv \mathbf{x}_{k}^{(n)} \big\|$ that stays constant up to some distance $d_0$ and then increases polynomially with some exponent $\kappa$. This is a simple model for time-of-arrival distance measurements \cite{garcia14tradeoff}. We set $\sigma^2_0 = 50$ and $\kappa = 2$ and, if not stated otherwise, $d_0 = 50$.

In the estimation layer, we use $J \rmv=\rmv 3.600$ samples. (Choosing $J$ below $3.000$ was observed in some rare cases to lead to a convergence to the wrong estimate.)
A resampling step is performed to avoid weight degeneracy \cite{douc05}. Resampling transforms weighted samples $\big\{ \big(\tilde{\mathbf{x}}_{k}^{(n)(j)} \!,w_{k}^{(n)(j)}\big) \big\}_{j=1}^J$ representing the belief $b\big(\mathbf{x}_{k}^{(n)}\big)$ into nonweighted samples $\big\{ \mathbf{x}_{k}^{(n)(j)} \big\}_{j=1}^J$.
(We note that weighted samples arise in the estimation layer, as discussed in \cite{meyer12,meyer2014coslat}.)
We use a somewhat nonorthodox type of resampling that helps move samples to positions with high probability mass, thereby reducing the number of samples needed. More specifically, at every $L$th time step $n$, we sample from a kernel approximation of the belief; at all other time steps, we perform standard systematic resampling \cite{douc05}. The kernel approximation of the belief $b\big(\mathbf{x}_{k}^{(n)}\big)$ is obtained as \cite{silverman}
\[
\tilde{b}\big(\mathbf{x}_{k}^{(n)}\big) \ist=\ist \sum_{j = 1}^J w_{k}^{(n)(j)} \ist K\big(\mathbf{x}_{k}^{(n)} \!-\! \tilde{\mathbf{x}}_{k}^{(n)(j)} \big) \,,
\]
with the Gaussian kernel $K(\mathbf{x}) = (2 \pi \sigma_{\!K}^2)^{-1} \exp\rmv \big({- \| \mathbf{x} \|^2}/$ \linebreak $(2 \sigma_{\!K}^2) \big)$. Here, the variance $\sigma_{\!K}^2$ is chosen as $\sigma_{\!K}^2 = J^{1/3} \ist\ist T_{k}^{(n)}/2$ if $T_{k}^{(n)} \rmv<\rmv 2\ist \sigma^2_0$ and $\sigma_{\!K}^2 \rmv\rmv=\rmv \sigma^2_0$ otherwise, where $T_{k}^{(n)}$ denotes the trace of the weighted sample covariance matrix defined as
\[
\bd{C}_{k}^{(n)} \ist=\ist \sum_{j=1}^J w_{k}^{(n)(j)} \ist \tilde{\mathbf{x}}_{k}^{(n)(j)} \tilde{\mathbf{x}}_{k}^{(n)(j)\text{T}} - \bm{\mu}_{k}^{(n)} \rmv \bm{\mu}_{k}^{(n)\text{T}} \ist,
\]
with $\bm{\mu}_{k}^{(n)} = \sum_{j=1}^J w_{k}^{(n)(j)} \ist \tilde{\mathbf{x}}_{k}^{(n)(j)}$. 
This case distinction in choosing
$\sigma_{\!K}^2$ is used since $\sigma_{\!K}^2 = J^{1/3} \ist\ist T_{k}^{(n)}/2$ 
is only accurate for a unimodal distribution \cite{silverman} whereas $\sigma_{\!K}^2 \rmv\rmv=\rmv \sigma^2_0$ is suitable for annularly shaped distributions
(here, the width of the annulus is determined by $\sigma^2_0$ \cite{ihler}). We choose $L \rmv=\rmv 40$ if $T_{k}^{(n)} \rmv<\rmv 80$,
$L \rmv=\rmv 20$ if $80 \leq T_{k}^{(n)} \rmv<\rmv 1000$, and $L \rmv=\rmv 10$ if $T_{k}^{(n)} \ge 1000$; this 
led to good results in our simulation setting. 

We employ a censoring scheme \cite{lien} to reduce the number of samples and avoid numerical problems during the first time steps, where the mobile CAs still have uninformative beliefs. More specifically, only CAs $l$ with $T_{k}^{(n)} \rmv<\rmv 10$ are used as localization partners by neighbor CAs and (in our third scenario) are involved in localizing the target. In the control layer, this censoring scheme corresponds to the following strategy: as long as CA $l$ is not localized (i.e., $T_{k}^{(n)} \geq 10$), its objective function is $\tilde{D}_h\big(\bd{u}^{(n+1)}\big) \triangleq -h\big(\bm{{\sf{x}}}_{l}^{(n+1)} \big|\ist {\sf{y}}_{l,1}^{(n+1)}; y_{l,1}^{(1:n)}\!, \bd{u}_{l}^{(1:n+1)}\big)$, i.e., the negative differential entropy of only the own state conditioned on only the own measurement relative to the anchor CA, ${\sf{y}}_{l,1}^{(n+1)}\rmv$. 

The local gradient ascents in the controller (see \eqref{eq:controller_output}) use the reference points $\bd{u}^{(n)}_{\text{r},l} \!=\rmv \bd{0}$, which are consistent with the state evolution model \eqref{eq:simu_state-evol_CA}, and step sizes $c_{l}^{(n)}$ chosen such that $\big\|\bd{u}^{(n)}_{l}\big\| = u^{\text{max}}_l$. Thus, each CA $l \rmv\in\rmv \cl{C}$ moves with maximum nominal speed (determined by $u^{\text{max}}_l$) in the direction of maximum local increase of the objective function. If not stated otherwise, the number of samples used in the control layer is $JJ' \!=\rmv 60.000$, with $J \rmv=\rmv 1.200$ and $J' \!=\rmv 50$. (The $J \rmv=\rmv 1.200$ samples are obtained by random selection from the $3.600$ samples produced by the estimation layer.) 
We note that a reduction of $J'$ to $J' \!=\! 1$ was observed to result in more jagged CA trajectories and a slightly slower reduction of the estimation error over time.

\vspace{-1mm}

\subsection{Noncooperative Self-Localization}
\label{sec:simNoncooperative}

\begin{figure}
\vspace{-1mm}
\centering
\psfrag{s01}[t][t][.85]{\color[rgb]{0,0,0}\setlength{\tabcolsep}{0pt}\begin{tabular}{c}\raisebox{-1mm}{$\!\rmv x_1$}\end{tabular}}
\psfrag{s02}[b][b][.85]{\color[rgb]{0,0,0}\setlength{\tabcolsep}{0pt}\begin{tabular}{c}\raisebox{1mm}{$x_2$}\end{tabular}}
\psfrag{s05}[l][l][.69]{\color[rgb]{0,0,0}CA 2 ($d_0 \rmv=\rmv 20$)}
\psfrag{s06}[l][l][.69]{\color[rgb]{0,0,0}CA 2 ($d_0 \rmv=\rmv 20$)}
\psfrag{s07}[l][l][.69]{\color[rgb]{0,0,0}CA 3 ($d_0 \rmv=\rmv 50$)}
\psfrag{s08}[l][l][.69]{\color[rgb]{0,0,0}CA 4 ($d_0 \rmv=\rmv 100$)}
\psfrag{s09}[l][l][.69]{\color[rgb]{0,0,0}CA 5 ($d_0 \rmv=\rmv 100$, no control)}
\psfrag{s11}[][]{\color[rgb]{0,0,0}\setlength{\tabcolsep}{0pt}\begin{tabular}{c} \end{tabular}}
\psfrag{s12}[][]{\color[rgb]{0,0,0}\setlength{\tabcolsep}{0pt}\begin{tabular}{c} \end{tabular}}
\psfrag{x01}[t][t][.81]{$-100$}
\psfrag{x02}[t][t][.81]{$-50$}
\psfrag{x03}[t][t][.81]{$0$}
\psfrag{x04}[t][t][.81]{$50$}
\psfrag{x05}[t][t][.81]{$100$}

\psfrag{v01}[r][r][.81]{$-100$}
\psfrag{v02}[r][r][.81]{$-50$}
\psfrag{v03}[r][r][.81]{$0$}
\psfrag{v04}[r][r][.81]{$50$}
\psfrag{v05}[r][r][.81]{$100$}
\includegraphics[scale=.48]{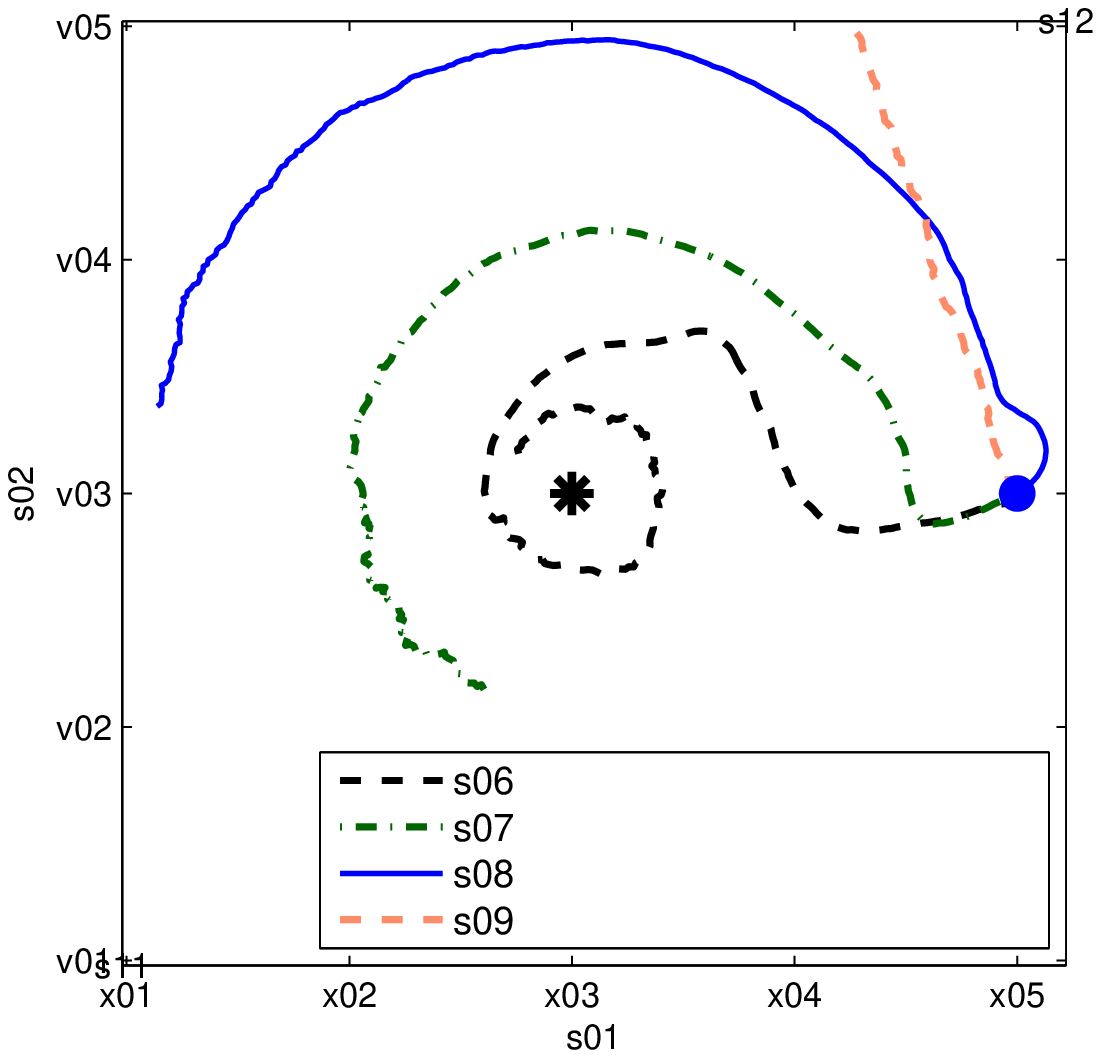}
\vspace{-.4mm}
\renewcommand{\baselinestretch}{1.05}\small\normalsize
\caption{Example trajectories for noncooperative self-localization with informa\-tion-seeking control (execpt CA 5). The initial CA position and the anchor position are indicated by a 
bullet and a star, respectively.}
\label{fig:nonCoopLoc}
\vspace{-1mm}
\end{figure}

To study the behavior of the controller, we consider four mobile CAs $l \rmv=\rmv 2,3,4,5$ that perform self-localization without any cooperation during 300 time steps $n$. The mobile CAs measure their distance to the static anchor CA ($l \rmv=\rmv 1$), which is located at position $[0\ist, 0]^{\text{T}}\rmv$, but they do not measure any distances between themselves.
Their measurement models use different values of $d_0$, namely, $d_0 \rmv=\rmv 20$, $50$, $100$, and $100$ for $l \rmv=\rmv 2$, $3$, $4$, and $5$, respectively.
The mobile CAs start at position $[100\ist ,0]^{\text{T}}$ and move with identical nominal speed determined by $u_l^{\text{max}} \rmv=\rmv 1$.
The objective function for the control of CAs $2$, $3$, and $4$ is $\tilde{D}_h\big(\bd{u}_{l}^{(n+1)}\big) \triangleq -h\big(\bm{{\sf{x}}}_{l}^{(n+1)} \big| \ist {\sf{y}}_{l,1}^{(n+1)} ; y_{l,1}^{(1:n)}\!, \bd{u}_{l}^{(1:n+1)}\big)$. CA 5 is not controlled; it randomly chooses a direction at time $n \rmv\rmv=\!\rmv 1$ and then moves in that direction with constant nominal speed determined by $u^{\text{max}}_l \rmv=\rmv 1$. Fig.\ \ref{fig:nonCoopLoc} shows an example of the trajectories of the four mobile CAs. These trajectories are quite different because of the different values of $d_0$ and the fact that CA 5 is not controlled. CA 4, after an initial turn, is roughly localized in the sense that the shape of its marginal posterior has changed from an annulus to only a segment of an annulus. Thereafter, CA 4 turns around the anchor, which is reasonable in view of the single distance measurement available at each time $n$ and the fact that, since $d_0 \rmv=\rmv 100$, the measurement noise cannot be decreased by approaching the anchor. CA 3 (with $d_0 \rmv=\rmv 50$) initially 
approaches the anchor. At a distance of 50 to the anchor, the measurement noise cannot be decreased any more, and thus CA 3 turns around the anchor without approaching it further. A similar behavior is exhibited by CA 2 (with $d_0 \rmv=\rmv 20$). 

\begin{figure}
\centering
\psfrag{s01}[t][t][.85]{\color[rgb]{0,0,0}\setlength{\tabcolsep}{0pt}\begin{tabular}{c}\raisebox{-1mm}{time step $n$}\end{tabular}}
\psfrag{s02}[b][b][.85]{\color[rgb]{0,0,0}\setlength{\tabcolsep}{0pt}\begin{tabular}{c}\raisebox{0mm}{RMSE}\end{tabular}}
\psfrag{s05}[l][l][.69]{\color[rgb]{0,0,0}CA 3 ($d_0 \rmv=\rmv 50$)}
\psfrag{s06}[l][l][.69]{\color[rgb]{0,0,0}CA 5 ($d_0 \rmv=\rmv 100$, no control)}
\psfrag{s07}[l][l][.69]{\color[rgb]{0,0,0}CA 2 ($d_0 \rmv=\rmv 20$)}
\psfrag{s08}[l][l][.69]{\color[rgb]{0,0,0}CA 3 ($d_0 \rmv=\rmv 50$)}
\psfrag{s09}[l][l][.69]{\color[rgb]{0,0,0}CA 4 ($d_0 \rmv=\rmv 100$)}
\psfrag{s11}[][]{\color[rgb]{0,0,0}\setlength{\tabcolsep}{0pt}\begin{tabular}{c} \end{tabular}}
\psfrag{s12}[][]{\color[rgb]{0,0,0}\setlength{\tabcolsep}{0pt}\begin{tabular}{c} \end{tabular}}
\psfrag{x01}[t][t][.81]{$50$}
\psfrag{x02}[t][t][.81]{$100$}
\psfrag{x03}[t][t][.81]{$150$}
\psfrag{x04}[t][t][.81]{$200$}
\psfrag{x05}[t][t][.81]{$250$}
\psfrag{x06}[t][t][.81]{$300$}
\psfrag{v01}[r][r][.81]{$0$}
\psfrag{v02}[r][r][.81]{$20$}
\psfrag{v03}[r][r][.81]{$40$}
\psfrag{v04}[r][r][.81]{$60$}
\psfrag{v05}[r][r][.81]{$80$}
\psfrag{v06}[r][r][.81]{$100$}
\psfrag{v07}[r][r][.81]{$120$}
\includegraphics[height=5cm,width=7.3cm]{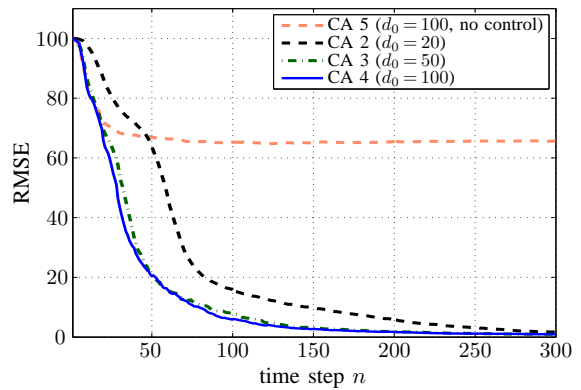}
\renewcommand{\baselinestretch}{1.05}\small\normalsize
\vspace{-.3mm}
\caption{Self-localization RMSE for noncooperative self-localization with informa\-tion-seeking control (execpt CA 5).}
\label{fig:nonCoopLocRMSEvsTime}
\vspace{.5mm}
\end{figure}

Fig.\ \ref{fig:nonCoopLocRMSEvsTime} shows the self-localization root-mean-square errors (RMSEs) of the four mobile CAs. These RMSEs were determined at each time $n$ 
by averaging over 300 simulation runs. As can be seen, the three CAs performing information-seeking control ($l = 2,3,4$) are fairly well localized after about 100 time steps. CA 2 (with $d_0=20$) takes longer to localize itself than CAs 3 and 4 since, prior to reaching a distance of 20 to the anchor, it has a larger 
noise variance (see \eqref{eq:simVariance}). The performance of CA 3 and CA 4 is almost identical; the larger noise variance of CA 3 during the initial time steps is compensated by a smaller turning radius once a distance of 50 to the anchor has been reached. CA 5 is unable to localize itself, due to the lack of intelligent control.

\vspace{-1mm}

\subsection{Cooperative Self-Localization}
\label{sec:simCooperative}

\begin{figure}
\vspace{-1mm}
\centering
\psfrag{s01}[t][t][.85]{\color[rgb]{0,0,0}\setlength{\tabcolsep}{0pt}\begin{tabular}{c}\raisebox{-.7mm}{$\,\,\ist x_1$}\end{tabular}}
\psfrag{s02}[b][b][.85]{\color[rgb]{0,0,0}\setlength{\tabcolsep}{0pt}\begin{tabular}{c}\raisebox{2mm}{$x_2$}\end{tabular}}
\psfrag{s05}[l][l][.69]{\color[rgb]{0,0,0}CA 4 ($u_4^{\text{max}}\rmv=\rmv 0.1$)}
\psfrag{s06}[l][l][.69]{\color[rgb]{0,0,0}CA 2 ($u_2^{\text{max}}\rmv=\rmv 1$)}
\psfrag{s07}[l][l][.69]{\color[rgb]{0,0,0}CA 3 ($u_3^{\text{max}}\rmv=\rmv 0.3$)}
\psfrag{s08}[l][l][.69]{\color[rgb]{0,0,0}CA 4 ($u_4^{\text{max}}\rmv=\rmv 0.1$)}
\psfrag{s10}[][]{\color[rgb]{0,0,0}\setlength{\tabcolsep}{0pt}\begin{tabular}{c} \end{tabular}}
\psfrag{s11}[][]{\color[rgb]{0,0,0}\setlength{\tabcolsep}{0pt}\begin{tabular}{c} \end{tabular}} 
\psfrag{x01}[t][t][.81]{$-100$}
\psfrag{x02}[t][t][.81]{$-50$}
\psfrag{x03}[t][t][.81]{$0$}
\psfrag{x04}[t][t][.81]{$50$}
\psfrag{x05}[t][t][.81]{$100$}
\psfrag{v01}[r][r][.81]{$-100$}
\psfrag{v02}[r][r][.81]{$-50$}
\psfrag{v03}[r][r][.81]{$0$}
\psfrag{v04}[r][r][.81]{$50$}
\psfrag{v05}[r][r][.81]{$100$}
\includegraphics[scale=.48]{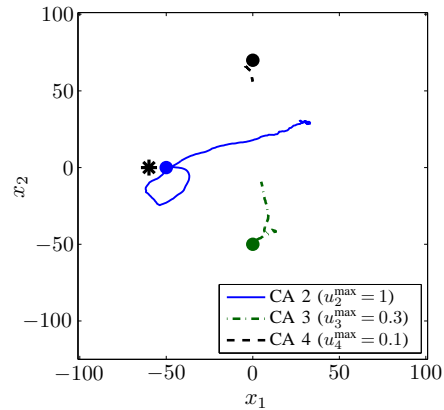}
\vspace{-.7mm}
\renewcommand{\baselinestretch}{1.05}\small\normalsize
\caption{Example trajectories for cooperative self-localization with information-seeking control (C--C scheme). The initial CA positions and the anchor position are indicated by bullets and a star, respectively.}
\label{fig:coopLoc}
\vspace{-.3mm}
\end{figure}

Next, we study the proposed method for cooperative self-localization with information-seeking control (abbreviated as\linebreak C--C). There are three mobile CAs $l \rmv=\rmv 2,3,4$ with different start points ($[-50\ist, 0]^{\text{T}}\rmv$, $[0\ist, -50]^{\text{T}}\rmv$, and $[0\ist, 70]^{\text{T}}$ for $l \rmv=\rmv 2$, $3$, and $4$, respectively) and different nominal speeds ($u_l^{\text{max}} \rmv=\rmv 1$, $0.3$, and $0.1$ for $l \rmv=\rmv 2$, $3$, and $4$, respectively). The mobile CAs measure their distances to a static anchor $l \rmv=\rmv 1$ located at $[-60\ist, 0]^{\text{T}}$ and to each other, using $d_0 \rmv=\rmv 50$. Example trajectories are shown in Fig.\ \ref{fig:coopLoc}. For comparison, we also consider noncooperative self-localization with information-seeking control as studied in Section \ref{sec:simNoncooperative} (abbreviated as N--C). Finally, we consider another scheme (abbreviated as C--N) where the CAs cooperate in the estimation layer but no intelligent control is performed. Here, each CA randomly chooses a direction and then moves in that direction with constant nominal speed determined by $u^{\text{max}}_l$.

\begin{figure}
\centering
\psfrag{s01}[t][t][.85]{\color[rgb]{0,0,0}\setlength{\tabcolsep}{0pt}\begin{tabular}{c}\raisebox{-1.1mm}{time step $n$}\end{tabular}}
\psfrag{s02}[b][b][.85]{\color[rgb]{0,0,0}\setlength{\tabcolsep}{0pt}\begin{tabular}{c}\raisebox{0mm}{RMSE}\end{tabular}}
\psfrag{s05}[l][l][0.69]{\color[rgb]{0,0,0}C--C (proposed)}
\psfrag{s06}[l][l][0.69]{\color[rgb]{0,0,0}C--N}
\psfrag{s07}[l][l][0.69]{\color[rgb]{0,0,0}\raisebox{.5mm}{N--C}}
\psfrag{s08}[l][l][0.69]{\color[rgb]{0,0,0}C--C (proposed)}
\psfrag{s10}[][]{\color[rgb]{0,0,0}\setlength{\tabcolsep}{0pt}\begin{tabular}{c} \end{tabular}}
\psfrag{s11}[][]{\color[rgb]{0,0,0}\setlength{\tabcolsep}{0pt}\begin{tabular}{c} \end{tabular}}
\psfrag{x01}[t][t][.81]{$250$}
\psfrag{x02}[t][t][.81]{$50$}
\psfrag{x03}[t][t][.81]{$100$}
\psfrag{x04}[t][t][.81]{$150$}
\psfrag{x05}[t][t][.81]{$200$}
\psfrag{v01}[r][r][.81]{$0$}
\psfrag{v02}[r][r][.81]{$10$}
\psfrag{v03}[r][r][.81]{$20$}
\psfrag{v04}[r][r][.81]{$30$}
\psfrag{v05}[r][r][.81]{$40$}
\psfrag{v06}[r][r][.81]{$50$}
\psfrag{v07}[r][r][.81]{$60$}
\psfrag{v08}[r][r][.81]{$70$}
\psfrag{v09}[r][r][.81]{$80$}
\psfrag{v10}[r][r][.81]{$90$}
\includegraphics[height=5cm,width=7.3cm]{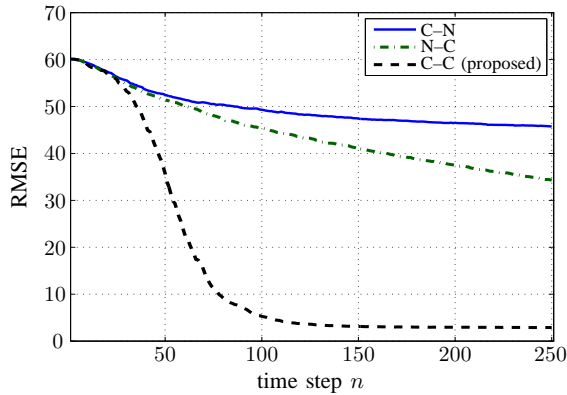}
\renewcommand{\baselinestretch}{1.05}\small\normalsize
\vspace{-.3mm}
\caption{Self-localization RMSE of the proposed estimation/control method and of two reference methods.}
\label{fig:coopLocRMSE}
\vspace{-1mm}
\end{figure}

Fig.\ \ref{fig:coopLocRMSE} shows the self-localization RMSEs of the three schemes, which were determined by averaging over the three mobile CAs and over 300 simulation runs. It is seen that the RMSEs of the two reference schemes N--C and C--N decrease only very slowly whereas, after about 100 time steps, the RMSE of the proposed C--C scheme has decreased to a low value. This behavior can be explained as follows. Without cooperation (N--C) or without intelligent control (C--N), CAs 3 and 4 need a long time to localize themselves because they are slow and initially far away from the anchor. On the other hand, CA 2 localizes itself very quickly because it is fast and initially close to the anchor. With cooperation and control (C--C), CA 2 moves in such a way that it supports the self-localization of the two other CAs. In fact, as shown by Fig.\ \ref{fig:coopLoc}, CA 2 first localizes itself by starting to turn around the anchor and then makes a sharp turn to approach CAs 3 and 4, which helps them localize themselves. This demonstrates the function and benefits of cooperative estimation and control.

\vspace{-1mm}

\subsection{Cooperative Self-Localization and Target Tracking}
\label{sec:simCoslat}

\begin{figure}
\centering
\psfrag{s01}[t][t][.85]{\color[rgb]{0,0,0}\setlength{\tabcolsep}{0pt}\begin{tabular}{c}\raisebox{-1mm}{$\,\,\ist x_1$}\end{tabular}}
\psfrag{s02}[b][b][.85]{\color[rgb]{0,0,0}\setlength{\tabcolsep}{0pt}\begin{tabular}{c}\raisebox{2mm}{$x_2$}\end{tabular}}
\psfrag{s05}[l][l][0.69]{\color[rgb]{0,0,0}CA 2}
\psfrag{s06}[l][l][0.69]{\color[rgb]{0,0,0}CA 2}
\psfrag{s07}[l][l][0.69]{\color[rgb]{0,0,0}CA 3}
\psfrag{s08}[l][l][0.69]{\color[rgb]{0,0,0}target}
\psfrag{s10}[][]{\color[rgb]{0,0,0}\setlength{\tabcolsep}{0pt}\begin{tabular}{c} \end{tabular}}
\psfrag{s11}[][]{\color[rgb]{0,0,0}\setlength{\tabcolsep}{0pt}\begin{tabular}{c} \end{tabular}}
\psfrag{x01}[t][t][.81]{$-100$}
\psfrag{x02}[t][t][.81]{$-50$}
\psfrag{x03}[t][t][.81]{$0$}
\psfrag{x04}[t][t][.81]{$50$}
\psfrag{x05}[t][t][.81]{$100$}
\psfrag{v01}[r][r][.81]{$-100$}
\psfrag{v02}[r][r][.81]{$-50$}
\psfrag{v03}[r][r][.81]{$0$}
\psfrag{v04}[r][r][.81]{$50$}
\psfrag{v05}[r][r][.81]{$100$}
\includegraphics[scale=.48]{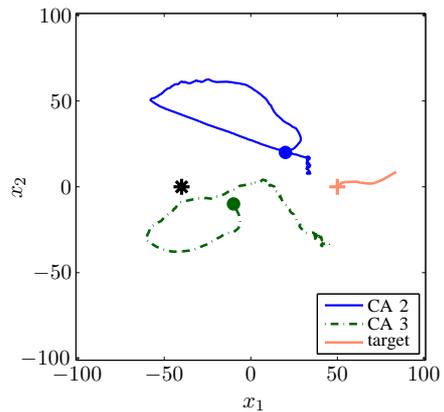}
\vspace{-.3mm}
\renewcommand{\baselinestretch}{1.05}\small\normalsize
\caption{Example trajectories for cooperative simultaneous self-localization and target tracking with information-seeking control (C--C scheme).
The initial CA positions are indicated by bullets, the initial target position by a cross, and the anchor position by a star.}
\label{fig:coslat}
\vspace{-1mm}
\end{figure}

Finally, we consider cooperative simultaneous self-locali\-za\-tion and target tracking. Two mobile CAs $l \rmv= 2,3$ starting at position $[20\ist, 20]^{\text{T}}$ and $[-10\ist, -10]^{\text{T}}$, respectively and with nominal speed determined by $u_l^{\text{max}} \rmv=\rmv 1$ cooperatively localize and track themselves and a mobile target. 
There is also a static anchor $l \rmv=\rmv 1$ at position $[-50\ist, 0]^{\text{T}}\rmv$. The target state $\mathbf{x}^{(n)}_{m} \rmv= \mathbf{x}^{(n)}_{4}$ consists of position and velocity, i.e., $\mathbf{x}^{(n)}_{4} \!\triangleq \big[ x^{(n)}_{4,1}\ist, x^{(n)}_{4,2}\ist, \dot{x}^{(n)}_{4,1}\ist, \dot{x}^{(n)}_{4,2} \big]^\text{T}\rmv$. The target state evolves according to \cite{rong}
\vspace{-1.5mm}
\[
\mathbf{x}_{4}^{(n)} \ist=\, \mathbf{G}\mathbf{x}^{(n-1)}_{4} + \mathbf{W}\mathbf{q}^{(n)}_{4} \ist, \quad\;\; n \!=\! 1,2,\dots \,,
\vspace*{-1mm}
\]
where 
\vspace*{-.5mm}
\[
\mathbf{G} = 
{ \begin{pmatrix}
   1 \!&\! 0 \!&\! 1 \!&\! 0 \\[0mm]
   0 \!&\! 1 \!&\! 0 \!&\! 1 \\[0mm]
   0  \!&\! 0  \!&\! 1 \!&\! 0 \\[0mm]
   0 \!&\! 0 \!&\! 0 \!&\! 1
  \end{pmatrix} } \ist,
\quad\, \mathbf{W} =
{ \begin{pmatrix}
   0.5 \!&\! 0 \\[0mm]
   0 \!&\! 0.5 \\[0mm]
   1  \!&\! 0  \\[0mm]
   0 \!& 1 
  \end{pmatrix} }\ist,
\vspace*{-.4mm}
\]
and $\mathbf{q}_{4}^{(n)} \!\rmv\in\rmv \mathbb{R}^2\rmv$ is zero-mean Gaussian with independent and identically distributed entries, i.e., $\mathbf{q}_{4}^{(n)} \!\sim \mathcal{N}(\mathbf{0},\tilde{\sigma}_q^2\mathbf{I})$ with $\tilde{\sigma}_q^2 \!=\! 10^{-5}\rmv$, and with $\mathbf{q}_{4}^{(n)}$ and $\mathbf{q}_{4}^{(n')}$ independent unless $n \!=\! n'\rmv$. The target trajectory is initialized with position $\big[x^{(0)}_{4,1} \ist,x^{(0)}_{4,2}\big]^{\mathrm{T}} \!= [50\ist ,0]^{\mathrm{T}}$ and velocity $\big[\dot{x}^{(0)}_{4,1}\ist,\dot{x}^{(0)}_{4,2}\big]^{\mathrm{T}} \!= [0.05\ist,0.05]^{\mathrm{T}}\rmv$. In the initialization of the algorithms, we use a target position prior that is uniform on $[-200,200] \!\times\! [-200,200]$ and a target velocity prior that is Gaussian with mean $[0 \ist, 0]^\text{T}$ and covariance matrix $\mathrm{diag}\ist\{10^{-1}\rmv, 10^{-1}\}$. 
The number of samples used in the estimation layer is $J = 120.000$; the number of samples used in the control layer is $JJ' \!=\rmv 6.000$, with $J \rmv=\rmv 1.200$ and $J' \!=\rmv 5$.
Fig.\ \ref{fig:coslat}\linebreak shows an example of CA and target trajectories obtained with the proposed method for cooperative localization with information-seeking control (C--C). One can observe that the two CAs first start turning around the anchor to localize themselves and then approach the target. Finally, at a distance of 50 to the target, where further approaching the target would no longer decrease the measurement noise, the CAs spread out to achieve a geometric formation that is favorable for cooperatively localizing and tracking the target.

\begin{figure*}[t!]	
\begin{minipage}[H!]{0.48\textwidth}
\vspace{-1mm}
\centering
\psfrag{s01}[t][t][.85]{\color[rgb]{0,0,0}\setlength{\tabcolsep}{0pt}\begin{tabular}{c}\raisebox{-1mm}{time step $n$}\end{tabular}}
\psfrag{s02}[b][b][.85]{\color[rgb]{0,0,0}\setlength{\tabcolsep}{0pt}\begin{tabular}{c}\raisebox{0mm}{RMSE}\end{tabular}}
\psfrag{s05}[l][l][.67]{\color[rgb]{0,0,0}C--C}
\psfrag{s06}[l][l][.67]{\color[rgb]{0,0,0}C--N}
\psfrag{s07}[l][l][.67]{\color[rgb]{0,0,0}\raisebox{.5mm}{N--C}}
\psfrag{s08}[l][l][.67]{\color[rgb]{0,0,0}C--C (proposed)}
\psfrag{s10}[][]{\color[rgb]{0,0,0}\setlength{\tabcolsep}{0pt}\begin{tabular}{c} \end{tabular}}
\psfrag{s11}[][]{\color[rgb]{0,0,0}\setlength{\tabcolsep}{0pt}\begin{tabular}{c} \end{tabular}}
\psfrag{s12}[l][l][.93]{\color[rgb]{0,0,0}\setlength{\tabcolsep}{0pt}\begin{tabular}{l}\raisebox{-23mm}{\hspace*{5mm}(a)}\end{tabular}}
\psfrag{x01}[t][t][.81]{$50$}
\psfrag{x02}[t][t][.81]{$100$}
\psfrag{x03}[t][t][.81]{$150$}
\psfrag{x04}[t][t][.81]{$200$}
\psfrag{x05}[t][t][.81]{$250$}
\psfrag{x06}[t][t][.81]{$300$}
\psfrag{x07}[t][t][.81]{$350$}
\psfrag{x08}[t][t][.81]{$400$}
\psfrag{v01}[r][r][.81]{$0$}
\psfrag{v02}[r][r][.81]{$10$}
\psfrag{v03}[r][r][.81]{$20$}
\psfrag{v04}[r][r][.81]{$30$}
\psfrag{v05}[r][r][.81]{$40$}
\psfrag{v06}[r][r][.81]{$50$}
\psfrag{v07}[r][r][.81]{$60$}
\hspace*{3.5mm}\includegraphics[height=5cm,width=7.3cm]{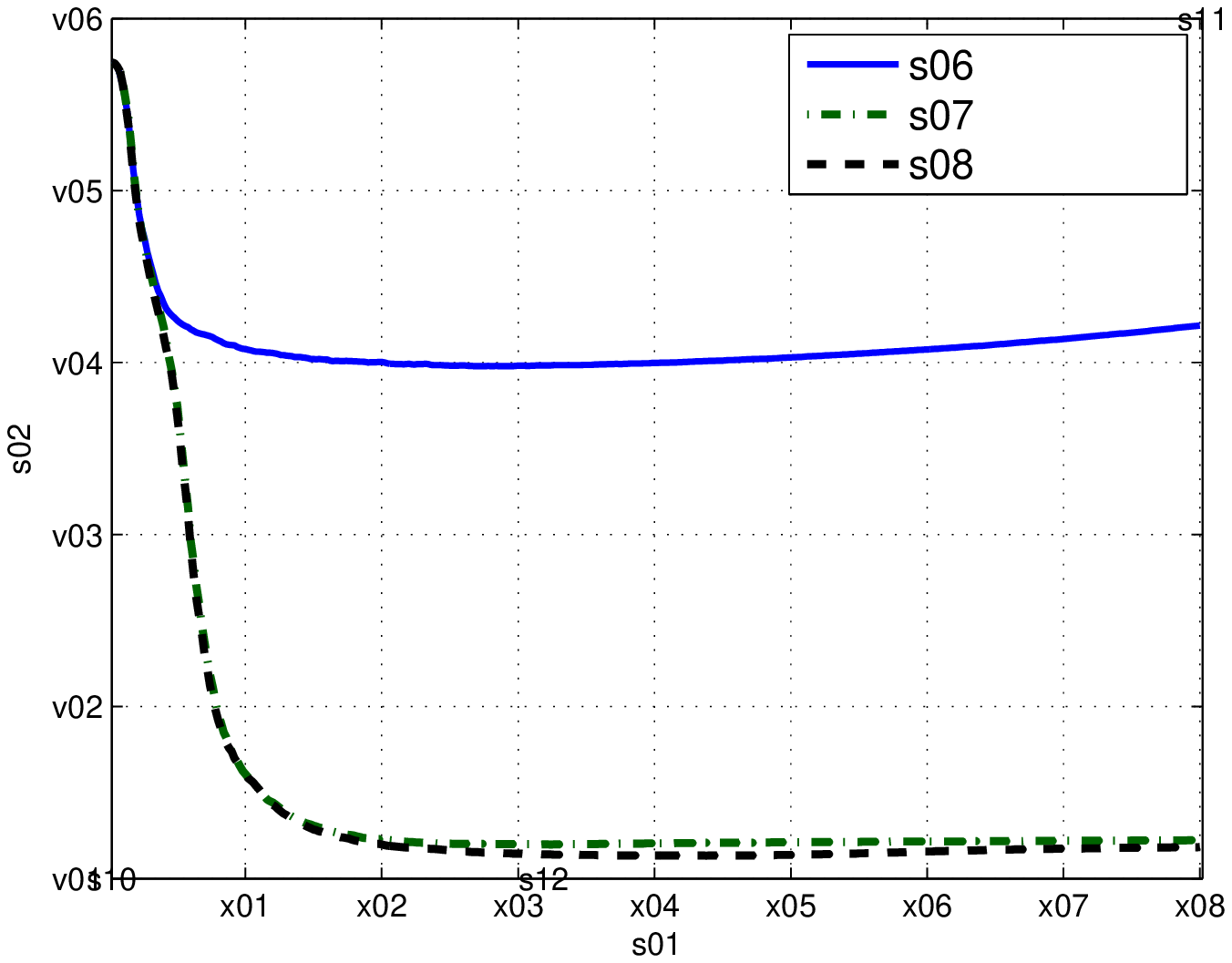}
\end{minipage}\hspace{3mm}
\begin{minipage}[H!]{0.48\textwidth}
\centering
\vspace{-1mm}
\psfrag{s01}[t][t][.85]{\color[rgb]{0,0,0}\setlength{\tabcolsep}{0pt}\begin{tabular}{c}\raisebox{-1mm}{time step $n$}\end{tabular}}
\psfrag{s02}[b][b][.85]{\color[rgb]{0,0,0}\setlength{\tabcolsep}{0pt}\begin{tabular}{c}\raisebox{0mm}{RMSE}\end{tabular}}
\psfrag{s05}[l][l][.67]{\color[rgb]{0,0,0}C--C}
\psfrag{s06}[l][l][.67]{\color[rgb]{0,0,0}C--N}
\psfrag{s07}[l][l][.67]{\color[rgb]{0,0,0}\raisebox{.5mm}{N--C}}
\psfrag{s08}[l][l][.67]{\color[rgb]{0,0,0}C--C (proposed)}
\psfrag{s10}[][]{\color[rgb]{0,0,0}\setlength{\tabcolsep}{0pt}\begin{tabular}{c} \end{tabular}}
\psfrag{s11}[][]{\color[rgb]{0,0,0}\setlength{\tabcolsep}{0pt}\begin{tabular}{c} \end{tabular}}
\psfrag{s12}[l][l][.93]{\color[rgb]{0,0,0}\setlength{\tabcolsep}{0pt}\begin{tabular}{l}\raisebox{-23mm}{\hspace*{5mm}(b)}\end{tabular}}
\psfrag{x01}[t][t][.81]{$50$}
\psfrag{x02}[t][t][.81]{$100$}
\psfrag{x03}[t][t][.81]{$150$}
\psfrag{x04}[t][t][.81]{$200$}
\psfrag{x05}[t][t][.81]{$250$}
\psfrag{x06}[t][t][.81]{$300$}
\psfrag{x07}[t][t][.81]{$350$}
\psfrag{x08}[t][t][.81]{$400$}
\psfrag{v01}[r][r][.81]{$0$}
\psfrag{v02}[r][r][.81]{$20$}
\psfrag{v03}[r][r][.81]{$40$}
\psfrag{v04}[r][r][.81]{$60$}
\psfrag{v05}[r][r][.81]{$80$}
\psfrag{v06}[r][r][.81]{$100$}
\psfrag{v07}[r][r][.81]{$120$}
\hspace*{2mm}\includegraphics[height=5cm,width=7.3cm]{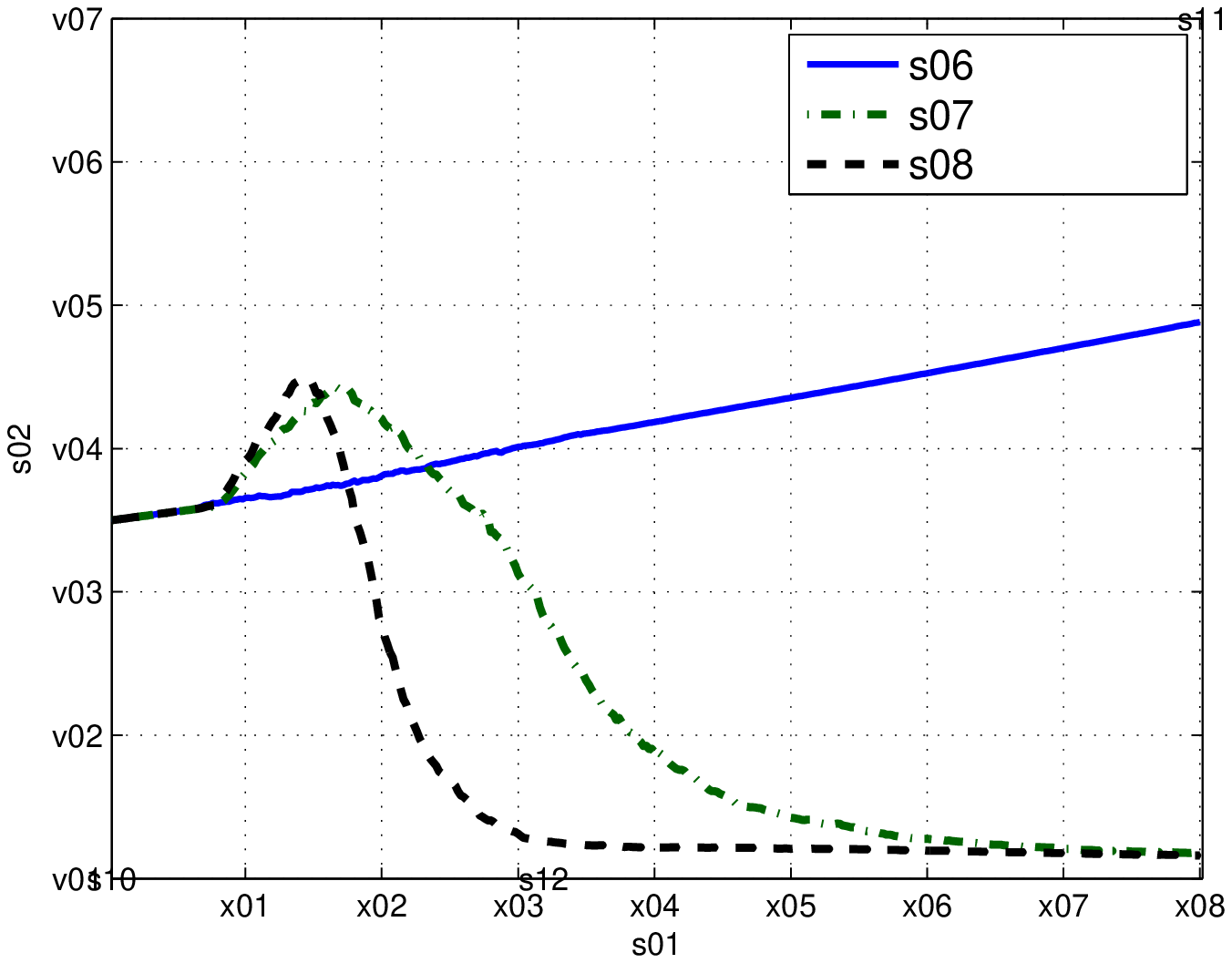}
\end{minipage}\hspace{3mm}
\vspace{5mm}
\renewcommand{\baselinestretch}{1.15}\small\normalsize
\caption{Performance of three different methods for simultaneous self-localization and target localization:
(a) Self-localization RMSE, (b) target-localization RMSE.}  
\label{fig:coslatRMSE}
\vspace{-2mm}
\end{figure*}

As before, we compare our C--C method with two reference methods, namely, noncooperative localization with information-seeking control (N--C) and cooperative localization with fixed, randomly chosen directions of movement (C--N). Fig.\ \ref{fig:coslatRMSE} shows the self-localization RMSEs and target localization RMSEs of the three schemes, which were determined by averaging over the two CAs and over 100 simulation runs. The following observations can be made:

\begin{itemize}

\vspace{1.5mm}

\item
The self-localization performance of C--N is very poor: after an initial decrease, the RMSE slowly increases. In fact, typically, no cooperation actually takes place, since the CAs are unable to localize themselves and thus each CA is censored by the respective other CA. The self-localization RMSEs of C--C and N--C decrease rather quickly to a low value. They are very similar, which can be explained as follows. Because both CAs move with the same nominal speed, they localize themselves approximately in the same manner. Therefore, as long as the CAs are not localized, no cooperation takes place due to censoring, and after they are localized, no further gain can be achieved by cooperation.

\vspace{1.5mm}

\item
The target localization RMSEs of the three methods are initially equal to 50 and slowly increase during the first 40 time steps. Indeed, due to the censoring scheme, the CAs start localizing the target only when they are localized themselves. Therefore, during the first 40 time steps, no measurements of the distance to the target are used by the CAs, and thus the CAs' target position estimation is solely based on the prior distribution, which is uniform. This leads to a target position estimate of $[0\ist, 0]^{\text{T}}$ and in turn (since the target is initially located at $[50\ist, 0]^{\text{T}}$) to an initial target localization RMSE of 50 at time $n=1$. During the first 40 time steps, the RMSE slowly increases since the target slowly moves away from  $[0\ist, 0]^{\text{T}}\rmv$. The RMSE of C--N continues to increase in this manner even after $n \rmv=\rmv 40$ since with C--N, the CAs are never localized and therefore never start localizing the target. For C--C and N--C (both employing information-seeking control), after $n \rmv=\rmv 40$, the RMSE first increases and then decreases. The RMSE of C--C decreases sooner and more quickly than that of N--C, which again shows the benefits of cooperative estimation.

\vspace{1.5mm}

The initial increase and subsequent decrease of the target localization RMSE observed with C--C and N--C\linebreak after $n \rmv=\rmv 40$ can be explained as follows. 
After the CAs localized themselves and start localizing the target, the target position posterior at a given CA is roughly annularly shaped, with the center of the annulus being the CA position. (This position is equal to the turning point of the respective CA trajectory in Fig.\ \ref{fig:coslat}.) The resulting target position estimate is located at that center. Thus, it is more distant from the true target position than the estimate $[0\ist, 0]^{\text{T}}$ that was obtained when the CA was not yet localized and the target position posterior was still uniform. As the CAs approach the target, the target position posterior becomes unimodal and the target can be localized, resulting in a decrease of the target localization RMSE.

\vspace{-1mm}

\end{itemize}

\section{Conclusion}

We proposed a Bayesian framework and method for distributed estimation with information-seeking control in agent networks. Distributed, cooperative estimation is performed for time-varying global states (related to noncooperative targets or features of the environment) and/or time-varying local states (related to individual cooperative agents), using a combination of belief propagation message passing and consensus. The distributed, cooperative control seeks to optimize the behavior of the cooperative agents by maximizing the negative joint posterior entropy of the agent states via a gradient ascent. A probabilistic information transfer from the estimation layer to the control layer enables effective control strategies and thus leads to excellent estimation performance. 

A major advantage of the proposed approach is its generality. Our method relies on general state evolution and measurement models, an information-theoretic objective function for control, and sample-based representations of probability distributions. These characteristics make it suitable for nonlinear and non-Gaussian systems, such as those arising in location-aware networks. Numerical simulations for a simultaneous agent self-localization and target tracking problem demonstrated intelligent behavior of the cooperative agents and a resulting improvement of estimation performance.

Possible directions for future research include an extension of the myopic controller (i.e., optimizing only one time step ahead) to a receding horizon \cite{mayne00}; this can be expected to improve the performance in scenarios with multiple time-varying global states. Furthermore, the complexity and communication cost of the proposed method can be reduced by introducing Gaussian or Gaussian mixture approximations \cite{alspach72} and using cubature points \cite{arasaratnam09} instead of random samples.

\appendices

\vspace{-1mm}

\section{Proof of Equation \eqref{eq:entropyStep1}}
\label{sec:proof}

\vspace{.7mm}

We will use the following transformation rule for differential entropy \cite[Eq.\ 18]{deville12}: For a continuous
random vector $\bd{a}$ and a transformed random vector of identical dimension $\bd{b} = g(\bd{a})$, where $g(\cdot)$ is a bijective differentiable function with Jacobian determinant 
\vspace*{-1mm}
$J_{g}(\bd{a}) = \det \frac{\partial g(\bd{a})}{\partial \bd{a}}$,
\be
h(\bm{{\sf{b}}}) \ist=\ist h(\bm{{\sf{a}}}) + e(\bm{{\sf{a}}}) \,, \quad \text{with} \;\; e(\bm{{\sf{a}}}) \triangleq\rmv \int \! f(\bd{a}) \ist \log |J_{g}(\bd{a})| \,\mathrm{d}\bd{a} \,.
\label{eq:entr-trafo}
\ee

The conditional differential entropy $h(\bm{{\sf{x}}}^{+} |\ist \bm{{\sf{y}}}^{+};\bd{u}^+)$ can be expanded as \cite[Chap.\ 8]{cover06}
\be
h(\bm{{\sf{x}}}^{+} |\ist \bm{{\sf{y}}}^{+};\bd{u}^+) \ist=\ist h(\bm{{\sf{x}}}^{+}\!,\bm{{\sf{y}}}^{+};\bd{u}^+) - h(\bm{{\sf{y}}}^{+};\bd{u}^+) \,.
\label{eq:condEntroy}
\ee
The vector $\bd{x}^{+}$ consists of $\bd{x}_{l}^+$ and $\bd{x}^{+}_{\cl{A} \setminus \{l\}} \!\triangleq\rmv \big[ \bd{x}_{k}^{+} \big]_{k \in \cl{A} \backslash \{l\}}$, and there is $\bd{x}_{l}^{+} \rmv= \tilde{g}_{l}(\bd{x}_{l},\bd{u}_{l}^{+})$ (see \eqref{eq:statrans_control}). Thus, the first term on the right-hand side of \eqref{eq:condEntroy} can be expressed as $h(\bm{{\sf{x}}}^{+}\!,\bm{{\sf{y}}}^{+};\bd{u}^+) = h\big( \tilde{g}_{l}(\bm{{\sf{x}}}_{l},\bd{u}_{l}^{+}), \bm{{\sf{x}}}^{+}_{\cl{A} \setminus \{l\}}, \bm{{\sf{y}}}^{+};\bd{u}^+\big)$. Applying the transformation rule \eqref{eq:entr-trafo} to the ``extended state evolution mapping'' 
$\tilde{g}_l^*: \big[ \bd{x}_{l}^{\text{T}}\rmv, \bd{x}^{+\text{T}}_{\cl{A} \setminus \{l\}} \ist, \bd{y}^{+\text{T}} \big]^\text{T} \!\mapsto\! \big[ \big( \tilde{g}_l(\bd{x}_{l},\bd{u}^+_{l}) \big)^{\text{T}}\!, \bd{x}^{+\text{T}}_{\cl{A} \setminus \{l\}}\ist , \bd{y}^{+\text{T}} \big]^\text{T}\rmv$, we then obtain
\pagebreak 
\begin{align}
h(\bm{{\sf{x}}}^{+}\!,\bm{{\sf{y}}}^{+};\bd{u}^+) 
&\ist= \ist h\big( \bm{{\sf{x}}}_{l}, \bm{{\sf{x}}}^{+}_{\cl{A} \setminus \{l\}},\bm{{\sf{y}}}^{+};\bd{u}^+\big) \nonumber\\[.4mm]
&\hspace{10mm}  \ist+\ist e\big(\bm{{\sf{x}}}_{l},\bm{{\sf{x}}}^{+}_{\cl{A} \setminus \{l\}},\bm{{\sf{y}}}^{+};\bd{u}_l^+\big) \,,\label{eq:condEntroy_e}\\[-9mm]
\nonumber
\end{align}
where 
\begin{align*}
&e\big(\bm{{\sf{x}}}_{l},\bm{{\sf{x}}}^{+}_{\cl{A} \setminus \{l\}},\bm{{\sf{y}}}^{+};\bd{u}_l^+\big) \\[.7mm] 
&\;\triangleq \int\!\! \int\!\! \int \! f\big(\bd{x}_{l},\bd{x}^+_{\cl{A} \setminus \{l\}},\bd{y}^{+}\big) 
  \ist \log \big|J_{\tilde{g}_l^*}\rmv\big(\bd{x}_{l},\bd{x}^+_{\cl{A} \setminus \{l\}},\bd{y}^{+};\bd{u}_{l}^+\big)\big| \\[-.5mm] 
&\hspace{57mm} \times \mathrm{d}\bd{x}_{l} \,\mathrm{d}\bd{x}^+_{\cl{A} \setminus \{l\}}\ist \mathrm{d}\bd{y}^{+} \rmv.
\end{align*}
Here, $J_{\tilde{g}_l^*}\rmv\big(\bd{x}_{l},\bd{x}^+_{\cl{A} \setminus \{l\}},\bd{y}^{+};\bd{u}_l^+\big)$ is the Jacobian determinant of 
$\tilde{g}_l^*\rmv\big(\bd{x}_{l},\bd{x}^+_{\cl{A} \setminus \{l\}},\bd{y}^{+};\bd{u}_{l}^+\big)$.
It is easily seen that 
$J_{\tilde{g}_l^*}\rmv\big(\bd{x}_{l},\bd{x}^+_{\cl{A} \setminus \{l\}},$\linebreak $\bd{y}^{+};\bd{u}^+\big) = J_{\tilde{g}_l}(\bd{x}_{l};\bd{u}^{+})$, 
and thus we obtain further
\begin{align}
&e\big(\bm{{\sf{x}}}_{l},\bm{{\sf{x}}}^{+}_{\cl{A} \setminus \{l\}},\bm{{\sf{y}}}^{+}; \bd{u}_l^+\big) \nonumber\\[.7mm]
&\hspace{10mm}= \int \bigg[ \int \!\!\int \! f\big(\bd{x}_{l},\bd{x}^+_{\cl{A} \setminus \{l\}},\bd{y}^{+}\big) 
  \,\mathrm{d}\bd{x}^+_{\cl{A} \setminus \{l\}}\ist \mathrm{d}\bd{y}^{+} \bigg]\nonumber\\[0mm]
&\hspace{44mm}\times\log|J_{\tilde{g}_l}(\bd{x}_{l};\bd{u}_{l}^{+})| \,\ist \mathrm{d}\bd{x}_{l} \nn\\[-.5mm]
&\hspace{10mm}= \int\! f(\bd{x}_{l}) \ist \log|J_{\tilde{g}_l}(\bd{x}_{l};\bd{u}_{l}^{+})| \,\ist \mathrm{d}\bd{x}_{l}  \nn\\[.4mm]
&\hspace{10mm}\eq G_l(\bd{u}_{l}^{+}) \,.
\label{eq:entropyDiff}
\end{align}
Inserting \eqref{eq:entropyDiff} into \eqref{eq:condEntroy_e} and the resulting expression of $h(\bm{{\sf{x}}}^{+}\!,\bm{{\sf{y}}}^{+};\bd{u}^+)$ into
\eqref{eq:condEntroy} gives
\begin{align}
\hspace{-2mm}h(\bm{{\sf{x}}}^{+} |\ist \bm{{\sf{y}}}^{+};\bd{u}^+) &\,= \, h\big(\bm{{\sf{x}}}_{l},\bm{{\sf{x}}}^{+}_{\cl{A} \setminus \{l\}},\bm{{\sf{y}}}^{+};\bd{u}^+\big) 
  \ist+\ist G_l(\bd{u}_{l}^{+}) \nonumber\\[0mm]
&\hspace{37mm}-\ist h(\bm{{\sf{y}}}^{+};\bd{u}^+) \,.
\label{eq:condEntroy_1}
\end{align}

Next, we repeat this transformation procedure but apply it to the term $h\big(\bm{{\sf{x}}}_{l},\bm{{\sf{x}}}^{+}_{\cl{A} \setminus \{l\}},\bm{{\sf{y}}}^{+};\bd{u}^+\big)$
in \eqref{eq:condEntroy_1} instead of $h(\bm{{\sf{x}}}^{+}\!,\bm{{\sf{y}}}^{+};\bd{u}^+)$. Consider an arbitrary $l' \!\in \cl{C} \ist\backslash \{l\}$, and note
that $\bm{{\sf{x}}}^{+}_{\cl{A} \setminus \{l\}}$ consists of $\bd{x}_{l'}^+$ and 
$\bd{x}^{+}_{\cl{A} \setminus \{l,l'\}} \!\triangleq\rmv \big[ \bd{x}_{k}^{+} \big]_{k \in \cl{A} \backslash \{l,l'\}}$, where
$\bd{x}_{l'}^{+} \rmv= \tilde{g}_{l'}(\bd{x}_{l'},\bd{u}_{l'}^{+})$ according to \eqref{eq:statrans_control}. Proceeding as above and inserting the resulting expression of 
$h\big(\bm{{\sf{x}}}_{l},\bm{{\sf{x}}}^{+}_{\cl{A} \setminus \{l\}},\bm{{\sf{y}}}^{+};\bd{u}^+\big)$ into \eqref{eq:condEntroy_1} yields
\begin{align*}
h(\bm{{\sf{x}}}^{+} |\ist \bm{{\sf{y}}}^{+};\bd{u}^+) &\,= \, h\big(\bm{{\sf{x}}}_{l}, \bm{{\sf{x}}}_{l'}, \bm{{\sf{x}}}^{+}_{\cl{A} \setminus \{l,\ist l'\}},\bm{{\sf{y}}}^{+};\bd{u}^+\big) 
  \ist+\ist G_{l'}(\bd{u}_{l'}^{+})\nonumber\\[.5mm]
&\hspace{28mm}+\ist G_l(\bd{u}_{l}^{+}) \ist-\ist h(\bm{{\sf{y}}}^{+};\bd{u}^+) \,.
\end{align*}
We continue this procedure in a recursive fashion, splitting off CA state vectors from $\bd{x}^{+}_{\cl{A} \setminus \{l,\ist l'\}}$
until only the target states (contained in $\bd{x}^{+}_{\cl{T}}$) are left, and applying the transformation rule at each recursion. In the end, we obtain
\begin{align*}
h(\bm{{\sf{x}}}^+|\ist \bm{{\sf{y}}}^{+};\bd{u}^+) &\eq h(\bm{{\sf{x}}}_{\cl{C}},\bm{{\sf{x}}}^{+}_{\cl{T}},\bm{{\sf{y}}}^{+};\bd{u}^+) 
\,+\ist \sum_{l \in \cl{C}} G_l(\bd{u}_{l}^{+})\nonumber\\[-.5mm]
&\hspace{38mm}-\ist h(\bm{{\sf{y}}}^{+};\bd{u}^+) \,.
\end{align*}
Finally, Equation \eqref{eq:entropyStep1} is obtained by noting that 
$h(\bm{{\sf{x}}}_{\cl{C}},\bm{{\sf{x}}}^{+}_{\cl{T}},\bm{{\sf{y}}}^{+};$\linebreak $\bd{u}^+) = h(\bm{{\sf{x}}}_{\cl{C}}, \bm{{\sf{x}}}_{\cl{T}}^{+} \ist|\ist \bm{{\sf{y}}}^{+};\bd{u}^+) + h(\bm{{\sf{y}}}^{+};\bd{u}^+)$.

\vspace{1mm}

\section{Derivation of \eqref{eq:MIderivative3} and \eqref{eq:fyparticle}}
\label{sec:sampleBasedCalcGradient-A}

\vspace{1mm}

\subsubsection{Derivation of \eqref{eq:MIderivative3}}
Let us first define the
vector $\breve{\bd{y}}_l^{+}  \triangleq \big[ \bd{y}_{l'\!,k}^{+} \big]_{l'\in\cl{C} \backslash \{l\}, \ist k\in\cl{A} \backslash \{l\}}$,
which contains all those measurements $\bd{y}_{l'\!,k}^{+}$ that are not contained in $\tilde{\bd{y}}_l^{+}$ (cf.\ \eqref{eq:comp-meas-D-tilde}).
The corresponding likelihood function is given by
\begin{equation}
f(\breve{\bd{y}}_l^{+}|\ist \bd{x}_{\cl{C}},\bd{x}_{\cl{T}}^{+}; \bd{u}^{+}\big) \,=\, \frac{f(\bd{y}^{+}|\ist \bd{x}_{\cl{C}},\bd{x}_{\cl{T}}^{+};\bd{u}^{+})}{f(\tilde{\bd{y}}_l^{+}|\ist \bd{x}_{l}, \bd{x}_{\cl{C}_l},\bd{x}_{\cl{T}_l}^{+}; \bd{u}^{+}_{\cl{C}_l} )} \,,
\label{eq:secondLikeL}
\end{equation}
which, according to \eqref{eq:CoSLATJointLikelihood} and \eqref{eq:firstLikeL}, involves all factors of $f(\bd{y}^{+}|\ist \bd{x}_{\cl{C}},\bd{x}_{\cl{T}}^{+};\bd{u}^{+})$ 
that do not depend on the local control vector $\bd{u}_{l}^{+}$.
Using 
\eqref{eq:secondLikeL} in \eqref{eq:MIderivative} yields
\vspace{.5mm}
\begin{align}
\hspace{-2mm}\frac{\partial D_I(\bd{u}^{+})}{\partial\bd{u}_{l}^{+}}
  &\,= \int\!\! \int \!\! \int \! f(\bd{x}_{\cl{C}},\bd{x}_{\cl{T}}^{+}) \, f(\breve{\bd{y}}_l^{+}|\ist \bd{x}_{\cl{C}},\bd{x}_{\cl{T}}^{+}; \bd{u}^{+})\nonumber\\[0mm]
&\hspace{6mm} \times \frac{\partial f(\tilde{\bd{y}}_l^{+}|\ist \bd{x}_{l}, \bd{x}_{\cl{C}_l},\bd{x}_{\cl{T}_l}^{+}; \bd{u}^{+}_{\cl{C}_l})}{\partial\bd{u}_{l}^{+}} \nn \\[0mm]
 &\hspace*{6mm}\times \log \frac{f(\bd{y}^{+}|\ist \bd{x}_{\cl{C}},\bd{x}_{\cl{T}}^{+};\bd{u}^{+})}
 {f(\bd{y}^{+};\bd{u}^{+})} \,\mathrm{d}\bd{x}_{\cl{C}} \, \mathrm{d}\bd{x}_{\cl{T}}^{+} \, \mathrm{d}\bd{y}^{+}.  \label{eq:MIderivativeLLf}
\end{align}
Setting $\bd{u}^{+} \!=\rmv \bd{u}_{\text{r}}^{+}\!$, and multiplying and dividing the integrand in \eqref{eq:MIderivativeLLf} by $f(\tilde{\bd{y}}_l^{+}|\ist \bd{x}_{l}, \bd{x}_{\cl{C}_l},\bd{x}_{\cl{T}_l}^{+}; \bd{u}^{+}_{{\text{r},\cl{C}_l}})$, we obtain \vspace{.5mm} further
\begin{align}
&\hspace{-1mm}\frac{\partial D_I(\bd{u}^{+})}{\partial \bd{u}_{l}^{+}}\bigg{|}_{\bd{u}^{+}=\ist \bd{u}^{+}_{\text{r}}} \nonumber\\[1mm]
&\hspace*{5mm}=\ist \int\!\! \int \!\! \int \rmv q(\bd{y}^{+}\!\rmv, \bd{x}_{\cl{C}},\bd{x}_{\cl{T}}^{+}) 
   \, \frac{1}{f(\tilde{\bd{y}}_l^{+}|\ist \bd{x}_{l}, \bd{x}_{\cl{C}_l},\bd{x}_{\cl{T}_l}^{+}; \bd{u}^{+}_{\text{r},\cl{C}_l})} \nonumber\\[.5mm]
&\hspace*{14mm}\times\frac{\partial f(\tilde{\bd{y}}_l^{+}|\ist \bd{x}_{l}, \bd{x}_{\cl{C}_l},\bd{x}_{\cl{T}_l}^{+}; \bd{u}^{+}_{{\cl{C}_l}})}{\partial\bd{u}_{l}^{+}}
  \bigg{|}_{\bd{u}_{\cl{C}_l}^{+}=\ist \bd{u}_{\text{r},\cl{C}_l}^{+}} \nonumber \\[-.5mm]
 &\hspace*{14mm}\times \log \frac{f(\bd{y}^{+}|\ist \bd{x}_{\cl{C}},\bd{x}_{\cl{T}}^{+};\bd{u}_{\text{r}}^{+}) }{ f(\bd{y}^{+}; \bd{u}^{+}_{\text{r}})} \,\mathrm{d}\bd{x}_{\cl{C}} \,\mathrm{d}\bd{x}_{\cl{T}}^{+} \, \mathrm{d}\bd{y}^{+} \rmv,
\label{eq:MIderivativeIS} \\[-7.5mm]
\nn
\end{align}
where 
\begin{align*}
q(\bd{y}^{+}\!\rmv, \bd{x}_{\cl{C}},\bd{x}_{\cl{T}}^{+}) &\,\triangleq\, f(\bd{x}_{\cl{C}},\bd{x}_{\cl{T}}^{+}) 
\, f(\tilde{\bd{y}}_l^{+}|\ist \bd{x}_{l}, \bd{x}_{\cl{C}_l},\bd{x}_{\cl{T}_l}^{+}; \bd{u}^{+}_{{\text{r},\cl{C}_l}}) \nn\\
&\hspace{28mm}\times f(\breve{\bd{y}}_l^{+}|\ist \bd{x}_{\cl{C}},\bd{x}_{\cl{T}}^{+}; \bd{u}_{\text{r}}^{+}) \,.
\end{align*}
Then, \eqref{eq:MIderivative3} is recognized to be a Monte Carlo approximation of \eqref{eq:MIderivativeIS} that is obtained by performing importance sampling \cite{doucet} using $q(\bd{y}^{+}\!\rmv, \bd{x}_{\cl{C}},\bd{x}_{\cl{T}}^{+})$ as importance density, i.e., the samples $\bd{y}^{+(j,j')}\rmv$, $\bd{x}_{\cl{C}}^{(j)}\!$, and $\bd{x}_{\cl{T}}^{+(j)}$ occurring in \eqref{eq:MIderivative3} are drawn from $q(\bd{y}^{+}\!\rmv, \bd{x}_{\cl{C}},\bd{x}_{\cl{T}}^{+})$.
Using \eqref{eq:secondLikeL}, this importance density can be expressed as
\begin{align*}
q(\bd{y}^{+}\!\rmv, \bd{x}_{\cl{C}},\bd{x}_{\cl{T}}^{+}) 
  &\ist=\ist f(\bd{x}_{\cl{C}},\bd{x}_{\cl{T}}^{+}) \ist f(\bd{y}^{+}| \ist \bd{x}_{\cl{C}},\bd{x}_{\cl{T}}^{+}; \bd{u}_{\text{r}}^{+})\\[.5mm]
  &\ist=\ist f(\bd{x}_{\cl{C}},\bd{x}_{\cl{T}}^{+}, \bd{y}^{+}; \bd{u}_{\text{r}}^{+}) \,.
\end{align*}
The first expression, $f(\bd{x}_{\cl{C}},\bd{x}_{\cl{T}}^{+}) \ist f(\bd{y}^{+}| \ist \bd{x}_{\cl{C}},\bd{x}_{\cl{T}}^{+}; \bd{u}_{\text{r}}^{+})$,
underlies the two-stage sampling procedure described in Section \ref{sec:firstGradient}.

\vspace{1.5mm}

\subsubsection{Derivation of \eqref{eq:fyparticle}}
We have
\begin{align}
f(\bd{y}^{+}; \bd{u}^{+}_{\text{r}}) &\,= \int \!\! \int \! f(\bd{y}^{+}|\ist \bd{x}_{\cl{C}},\bd{x}_{\cl{T}}^{+}; \bd{u}^{+}_{\text{r}}) \,f(\bd{x}_{\cl{C}},\bd{x}_{\cl{T}}^{+}) \, \mathrm{d}\bd{x}_{\cl{C}} \,\mathrm{d}\bd{x}_{\cl{T}}^{+} \,.\nonumber\\[-2mm]
&\label{eq:f_y_u}\\[-8mm]
\nn
\end{align}
Using samples $\big\{ \big(\bd{x}_{\cl{C}}^{(j)}\!,\bd{x}_{\cl{T}}^{+(j)} \big) \big\}_{j=1}^J \!\rmv\sim\! f(\bd{x}_{\cl{C}},\bd{x}_{\cl{T}}^{+})$ 
(see Section \ref{sec:firstGradient}),
a Monte Carlo approximation of \eqref{eq:f_y_u} 
is obtained as
\[
f(\bd{y}^{+}; \bd{u}^+_{\text{r}}) 
  \,\approx\ist \frac{1}{J} \sum_{j''=1}^J f\big(\bd{y}^{+} \big|\ist \bd{x}^{(j'')}_{\cl{C}} \!,\bd{x}^{+(j'')}_{\cl{T}}; \bd{u}^{+}_{\text{r}}\big) \,.
\vspace{-.8mm}
\]
Evaluating this for $\bd{y}^{+} \!= \bd{y}^{+(j,j')}$ (again see Section \ref{sec:firstGradient}) yields \eqref{eq:fyparticle}.

\vspace{-1mm}

\section{Drawing Samples from $f\big(\bd{y}^{+} \big|\ist \bd{x}^{(j)}_{\cl{C}}\!,\bd{x}_{\cl{T}}^{+(j)}; \bd{u}^{+}_{\rm r} \big)$}
\label{sec:drawingCen}

\vspace{1.5mm}

We consider the setting of Section \ref{sec:quasiCentral}. As discussed there, samples $\big\{ \bd{x}_{l'}^{(j)} \big\}_{j=1}^J \!\sim f(\bd{x}_{l'})$, $l' \!\in\rmv \cl{C}$ and $\big\{ \bd{x}_{m}^{+(j)} \big\}_{j=1}^J \!\sim f(\bd{x}_{m}^{+})$, $m \rmv\in\rmv \cl{T}$ are available at CA $l$, and it is assumed that the state evolution and measurement models of all CAs $l' \!\in\rmv \cl{C}$ are known to CA $l$. We start by noting that by combining \eqref{eq:futuremeas} and \eqref{eq:meas_mod}, the composite measurement vector $\bd{y}^{+}$ can be written \vspace*{-1mm} as
\be
\bd{y}^{+} =\ist 
\big[ d_l(\bd{x}_{l}^{+}\!, \bd{x}_{k}^{+}, \bd{v}_{l,k}^{+}) \big]_{l\in\cl{C},\ist k\in\cl{A}_{l} } \ist.
\label{eq:comp-meas-D}
\ee

First, CA $l$ obtains samples $\big\{\bd{x}_{l'}^{+(j)}\big\}_{j=1}^J \rmv\sim \tilde{f}(\bd{x}^+_{l'}) \triangleq f(\bd{x}^+_{l'}) \big|_{\bd{x}^+_{l'} \ist=\, \tilde{g}_{l'}(\bd{x}_{l'},\bd{u}^{+}_{\text{r},l'} )}$
(see \eqref{eq:statrans_control}) for all $l' \!\in\rmv \cl{C}$ by evaluating $\tilde{g}_{l'}(\bd{x}_{l'},\bd{u}_{l'}^{+})$ 
at $\bd{x}_{l'} \rmv=\rmv \bd{x}_{l'}^{(j)}$ and $\bd{u}_{l'}^{+} \!=\rmv \bd{u}^{+}_{\text{r},l'}$, i.e., 
\be
\bd{x}_{l'}^{+(j)} =\ist \tilde{g}_{l'}\big(\bd{x}_{l'}^{(j)}\!,\bd{u}^{+}_{\text{r},l'} \big) \,, \quad\! j \rmv=\rmv 1,\ldots,J \ist.
\label{eq:sampleY}
\ee
Thus, at this point, samples $\big\{\bd{x}^{+ (j)}_k\big\}^{J}_{j=1}$ for all $k \in\cl{A}$ are available at CA $l$. Next, for each $j \rmv\in \{1,\dots,J\}$, CA $l$ draws samples 
$\big\{ \bd{v}_{l'\!,k}^{+(j,j')} \big\}_{j'=1}^{J'} \!\sim\rmv f(\bd{v}_{l'\!,k}^{+})$ for $l' \!\in\rmv \cl{C}$ and $k \rmv\in\rmv \cl{A}_{l'}$. 
Finally, CA $l$ obtains samples 
$\big\{ \bd{y}^{+(j,j')} \big\}_{j'=1}^{J'} \!\sim\rmv f\big(\bd{y}^{+} \big|\ist \bd{x}^{(j)}_{\cl{C}}\!,$\linebreak 
$\bd{x}_{\cl{T}}^{+(j)}; \bd{u}^{+}_{\text{r}} \big)$ by evaluating \eqref{eq:comp-meas-D} using 
the appropriate samples, i.e., 
\begin{align*}
\bd{y}^{+(j,j')} &= \big[ d_{l'}\big(\bd{x}_{l'}^{+(j)}\!, \bd{x}_{k}^{+(j)}\!, \bd{v}_{l'\!,k}^{+(j,j')} \big) \big]_{l'\rmv\in\cl{C},\ist k\in\cl{A}_{l'} } \ist, \\[.5mm]
&\hspace*{50mm} j' \!=\rmv 1,\dots,J' \rmv.
\end{align*}

\vspace{-3mm}

\section{Drawing Samples from 
$f\big(\tilde{\bd{y}}_l^{+} \big|\ist \bd{x}^{(j)}_{l}\!, \bd{x}^{(j)}_{\cl{C}_l}\!,\bd{x}_{\cl{T}_l}^{+ (j)}; \bd{u}^{+}_{{\rm r},\cl{C}_l}\big)$}
\label{sec:drawingDis}

\vspace{1.2mm}

In the setting of Section \ref{sub:Decentralized-processing}, samples $\big\{ \bd{x}_{l'}^{(j)} \big\}_{j=1}^J \!\sim f(\bd{x}_{l'})$, $l' \!\in\rmv \{ l \} \cup\ist \cl{C}_l$ and $\big\{ \bd{x}_{m}^{+(j)} \big\}_{j=1}^J \!\sim\rmv f(\bd{x}_{m}^{+})$, $m \rmv\in\rmv \cl{T}_l$ are available at CA $l$.
We start by noting that combining \eqref{eq:comp-meas-D-1} and \eqref{eq:meas_mod} yields
\vspace{-.5mm}
\be
\bd{y}_l^{+} =\ist 
\big[ d_l(\bd{x}_{l}^{+}\!, \bd{x}_{k}^{+}, \bd{v}_{l,k}^{+})\big]_{k\in\cl{A}_{l}} \ist.
\label{eq:comp-meas-D_2}
\vspace{.5mm}
\ee
Based on the analogy of this expression to \eqref{eq:comp-meas-D}, CA $l$ first obtains samples 
$\big\{ \bd{y}_l^{+(j,j')} \big\}_{j'=1}^{J'} \!\sim\rmv f\big(\bd{y}_l^{+} \big|\ist \bd{x}^{(j)}_{l}\!,\bd{x}^{(j)}_{\cl{C}_l}\!,\bd{x}_{\cl{T}_l}^{+ (j)};\bd{u}^{+}_{{\rm r},\cl{C}_l}\big)$
by carrying out the steps of Appendix \ref{sec:drawingCen} with obvious modifica\-tions---in particular, $\bd{y}^+$ is replaced by $\bd{y}_l^+\rmv$, $\cl{C}$ by $\{ l \} \cup\ist \cl{C}_l$, and $\cl{T}$ by $\cl{T}_l$\ist. More specifically, CA $l$ obtains samples $\big\{\bd{x}_{l'}^{+(j)}\big\}_{j=1}^J$ for $l' \!\in\rmv \{ l \} \cup \cl{C}_l$ according to \eqref{eq:sampleY}.
Then, for each $j \rmv\in \{1,\dots,J\}$, CA $l$ draws samples $\big\{ \bd{v}_{l,k}^{+(j,j')} \big\}_{j'=1}^{J'} \!\sim\rmv f(\bd{v}_{l,k}^{+})$ for $k \rmv\in\rmv \cl{A}_l$ and, 
in turn, obtains samples $\big\{ \bd{y}_l^{+(j,j')} \big\}_{j'=1}^{J'}$ by evaluating \eqref{eq:comp-meas-D_2} using the appropriate samples, 
\vspace{.5mm}
i.e.,
\[
\bd{y}_l^{+(j,j')} \ist=\ist 
\big[ d_l\big(\bd{x}_{l}^{+(j)}\!, \bd{x}_{k}^{+(j)}\!, \bd{v}_{l,k}^{+(j,j')} \big) \big]_{k\in\cl{A}_{l}} \ist, \quad\!\rmv j' \!=\rmv 1,\dots,J' \rmv.
\vspace{-.5mm}
\]

It remains to obtain samples of those entries of $\tilde{\bd{y}}_l^{+}$ that are not contained in $\bd{y}_l^{+}$ (cf. \eqref{eq:comp-meas-D-tilde} and \eqref{eq:comp-meas-D-1}).
More specifically, for each sample $\bd{y}_{l}^{+(j,j')}\!$, CA $l$ needs to obtain samples $\bd{y}_{l'\!,l}^{+(j,j')}\!$, $l' \in \cl{C}_l$
.
This is done through communication with neighbor CAs: CA $l$ transmits to each neighbor CA $l' \!\in \cl{C}_l$ the samples $\big\{ \bd{y}_{l,l'}^{+(j,j')} \big\}_{j'=1}^{J'}$, $j = 1,\dots,J$,
and it receives from CA $l' \!\in \cl{C}_l$ the samples $\big\{ \bd{y}_{l'\!,l}^{+(j,j')} \big\}_{j'=1}^{J'}$, $j = 1,\dots,J$.
Thus, finally, samples $\big\{ \tilde{\bd{y}}_l^{+(j,j')} \big\}_{j'=1}^{J'} \!\sim\rmv f\big(\tilde{\bd{y}}_l^{+} \big|\ist \bd{x}^{(j)}_{l}\!, \bd{x}^{(j)}_{\cl{C}_l}\!,\bd{x}_{\cl{T}_l}^{+ (j)};\bd{u}^{+}_{{\rm r},\cl{C}_l}\big)$ are locally available at CA $l$.


\section*{Acknowledgment} 

The authors would like to thank Dr. G\"unther Koliander for illuminating discussions.

\vspace*{1mm}

\bibliographystyle{ieeetr_noParentheses}
\bibliography{references}

\begin{thebibliography}{10}

\bibitem{shima2009uav}
T.~Shima and S.~Rasmussen, {\em UAV Cooperative Decision and Control:
  Challenges and Practical Approaches}.
\newblock Philadelphia, PA: SIAM, 2009.

\bibitem{bullo2009distributed}
F.~Bullo, J.~Cort{\'e}s, and S.~Mart{\'\i}nez, {\em Distributed Control of
  Robotic Networks: A Mathematical Approach to Motion Coordination Algorithms}.
\newblock Princeton, NJ: Princeton University Press, 2009.

\bibitem{nayak2010wireless}
A.~Nayak and I.~Stojmenovi{\'c}, {\em {Wireless Sensor and Actuator Networks:
  Algorithms and Protocols for Scalable Coordination and Data Communication}}.
\newblock Hoboken, NJ: Wiley, 2010.

\bibitem{zhao2004wsn}
F.~Zhao and L.~J. Guibas, {\em Wireless Sensor Networks: An Information
  Processing Approach}.
\newblock Amsterdam, The Netherlands: Morgan Kaufmann, 2004.

\bibitem{corke10}
P.~Corke, T.~Wark, R.~Jurdak, W.~Hu, P.~Valencia, and D.~Moore,
  ``{Environmental wireless sensor networks},'' {\em Proc. IEEE}, vol.~98,
  no.~11, pp.~1903--1917, 2010.

\bibitem{ko10}
J.~Ko, C.~Lu, M.~B. Srivastava, J.~A. Stankovic, A.~Terzis, and M.~Welsh,
  ``{Wireless sensor networks for healthcare},'' {\em Proc. IEEE}, vol.~98,
  no.~11, pp.~1947--1960, 2010.

\bibitem{hlinkaMag13}
O.~Hlinka, F.~Hlawatsch, and P.~M. Djuric, ``{Distributed particle filtering in
  agent networks: A survey, classification, and comparison},'' {\em IEEE Signal
  Process. Mag.}, vol.~30, no.~1, pp.~61--81, 2013.

\bibitem{zhao07}
T.~Zhao and A.~Nehorai, ``{Distributed sequential Bayesian estimation of a
  diffusive source in wireless sensor networks},'' {\em IEEE Trans. Signal
  Process.}, vol.~55, no.~4, pp.~1511--1524, 2007.

\bibitem{aghajan2009multi}
H.~Aghajan and A.~Cavallaro, {\em Multi-Camera Networks: Principles and
  Applications}.
\newblock Burlington, MA: Academic Press, 2009.

\bibitem{kim12}
K.-D. Kim and P.~R. Kumar, ``{Cyber physical systems: A perspective at the
  centennial},'' {\em Proc. IEEE}, vol.~100, pp.~1287--1308, 2012.

\bibitem{haykin12}
S.~Haykin, {\em {Cognitive Dynamic Systems: Perception-Action Cycle, Radar and
  Radio}}.
\newblock New York, NY: Cambridge University Press, 2012.

\bibitem{ryan07}
A.~D. Ryan, H.~Durrant-Whyte, and J.~K. Hedrick, ``Information-theoretic sensor
  motion control for distributed estimation,'' in {\em Proc. IMECE '07},
  Seattle, WA, Nov. 2007.

\bibitem{hoffmann10}
G.~M. Hoffmann and C.~J. Tomlin, ``{Mobile sensor network control using mutual
  information methods and particle filters},'' {\em IEEE Trans. Autom.
  Control}, vol.~55, no.~1, pp.~32--47, 2010.

\bibitem{schwager11}
M.~Schwager, P.~Dames, D.~Rus, and V.~Kumar, ``{A multi-robot control policy
  for information gathering in the presence of unknown hazards},'' in {\em
  Proc. ISRR-11}, Flagstaff, AZ, Aug. 2011.

\bibitem{julian12}
B.~J. Julian, M.~Angermann, M.~Schwager, and D.~Rus, ``{Distributed robotic
  sensor networks: An information-theoretic approach},'' {\em Int. J. Robot.
  Res.}, vol.~31, pp.~1134--1154, Sep. 2012.

\bibitem{atanasov14}
N.~A. Atanasov, J.~L. Ny, and G.~J. Pappas, ``Distributed algorithms for
  stochastic source seeking with mobile robot networks,'' {\em J. Dyn. Sys.,
  Meas., Control}, vol.~137, Mar. 2015.

\bibitem{wymeersch}
H.~Wymeersch, J.~Lien, and M.~Z. Win, ``{Cooperative localization in wireless
  networks},'' {\em Proc. IEEE}, vol.~97, pp.~427--450, Feb. 2009.

\bibitem{sathyan13}
T.~Sathyan and M.~Hedley, ``{Fast and accurate cooperative tracking in wireless
  networks},'' {\em IEEE Trans. Mobile Comput.}, vol.~12, no.~9,
  pp.~1801--1813, 2013.

\bibitem{wu11}
Y.-C. Wu, Q.~M. Chaudhari, and E.~Serpedin, ``{Clock synchronization of
  wireless sensor networks},'' {\em IEEE Signal Process. Mag.}, vol.~28, Jan.
  2011.

\bibitem{meyer13sync}
F.~Meyer, B.~Etzlinger, F.~Hlawatsch, and A.~Springer, ``{A distributed
  particle-based belief propagation algorithm for cooperative simultaneous
  localization and synchronization},'' in {\em Proc. Asilomar Conf. Sig.,
  Syst., Comput.}, Pacific Grove, CA, pp.~527--531, Nov. 2013.

\bibitem{etzlinger13asilomar}
B.~Etzlinger, F.~Meyer, A.~Springer, F.~Hlawatsch, and H.~Wymeersch,
  ``{Cooperative simultaneous localization and synchronization: A distributed
  hybrid message passing algorithm},'' in {\em Proc. Asilomar Conf. Sig.,
  Syst., Comput.}, Pacific Grove, CA, pp.~1978--1982, Nov. 2013.

\bibitem{etzlinger14}
B.~Etzlinger, H.~Wymeersch, and A.~Springer, ``Cooperative synchronization in
  wireless networks,'' {\em IEEE Trans. Signal Process.}, vol.~62,
  pp.~2837--2849, Jun. 2014.

\bibitem{teng2012distr}
J.~Teng, H.~Snoussi, C.~Richard, and R.~Zhou, ``{Distributed variational
  filtering for simultaneous sensor localization and target tracking in
  wireless sensor networks},'' {\em IEEE Trans. Veh. Technol.}, vol.~61, no.~5,
  pp.~2305--2318, 2012.

\bibitem{meyer12}
F.~Meyer, E.~Riegler, O.~Hlinka, and F.~Hlawatsch, ``{Simultaneous distributed
  sensor self-localization and target tracking using belief propagation and
  likelihood consensus},'' in {\em Proc. 46th Asilomar Conf. Sig., Syst.,
  Comp.}, Pacific Grove, CA, pp.~1212--1216, Nov. 2012.

\bibitem{meyer2014coslat}
F.~Meyer, O.~Hlinka, H.~Wymeersch, E.~Riegler, and F.~Hlawatsch, ``Cooperative
  simultaneous localization and tracking in mobile agent networks.'' \emph{IEEE
  Trans.\ Signal Process.}, 2014, submitted.
\newblock Available online: http://arxiv.org/abs/1403.1824.

\bibitem{olfati07}
R.~Olfati-Saber, J.~A. Fax, and R.~M. Murray, ``{Consensus and cooperation in
  networked multi-agent systems},'' {\em Proc. IEEE}, vol.~95, no.~1,
  pp.~215--233, 2007.

\bibitem{dimakis10}
A.~G. Dimakis, S.~Kar, J.~M.~F. Moura, M.~G. Rabbat, and A.~Scaglione,
  ``{Gossip algorithms for distributed signal processing},'' {\em Proc. IEEE},
  vol.~98, pp.~1847--1864, Nov. 2010.

\bibitem{kschischang}
F.~R. Kschischang, B.~J. Frey, and H.-A. Loeliger, ``{Factor graphs and the
  sum-product algorithm},'' {\em IEEE Trans. Inf. Theory}, vol.~47,
  pp.~498--519, Feb. 2001.

\bibitem{bishop2006pattern}
C.~M. Bishop, {\em {Pattern Recognition and Machine Learning}}.
\newblock New York, NY: Springer, 2006.

\bibitem{burgard97}
W.~Burgard, D.~Fox, and S.~Thrun, ``Active mobile robot localization by entropy
  minimization,'' in {\em Proc. EUROBOT '97}, Brescia, Italy, pp.~155--162,
  Oct. 1997.

\bibitem{grocholsky02}
B.~Grocholsky, {\em {Information-Theoretic Control of Multiple Sensor
  Platforms}}.
\newblock PhD thesis, University of Sydney, Sydney, Australia, 2002.

\bibitem{cover06}
T.~M. Cover and J.~A. Thomas, {\em Elements of Information Theory}.
\newblock New York, NY: Wiley, 2006.

\bibitem{kay}
S.~M. Kay, {\em {Fundamentals of Statistical Signal Processing: Estimation
  Theory}}.
\newblock Upper Saddle River, NJ: Prentice-Hall, 1993.

\bibitem{morbidi13}
F.~Morbidi and G.~L. Mariottini, ``{Active target tracking and cooperative
  localization for teams of aerial vehicles},'' {\em IEEE Trans. Control Syst.
  Technol.}, vol.~21, no.~5, pp.~1694--1707, 2013.

\bibitem{rong}
X.~R. Li and V.~P. Jilkov, ``{Survey of maneuvering target tracking. Part I:
  Dynamic models},'' {\em IEEE Trans. Aerosp. Electron. Syst.}, vol.~39,
  pp.~1333--1364, Oct. 2003.

\bibitem{barShalom01}
Y.~Bar-Shalom, X.-R. Li, and T.~Kirubarajan, {\em {Estimation with applications
  to tracking and navigation}}.
\newblock New York, NY: Wiley, 2001.

\bibitem{lien}
J.~Lien, J.~Ferner, W.~Srichavengsup, H.~Wymeersch, and M.~Z. Win, ``A
  comparison of parametric and sample-based message representation in
  cooperative localization.'' \emph{Int.\ J.\ Navig.\ Observ.}, 2012.

\bibitem{farahmand}
S.~Farahmand, S.~I. Roumeliotis, and G.~B. Giannakis, ``Set-member\-ship
  constrained particle filter: Distributed adaptation for sensor networks,''
  {\em IEEE Trans. Signal Process.}, vol.~59, pp.~4122--4138, Sep. 2011.

\bibitem{savic14}
V.~Savic, H.~Wymeersch, and S.~Zazo, ``{Belief consensus algorithms for fast
  distributed target tracking in wireless sensor networks},'' {\em Signal
  Processing}, vol.~95, pp.~149--160, 2014.

\bibitem{hlinka14adaptation}
O.~Hlinka, F.~Hlawatsch, and P.~M. Djuric, ``Consensus-based distributed
  particle filtering with distributed proposal adaptation,'' {\em IEEE Trans.
  Signal Process.}, vol.~62, pp.~3029--3041, Jun. 2014.

\bibitem{lindberg13}
C.~Lindberg, L.~S. Muppirisetty, K.-M. Dahlen, V.~Savic, and H.~Wymeersch,
  ``{MAC delay in belief consensus for distributed tracking},'' in {\em Proc.
  IEEE WPNC-13}, Dresden, Germany, Mar. 2013.

\bibitem{fletcher87}
R.~Fletcher, {\em Practical Methods of Optimization}.
\newblock New York, NY: Wiley, 1987.

\bibitem{doucet}
A.~Doucet, N.~de~Freitas, and N.~Gordon, {\em {Sequential Monte Carlo Methods
  in Practice}}.
\newblock New York, NY: Springer, 2001.

\bibitem{lim00}
H.~Lim and C.~Kim, ``{Multicast tree construction and flooding in wireless ad
  hoc networks},'' in {\em Proc. MSWIM '00}, New York, NY, pp.~61--68, Aug.
  2000.

\bibitem{xiao05}
L.~Xiao, S.~Boyd, and S.~Lall, ``A scheme for robust distributed sensor fusion
  based on average consensus,'' in {\em Proc. IPSN '05}, Los Angeles, CA,
  pp.~63--70, Apr. 2005.

\bibitem{bartle95}
R.~G. Bartle, {\em {The Elements of Integration and Lebesgue Measure}}.
\newblock New York, NY: Wiley, 1995.

\bibitem{thrun05}
S.~Thrun, W.~Burgard, and D.~Fox, {\em Probabilistic Robotics}.
\newblock Cambridge, MA: MIT Press, 2005.

\bibitem{ihler}
A.~T. Ihler, J.~W. Fisher, R.~L. Moses, and A.~S. Willsky, ``{Nonparametric
  belief propagation for self-localization of sensor networks},'' {\em IEEE J.
  Sel. Areas Commun.}, vol.~23, pp.~809--819, Apr. 2005.

\bibitem{ristic}
B.~Ristic, S.~Arulampalam, and N.~Gordon, {\em {Beyond the Kalman Filter:
  Particle Filters for Tracking Applications}}.
\newblock Norwood, MA: Artech House, 2004.

\bibitem{garcia14tradeoff}
G.~E. Garcia, L.~S. Muppirisetty, E.~M. Schiller, and H.~Wymeersch, ``{On the
  trade-off between accuracy and delay in cooperative UWB localization:
  Performance bounds and scaling laws},'' {\em IEEE Trans. Wireless Commun.},
  vol.~13, pp.~4574--4585, Aug. 2014.

\bibitem{douc05}
R.~Douc and O.~Cappe, ``{Comparison of resampling schemes for particle
  filtering},'' in {\em Proc. ISPA-05}, Zagreb, Croatia, pp.~64--69, Sep. 2005.

\bibitem{silverman}
B.~W. Silverman and P.~J. Green, {\em {Density Estimation for Statistics and
  Data Analysis}}.
\newblock London, UK: Chapman and Hall, 1986.

\bibitem{mayne00}
D.~Q. Mayne, J.~B. Rawlings, C.~V. Rao, and P.~O.~M. Scokaert, ``{Constrained
  model predictive control: Stability and optimality},'' {\em Automatica},
  vol.~36, pp.~789--814, Jun. 2000.

\bibitem{alspach72}
D.~L. Alspach and H.~W. Sorenson, ``{Nonlinear Bayesian estimation using
  Gaussian sum approximations},'' {\em IEEE Trans. Autom. Control}, vol.~17,
  pp.~439--448, Aug. 1972.

\bibitem{arasaratnam09}
I.~Arasaratnam and S.~Haykin, ``{Cubature Kalman filters},'' {\em IEEE Trans.
  Autom. Control}, vol.~54, pp.~1254--1269, Jun. 2009.

\bibitem{deville12}
Y.~Deville and A.~Deville, ``Exact and approximate quantum independent
  component analysis for qubit uncoupling,'' in {\em Proc. LVA/ICA '12}, Tel
  Aviv, Israel, pp.~58--65, Mar. 2012.

\end{thebibliography}

\end{document}